\documentclass[fleqn,usenatbib,usedcolumn]{mnras}

\usepackage[T1]{fontenc}
\usepackage{ae,aecompl}
\usepackage{graphicx}

\newcommand{\kms}{$\,\rm{km\,s^{-1}}$}
\newcommand{\msolar}{$\rm{M}_{\sun}$}
\newcommand{\rtwo}{$\rm{R}_{200}$}

\usepackage{graphicx}	
\usepackage{amsmath}	
\usepackage{amssymb}	
\title[The SAMI Galaxy Survey: The cluster redshift survey and target selection]{The SAMI Galaxy Survey: The cluster redshift survey, target selection and cluster properties. }

\author[M. S. Owers et al.]{ M. S. Owers$^{1,2}$\thanks{E-mail: matt.owers@mq.edu.au},
J. T. Allen$^{3,4}$,
I. Baldry$^{5}$,
J. J. Bryant$^{2,3,4}$,
G. N. Cecil$^{6}$,
L. Cortese$^{7}$,
\newauthor{
S. M. Croom$^{3,4}$,
S. P. Driver$^{7}$,
L. M. R. Fogarty$^{3,4}$,
A. W. Green$^{2}$,
E. Helmich$^{8}$, 
}
\newauthor{
J. T. A. de Jong$^{8}$,
K. Kuijken$^{8}$,
S. Mahajan$^{9}$,
J. McFarland$^{10}$,
M. B. Pracy$^{3}$,
}
\newauthor{
A. G. S. Robotham$^{7}$,
G. Sikkema$^{10}$,
S. Sweet$^{11}$,
E. N. Taylor$^{12}$, 
G. Verdoes Kleijn$^{8}$,
}
\newauthor{
A. E. Bauer$^{2}$,
J. Bland-Hawthorn$^{3}$,
S. Brough$^{2}$,
M. Colless$^{11}$,
W. J. Couch$^{2}$,
}
\newauthor{
R. L Davies$^{13}$,
M. J. Drinkwater$^{4,14}$,
M. Goodwin$^{2}$, 
A. M. Hopkins$^{2}$,
}
\newauthor{
I. S. Konstantopoulos$^{2,15}$,
C. Foster$^{2}$,
J. S. Lawrence$^{2}$,
N. P. F Lorente$^{2}$,
}
\newauthor{
A. M. Medling$^{11,16,17}$,
N. Metcalfe$^{18}$,
S. N. Richards$^{2,3,4}$,
J. van de Sande$^{4}$,
N. Scott$^{4}$,
}
\newauthor{
T. Shanks$^{18}$,
R. Sharp$^{11}$,
A. D. Thomas$^{11}$
and C. Tonini$^{19}$ 
}
\\
\\
Author affiliations are listed at the end of the paper
}

\date{Accepted XXX. Received YYY; in original form ZZZ}

\pubyear{2016}

\begin{document}
\label{firstpage}
\pagerange{\pageref{firstpage}--\pageref{lastpage}}
\maketitle

\begin{abstract}
We describe the selection of galaxies targeted in eight low redshift clusters (APMCC0917, A168, A4038, EDCC442, A3880, A2399, A119 and A85; $0.029 < z < 0.058$) as part of the Sydney-AAO Multi-Object integral field Spectrograph Galaxy Survey (SAMI-GS). We have conducted a redshift survey of these clusters using the AAOmega multi-object spectrograph on the 3.9m Anglo-Australian Telescope. The redshift survey is used to determine cluster membership and to characterise the dynamical properties of the clusters. In combination with existing data, the survey resulted in 21,257 reliable redshift measurements and 2899 confirmed cluster member galaxies. Our redshift catalogue has a high spectroscopic completeness ($\sim 94\%$) for $r_{\rm petro} \leq 19.4$ and clustercentric distances $R< 2$\rtwo. We use the confirmed cluster member positions and redshifts to determine cluster velocity dispersion, \rtwo, virial and caustic masses, as well as cluster structure. The clusters have virial masses $14.25 \leq {\rm log }({\rm M}_{200}/$\msolar$) \leq 15.19$. The cluster sample exhibits a range of dynamical states, from relatively relaxed-appearing systems, to clusters with strong indications of merger-related substructure. Aperture- and PSF-matched photometry are derived from SDSS and VST/ATLAS imaging and used to estimate stellar masses. These estimates, in combination with the redshifts, are used to define the input target catalogue for the cluster portion of the SAMI-GS. The primary SAMI-GS cluster targets have $R< $\rtwo, velocities $|v_{\rm pec}| < 3.5\sigma_{200}$ and stellar masses  $9.5 \leq {\rm log(M}^*_{approx}/$\msolar)$ \leq 12$. Finally, we give an update on the SAMI-GS progress for the cluster regions.
\end{abstract}

\begin{keywords}
galaxies: clusters: individual (APMCC0917, A168, A4038, EDCC442, A3880, A2399, A119, A85) -- surveys -- galaxies
\end{keywords}



\section{Introduction}

Toward the end of the last century, large-area redshift surveys of statistically representative volumes of the nearby Universe were enabled by the advent of wide-field, highly multiplexed fibre-fed spectrographs capable of simultaneously collecting several hundred spectra. Surveys such as the 2-degree Field Galaxy Redshift Survey \citep[2dFGRS;][]{colless2001} and the Sloan Digital Sky Survey \citep[SDSS;][]{york2000} have been pivotal both in characterising galaxy environment and in precisely defining how fundamental galaxy properties such as luminosity, morphology, level of star formation, colour, gas-phase metallicity, stellar mass and nuclear activity correlate with the external environment on both large ($\sim$Mpc) and small ($\sim$kpc) scales \citep{lewis2002, norberg2002, bell2003, brinchmann2004, tremonti2004, kauffmann2003a, kauffmann2003b, croton2005,  baldry2006, peng2010}. The dominant physical mechanisms governing these correlations have to date remained elusive. 

Massive galaxy clusters are critical to understanding correlations between galaxy properties and environment; they host the densest environments where the effects of many of the physical mechanisms capable of galaxy transformation are strongest and, therefore, are expected to be more readily observed. The potential mechanisms that can act to transform a cluster galaxy are well known \citep[for an overview, see][]{boselli2006}. Interactions with the hot intracluster medium (ICM), such as ram-pressure and viscous stripping \citep{gunn1972,nulsen1982} can remove the cold HI gas that fuels star formation or the hot gas halo reservoir \citep[strangulation; ][]{larson1980, bekki2002}, thereby leading to quenching of star formation with little impact on stellar structure. The effect of gravitational interactions, through either tides due to the cluster potential \citep{byrd1990, bekki1999}, high-speed interactions between other cluster galaxies, or the combination of both \citep[harrassment; ][]{moore1996}, can impact both the distribution of old stars and the gas in a cluster galaxy, leading to transformations in morphological, kinematical, star-forming, and AGN properties of cluster galaxies \citep{byrd1990, bekki1999}. A large fraction of galaxies accreted onto clusters arrive in group-scale halos ($M_{200} < 10^{14}$\msolar)  \citep{mcgee2009}, where galaxy mergers and interactions can pre-process a galaxy before it falls into a cluster. The amplitude of the effect of these mechanisms is likely a function of parameters related to environment including cluster halo mass, ICM properties, and cluster merger activity, as well as intrinsic galaxy properties such as mass, morphology and gas content. 

Deep, complete multi-object spectroscopic observations of galaxy clusters allow the efficient collection of a large number of spectroscopically confirmed cluster members. These member galaxies are important kinematical probes of the cluster potential, allowing for relatively reliable dynamical mass determinations based on common estimators such as the velocity dispersion-based virial estimator \citep[][]{girardi1998}, the escape velocity profile-based caustic technique \citep[][]{diaferio1999} and by fitting the 2D projected-phase-space distribution \citep[][]{mamon2013} to name a few \citep[for a comprehensive analysis of different estimators, see;][]{old2014, old2015}. Many dynamical mass estimators assume spherical symmetry and dynamical equilibrium; these assumptions are violated during major cluster mergers, thereby affecting the accuracy of mass measurements. Substructure related to cluster merger activity is routinely detected and characterised using the combined redshift and position information for cluster members \citep[][]{dressler1988,colless1996,pinkney1996,pisani1996,ramella2007,owers2009a,owers2009b,owers2011a,owers2011b,owers2013}. Multi-object spectroscopic observations of clusters are therefore an important part of the tool-kit for characterising the global cluster environment, as well as the local environmental properties surrounding a galaxy. 

The observable imprint of the processes responsible for {\it environment-driven} galaxy transformation can reveal itself through spatially resolved spectroscopic observations \citep[e.g.,][]{pracy2012,merluzzi2013,brough2013,bekki2014, schaefer2017}. Therefore, crucial to understanding which of the environment-related physical mechanisms are at play is knowledge of the resolved properties of galaxies spanning a range in mass, in combination with a detailed description of the galaxy environment. The ongoing SAMI Galaxy Survey \citep[SAMI-GS;][]{Bland2011,croom2012, Bryant2014} is, for the first time, addressing this issue by obtaining resolved spectroscopy for a large sample of galaxies \citep{bryant2015, allen2015}. The SAMI-GS is primarily targeting galaxies selected from the Galaxy And Mass Assembly survey  \citep[GAMA; ][]{driver2009,driver2011, liske2015}, where deep, highly complete spectroscopy allows high fidelity environment metrics to be formulated \citep[e.g., local density and group membership][]{robotham2011,brough2013}. The SAMI-GS will collect resolved spectroscopy for $\sim 2700$ galaxies residing in the GAMA regions. However, at the low redshifts targeted for the SAMI-GS, the volume probed by the GAMA regions contain few rare, rich cluster-scale halos found in the high mass portion of the mass function. To probe the {\it full} range of galaxy environments, the SAMI-GS is also targeting $\sim 900$ galaxies in the eight massive ($M > 10^{14}$\msolar) clusters APMCC0917, A168, A4038, EDCC442, A3880, A2399, A119 and A85. For the majority of these clusters, only relatively shallow ($r <17.77,\,b_{\rm J} < 19.45$ for the SDSS and 2dFGRS, respectively), intermediate completeness ($\sim 80-90\%$) spectroscopy was available \citep[][]{depropris2002,rines2006}. To address the disparity in redshift depth and completeness between the GAMA regions and the dense cluster regions, we have conducted a redshift survey of the cluster regions using the AAOmega multi-object spectrograph on the 3.9m Anglo-Australian Telescope. The results and analysis of this survey are presented in this paper.

This paper provides details on the densest regions probed in the SAMI-GS: the cluster regions. In Section~\ref{cluster_sel} we outline the selection of the 8 cluster regions. In Section~\ref{sami_crs} we outline the SAMI Cluster Redshift Survey (SAMI-CRS) which we use to define cluster properties (Section~\ref{cluster_properties}). We then outline the updated photometry for cluster galaxies and the selection process for SAMI targets in the cluster regions. Finally, in 
Section~\ref{survey_progress}, we outline the SAMI-GS progress in the cluster regions. Throughout this paper, we assume a standard $\Lambda{\rm CDM}$ cosmology with $\rm{H}_0 = 70\,{\rm km\, s}^{-1}\, {\rm Mpc}^{-1}, \Omega_m=0.3, \Omega_{\Lambda}=0.7$.

\section{Selection of SAMI clusters}\label{cluster_sel}

Because the space density of massive clusters is low \citep[n($M > 1\times 10^{14}\,$\msolar)$ \,\sim 10^{-5}\,{\rm Mpc}^{-3}$;][]{murray2013}\footnote{http://hmf.icrar.org} and the equatorial GAMA regions targeted by the SAMI Galaxy Survey (hereafter SAMI-GS) probes $3.6 \times 10^5 \, {\rm Mpc}^3$ for $z <0.1$\footnote{http://cosmocalc.icrar.org}, there will be too few massive clusters in the main SAMI-GS volume to probe the densest galaxy environments. Therefore, we utilise the wide-area 2dFGRS and SDSS to select a number of cluster regions to include in the main survey.  The clusters are drawn from clusters within the 2dFGRS from the catalogue of \citet{depropris2002}, and also from clusters used in the Cluster Infall Regions in the SDSS (CIRS) survey of \citet{rines2006}. The initial selection of the clusters was based on the following criteria:
\begin{itemize}
\item $z \leq 0.06\,$ so that a significant portion of the galaxy luminosity/mass function can be probed in the cluster regions. For the limiting magnitude of the SAMI-CRS ($r$=19.4; Section~\ref{sami_crs}) we  probe $\sim 3\,$mag fainter than the knee in the cluster luminosity function \citep[$M^*_r = -20.6$;][]{popesso2005}. At this redshift, the stellar mass limit for the SAMI-GS is log$_{10} ({\rm M}^*/$\msolar)$\,>$ 10 (Section~\ref{SAMI_TS}). We therefore probe at least a factor of 50 in stellar mass when compared with the most massive cluster galaxies (log$_{10} ({\rm M}^*/$\msolar) $\sim 11.6$);

\item Sufficient spectroscopy to clearly define boundaries in the peculiar velocity-radius phase-space diagrams (Figure~\ref{mem_allocation}). For the clusters selected from the \citet{depropris2002} catalogue, this criterion was achieved by selecting only clusters with more than 50 members and where the spectroscopic completeness of the tile was $>70\%$. For the clusters selected from CIRS, we require that the infall pattern in the cluster velocity-radius phase-space diagram be classified as ``clean'' in the visual classification scheme provided by \citet{rines2006}.

\item  R.A. in the range $20-10$hr and declination $<5\deg$. This requirement meant that the clusters were observable for a significant portion of the night from the AAT during Semester B, which runs August to January. This constraint meant that the clusters did no overlap in R.A. with the GAMA portion of the SAMI-GS that is observed during Semester A.

\end{itemize}
The above selection criteria resulted in 18 clusters in the 2dFGRS Southern Galactic Pole region and 7 clusters from CIRS. We re-analyse the 2dFGRS clusters by selecting members using the caustic technique (see Section~\ref{subsect:memsel}), defining \rtwo\, and velocity dispersion of galaxies within \rtwo, $\sigma_{200}$, (as described in Section~\ref{subsect:memsel}). We make a further cut of clusters with $\sigma_{200} < 450$\kms, which according to the scaling relation of \citet{evrard2008} are likely to have ${\rm log}_{10}({\rm M}_{200}/$\msolar) $< 14$. The SAMI-GS is already well-populated in this mass range \citep[see Figure~11 in][]{bryant2015}. This leaves six 2dFGRS clusters; two of these appeared to have irregular and non-Gaussian velocity distributions within \rtwo\, that may affect their dispersion measurements and so they were removed from the final sample. We also remove a further 3 CIRS clusters: two with $\sigma_{200} <450$\kms\, \citep[where the $\sigma_{200}$ values are given in ][]{rines2006}, and one for which all of the \cite{rines2006} mass measures are ${\rm log}_{10}({\rm M}/$\msolar) $< 14$. The remaining eight clusters make up the final cluster sample for the SAMI-GS: four from the 2dFGRS region and four from CIRS. The selected clusters are listed in Table~\ref{clus_table}.

\begin{table*}
 \centering
  \caption{Clusters selected for SAMI observations. Clusters are ordered in increasing mass (per the caustic estimate). The ${\rm N_{mem}}$ and ${\rm N_Z}$ values give the number of cluster member redshifts and total number of redshifts, respectively. The completeness is given for the limiting magnitude $r_{petro} < 19.4$. The ${\rm N_{mem},\, N_Z}$ and completeness values are presented for the limiting radii $R<$\rtwo\, and $R< 2$\rtwo, and the values areseparated by a "/".\label{clus_table}}
  \begin{tabular}{@{}lcccccccccccc@{}}
  \hline
   Name     &     R.A.      & Decl.  & z & $\sigma_{200}$  & $R_{200}$ &M$_{200}$ & M$_{200}$&${\rm N_{mem}}$ & $\rm{N_Z}$ & Compl.\\
            & (J2000)             & (J2000)      &          &  (r$<R_{200}$)   & (Mpc) &   ($10^{14}$\msolar) &  ($10^{14}$\msolar)  &  && per cent\\
             &    (deg.)     & (deg.)     &          & (km/s) &   & Caustic  & Virial  &  && \\
\hline

APMCC 917 & 355.397880 & -29.236351 & 0.0509 &  492$\pm$47 & 1.19 &  1.8$\pm$0.7 &  2.1$\pm$0.6 &      86/119& 255/654  & 96/92\\
Abell 168 &  18.815777 &   0.213486 & 0.0449 &  546$\pm$29 & 1.32 &  1.9$\pm$1.1 &  3.0$\pm$0.4 &     192/276& 505/1382 & 94/95\\
Abell 4038 & 356.937810 & -28.140661 & 0.0293 &  597$\pm$29 & 1.46 &  2.3$\pm$1.4 &  2.9$\pm$0.5 &     164/263 & 885/2408 & 97/91\\
EDCC 442 &   6.380680 & -33.046570 & 0.0498 &  583$\pm$39 & 1.41 &  2.8$\pm$1.7 &  3.6$\pm$0.7 &     123/243 & 279/927 & 91/94\\
Abell 3880 & 336.977050 & -30.575371 & 0.0578 &  660$\pm$46 & 1.59 &  4.4$\pm$1.3 &  4.6$\pm$1.1 &     160/307& 356/1151 & 99/99\\
Abell 2399 & 329.372605 &  -7.795692 & 0.0580 &  690$\pm$32 & 1.66 &  4.7$\pm$1.5 &  6.1$\pm$0.8 &     254/343& 544/1394 & 99/99\\
Abell 119 &  14.067150 &  -1.255370 & 0.0442 &  840$\pm$36 & 2.04 &  8.6$\pm$3.1 &  9.7$\pm$1.1 &     372/578& 835/2341 & 89/85\\
Abell  85 &  10.460211 &  -9.303184 & 0.0549 & 1002$\pm$28 & 2.42 & 15.5$\pm$3.7 & 17.0$\pm$1.3 &     590/772 & 1736/3132 & 98/94\\

\hline
\end{tabular}
\end{table*}

\section{The SAMI Cluster Redshift Survey}\label{sami_crs}

The target selection for the GAMA portion of the SAMI-GS sample benefits greatly from the deep, highly complete spectroscopy provided by the GAMA survey which was conducted on the 3.9m Anglo-Australian Telescope \citep{driver2011,liske2015}. This spectroscopy probes galaxies with much lower stellar mass when compared with the SAMI survey limits, allowing for a robust definition of the environment surrounding the SAMI survey galaxies \citep[e.g., the GAMA group catalogue provided by][]{robotham2011}. While the selection of the clusters for the SAMI survey required some level of spectroscopy to be available from the SDSS and 2dFGRS, both of these surveys only probe down to galaxies $\sim 2$ magnitude brighter than the GAMA survey limits, and do not have the same level of spectroscopic completeness, particularly in the dense cluster cores. In order to provide a similar level of high-fidelity spectroscopy for the dense cluster regions, we conducted a redshift survey of the eight regions: the SAMI Cluster Redshift Survey (SAMI-CRS). In this section, we describe the SAMI-CRS. 

\subsection{Input catalogue for spectroscopic follow-up}\label{input_cat}

\subsubsection{VST/ATLAS survey photometry (APMCC0917, EDCC0442, A3880 and A4038)}\label{old_vst}

Targets for the four clusters selected from the 2dFGRS catalogue \citep{depropris2002} were selected from photometry provided by the VLT Survey Telescope's ATLAS (VST/ATLAS) survey which is described in detail in \citet{shanks2013,shanks2015}. Briefly, $u, g, r, i\, {\rm and}\, z$-band photometric catalogues for fields with centres within 4.5\rtwo\ of the cluster centres were retrieved from the VST archive at the Cambridge Astronomy Survey Unit\footnote{http://casu.ast.cam.ac.uk/vstsp/imgquery/search} (CASU). These data were obtained prior to the public data release of the VST/ATLAS survey and the second-order corrections to the night-to-night photometric zeropoints of the different pointings \citep[described in][]{shanks2015}  had not yet been applied. To apply these corrections, we followed a method similar to that described in \citet{shanks2015}; we cross-match unsaturated stars detected in the VST/ATLAS data with stars in the APASS\footnote{https://www.aavso.org/apass} photometric survey of bright stars that have $10 < V < 17$ and compare their magnitudes in the $g,r$ and $i-$bands. Each of the separate $gri$ VST/ATLAS catalogues is then corrected by the mean difference between the APASS and VST/ATLAS star magnitudes. This accounts for both the night-to-night variations in zeropoints, as well as converting VST/ATLAS Vega magnitudes onto the AB magnitude system used by APASS. Since there are no corresponding $u$ and $z-$band measurements in APASS, we determine the corrections in those bands by minimising the offset between the stellar locus of the VST/ATLAS $(u-g)$ vs $(g-r)$ and $(r-i)$ vs $(i-z)$ colour-colour diagrams and that of SDSS-selected stars. The $u$- and $z$-band photometry were only used in the selection of spectrophotometric standards described in \citet{bryant2015}. The final parent photometric catalogues selected are based on the $r$-band detections and include only unsaturated, extended (non-stellar) sources with $r_{kron} < 20.5$. The VST/ATLAS survey strategy included a $\sim 2\arcmin$ overlap between adjacent pointings, which meant that many objects had duplicate photometric measurements. Those duplicates were identified as sources whose positions are within $1\arcsec$ and the measurement with the highest S/N ratio was retained. 

The VST/ATLAS photometric measurements include Kron and Petrosian fluxes measured in circular apertures, as well as a series of 13 fixed-aperture fluxes. The selection of our targets is based on the $r$-band Petrosian magnitudes (as an estimate of the total magnitude) and on their position on the $g-i$ versus $r_{petro}$ colour-magnitude diagram (Section~\ref{CMR_selection}). The flexible apertures used to determine the Kron and Petrosian fluxes are measured separately in each of the different image bands; therefore, the aperture sizes may differ between the bands, leading to biases in the colour measurements. To mitigate this, from the 13 fixed-aperture fluxes we select the one with aperture size closest to the $r$-band Kron radius and use that to determine the fixed-aperture magnitude in the $g,\, r\, {\rm and}\, i$-bands (although note that here we do not attempt to correct for the different seeing conditions for the different bands; this is  addressed in Section~\ref{new_photo}). There are several further shortcomings of the VST/ATLAS photometry that impact the photometric measurements of extended, bright objects in particular. The first is that the maximum aperture size through which fluxes are measured has a radius of $12\arcsec$, meaning that objects larger than this limit will have their flux measurements underestimated. Further to this, we found that the local sky background measurement around large objects (i.e., those with Kron radii larger than $\sim 4\arcsec$) is systematically higher than the median sky background measurement, indicating that source flux is included in the subtracted background for these objects. This leads to an over-subtraction of the background for large objects. While these systematics effects lead to an underestimation of the object flux, they do not greatly impact the selection of targets for the spectroscopic follow-up. In Section~\ref{new_photo} we address these issues for the selection of targets for the SAMI-GS, where more care is required in determining the total magnitudes and accurate colours.

\subsubsection{SDSS photometry (A85, A168, A119 and A2399)}\label{old_sdss}

The photometry used for the input catalogues for the clusters in the SDSS regions is taken from SDSS DR9 photometry \citep{ahn2012}. For each cluster, positions and photometry for objects classified as either a galaxy or star with $r<22$ and within $4^{\circ}$ of the cluster centre were retrieved from the {\sf CasJobs} server\footnote{http://skyserver.sdss.org/CasJobs}. As an estimate for total flux, we utilise the SDSS Petrosian magnitudes, while for colour estimates we use the {\sf model} magnitudes. These measurements are suitable for the purpose of target selection for the spectroscopic follow-up, although we note that the {\sf model} magnitudes may produce biased colour measurements in the presence of strong colour gradients \citep{taylor2011}. We also note that while A85 has coverage with both VST/ATLAS and SDSS photometry, we use the SDSS photometry for this cluster and present a comparison of the final photometric measurements between the two surveys in Section~\ref{sdss_vst_comp}.

\subsubsection{Selection of targets for spectroscopic follow-up}\label{CMR_selection}

\begin{figure*}
\includegraphics[angle=90,width=.45\textwidth]{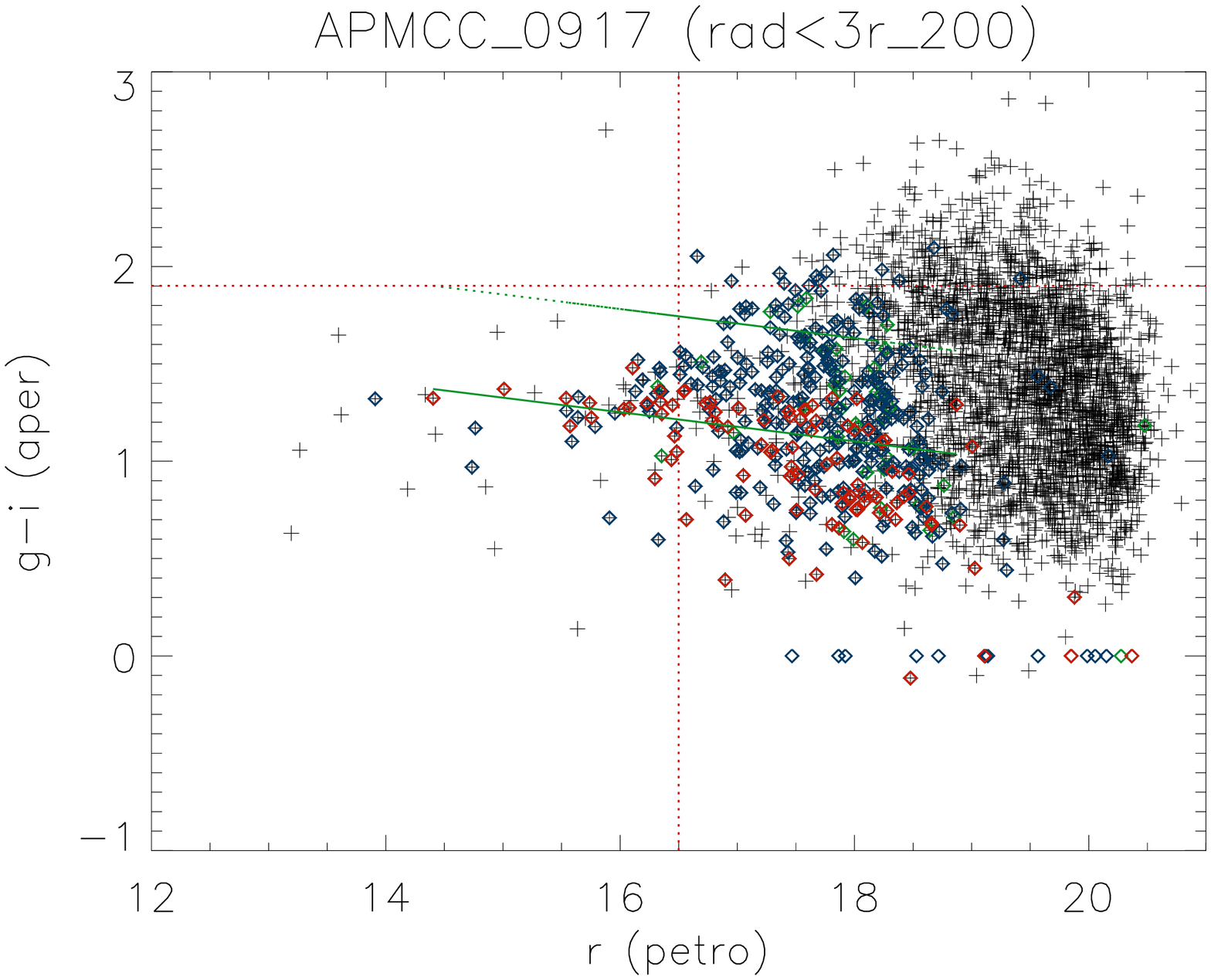}
\includegraphics[angle=90,width=.45\textwidth]{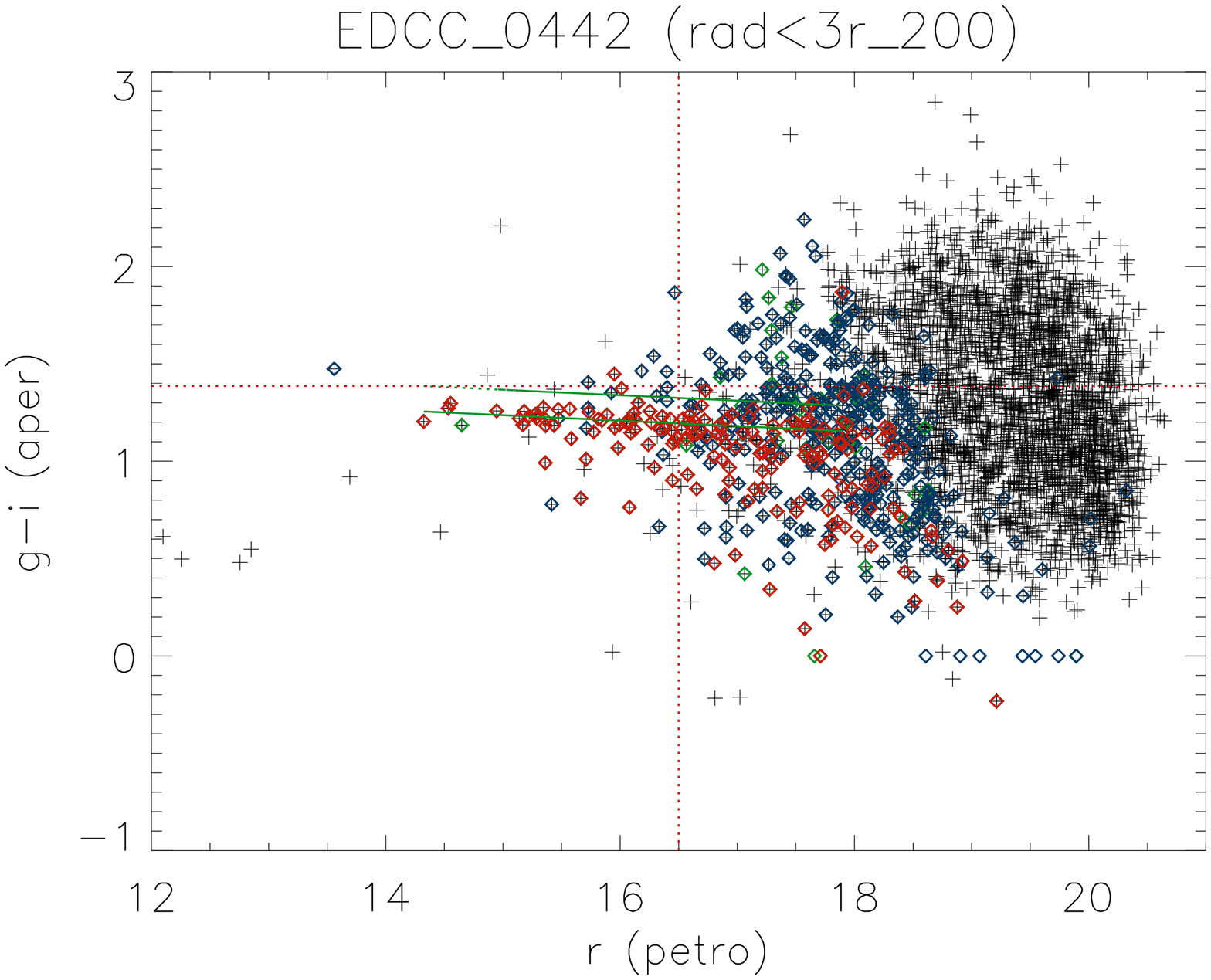}\\
\includegraphics[angle=90,width=.45\textwidth]{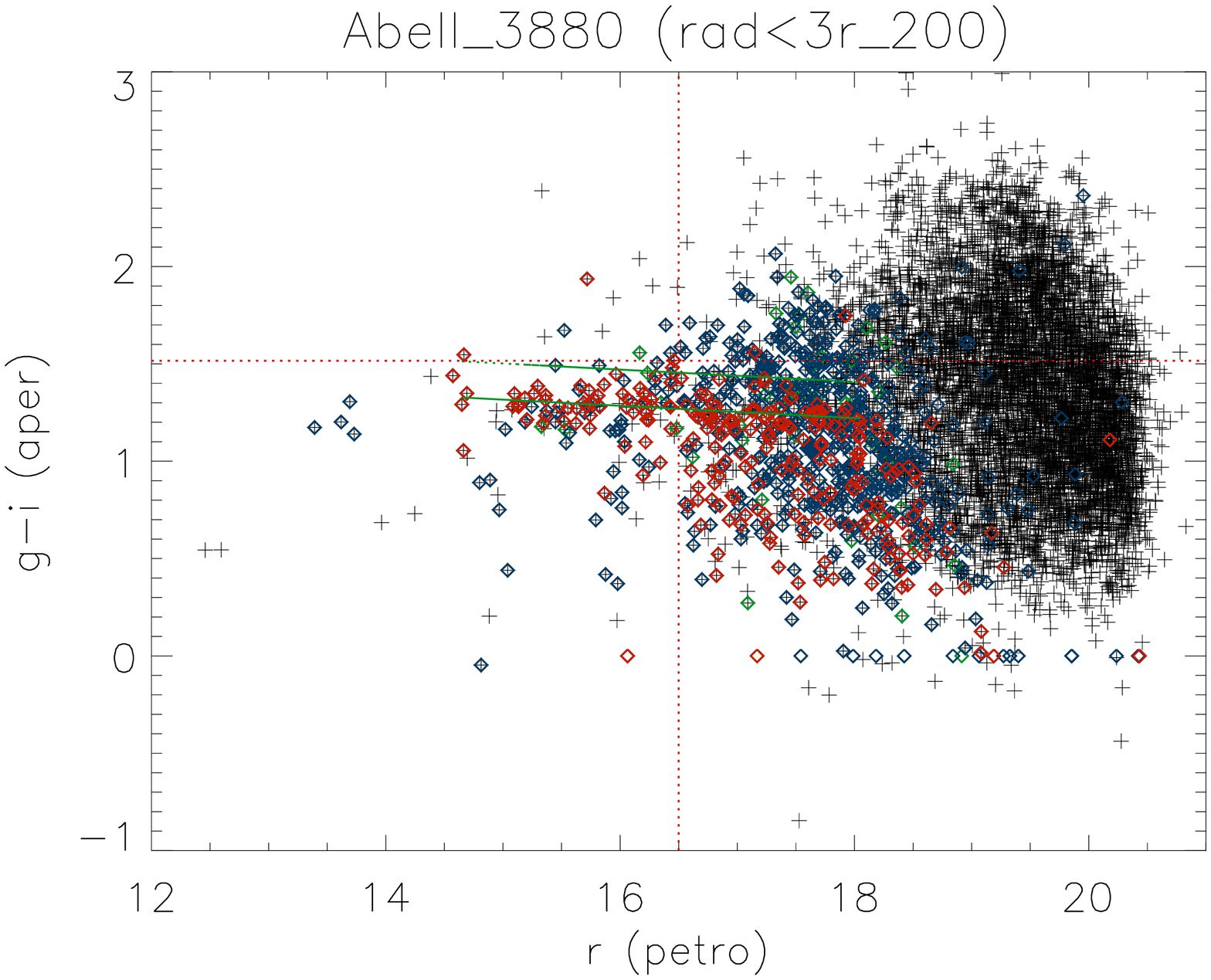}
\includegraphics[angle=90,width=.45\textwidth]{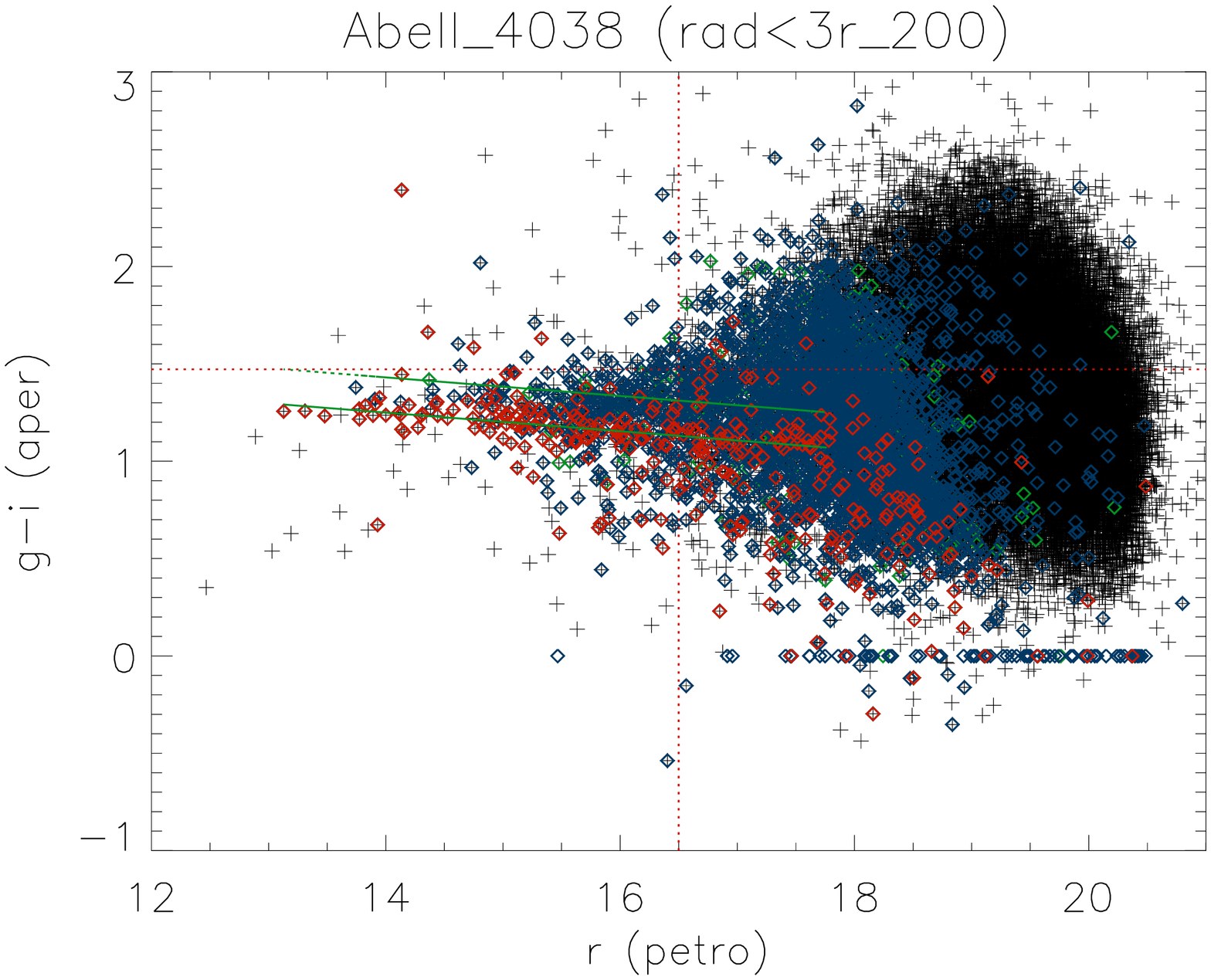}
\caption{Colour-magnitude diagrams for the SAMI-CRS clusters with VST/ATLAS photometry. The black plus symbols show all objects classified as galaxies within the field. The red open  diamond points show confirmed cluster members. Blue open diamonds show fore- and background galaxies with existing spectra from the 2dFGRS or 6dFGS. The lower green line shows the fit to the red sequence, while the upper shows the $3\sigma_{RS}$ upper limit to the envelope, where $\sigma_{RS}$ is determined from the scatter around the best fit. The horizontal red line shows the upper limit in colour used for selection of AAOmega targets. The vertical dashed line in the ATLAS clusters shows the upper limit in magnitude where galaxies of any colour are included as potential AAOmega targets.
\label{CMR_VST}}
\end{figure*}

\begin{figure*}
\includegraphics[angle=90,width=.45\textwidth]{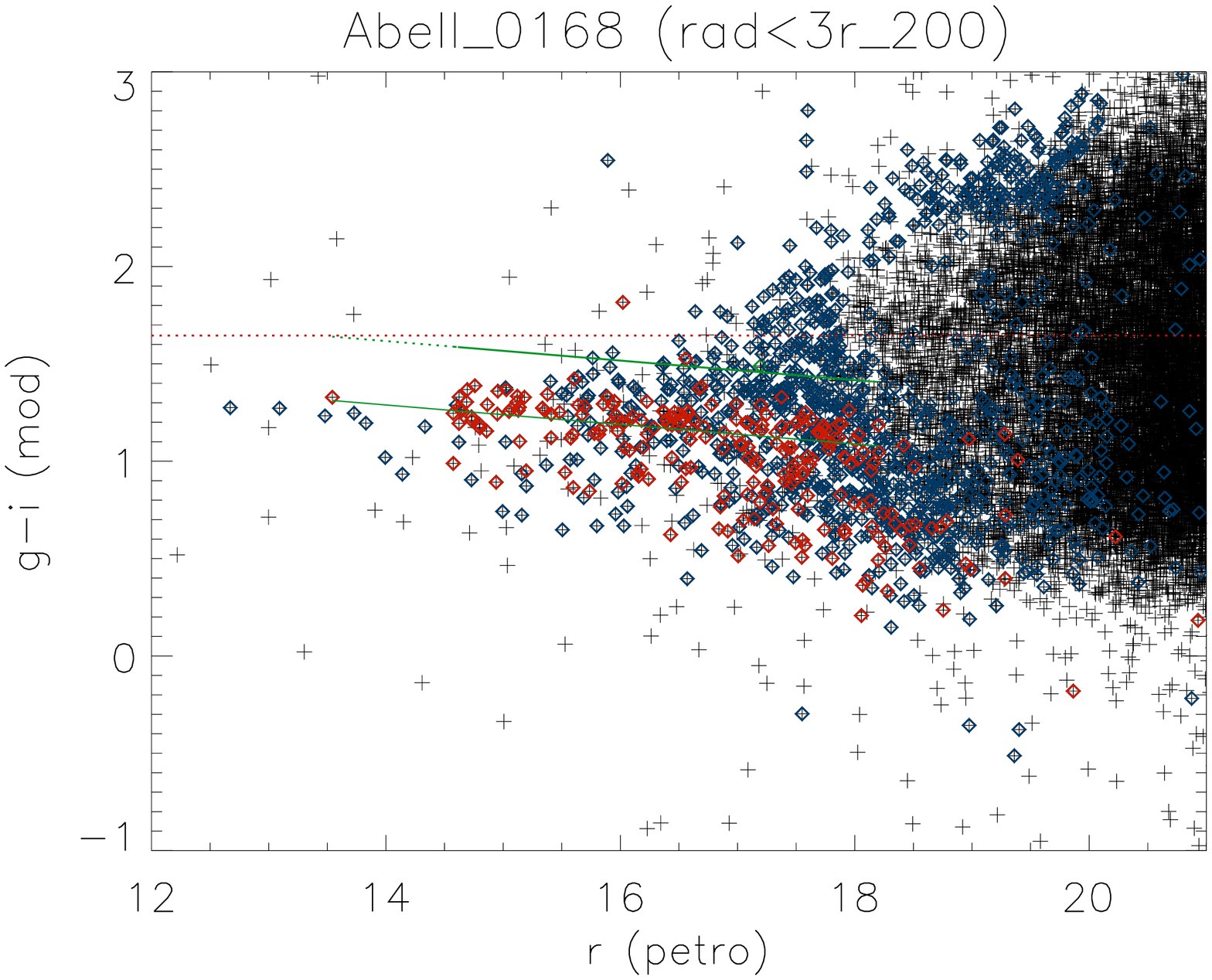}
\includegraphics[angle=90,width=.45\textwidth]{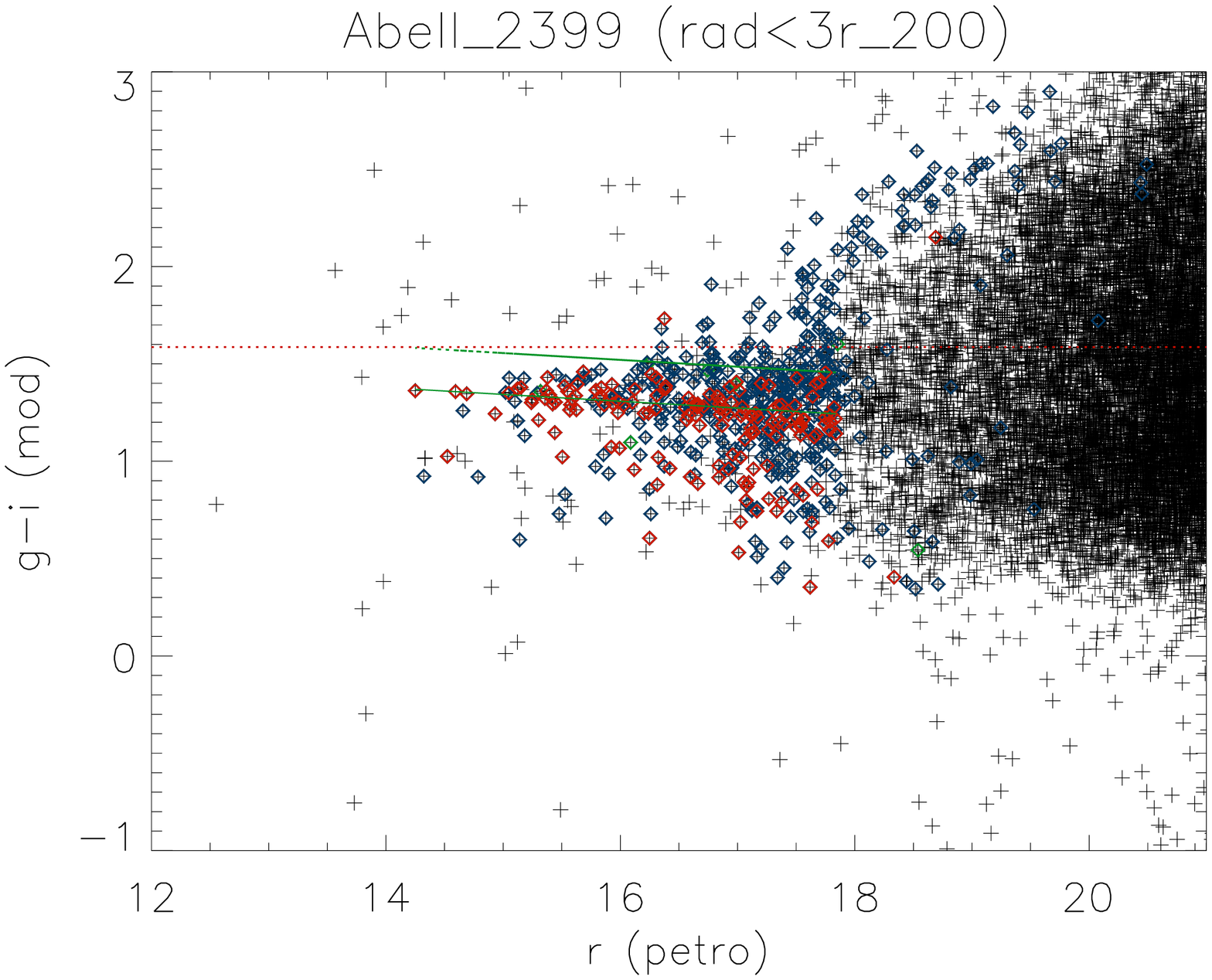}\\
\includegraphics[angle=90,width=.45\textwidth]{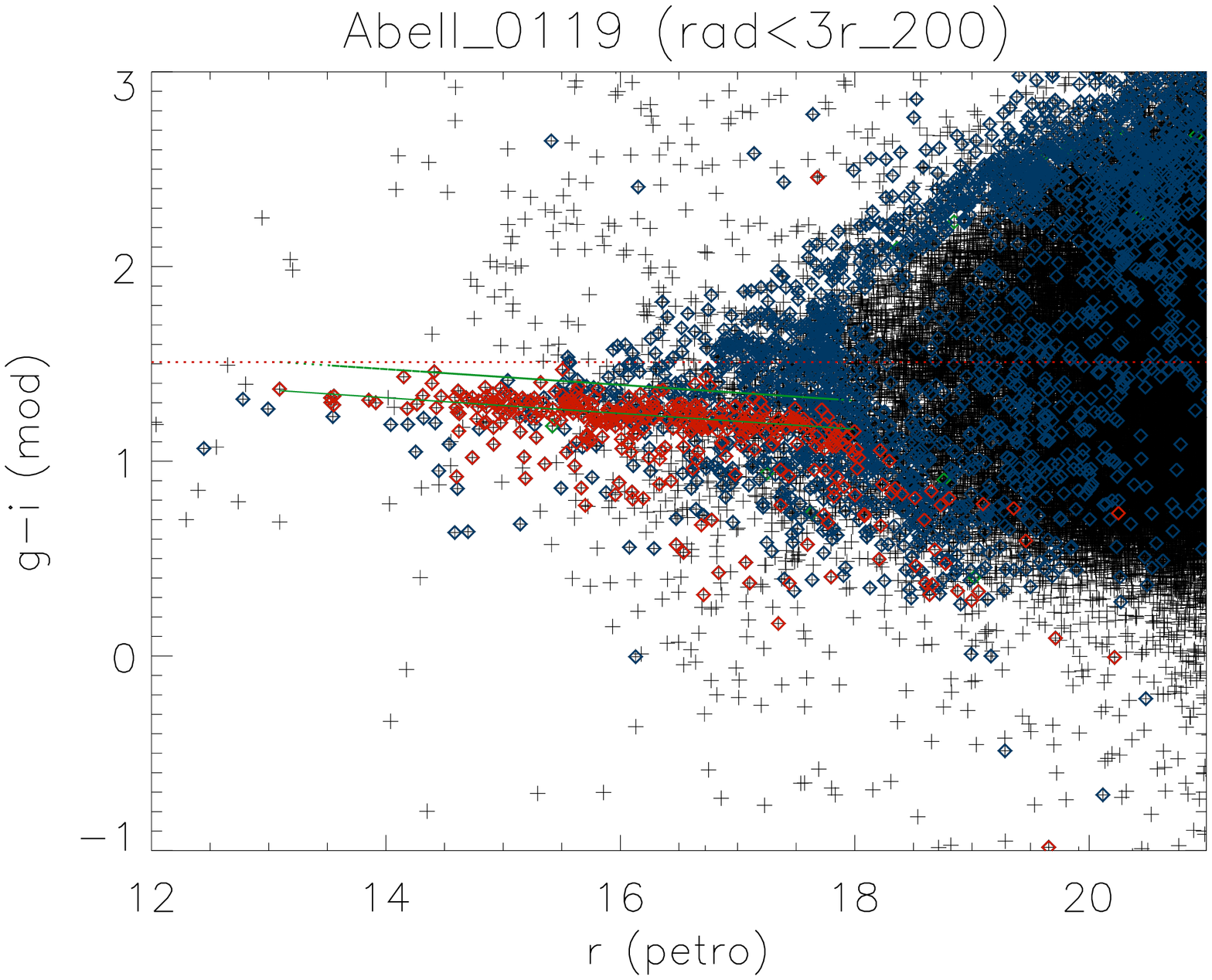}
\includegraphics[angle=90,width=.45\textwidth]{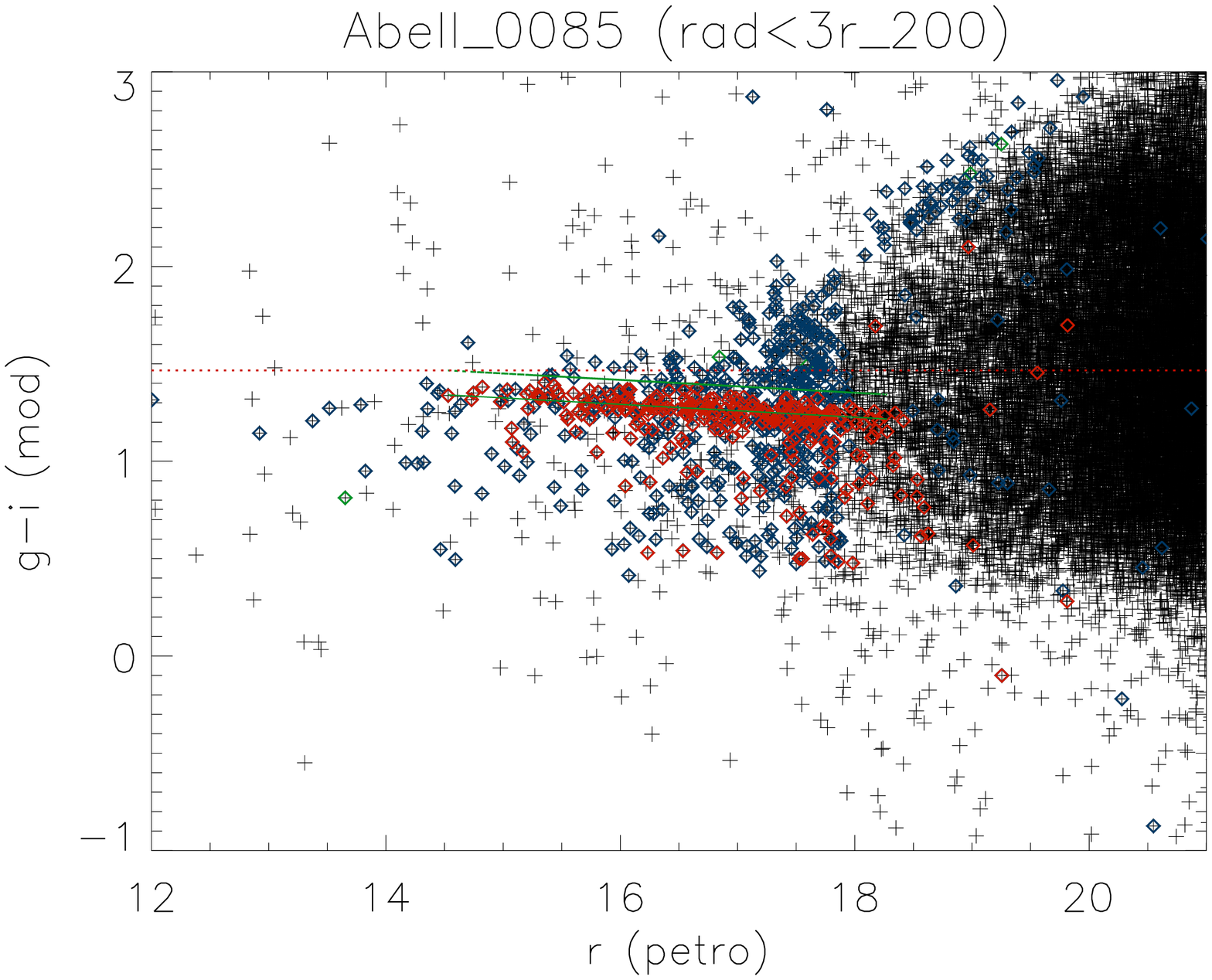}
\caption{Colour-magnitude diagrams for the SAMI-CRS clusters with SDSS photometry. The black plus symbols show all objects classified as galaxies within the field. The red open diamond points show confirmed cluster members. Blue open diamonds show fore- and background galaxies with existing spectra from the 2dFGRS, SDSS or 6dFGS. The lower green line shows the fit to the red sequence, while the upper shows the $3\sigma_{RS}$ upper limit to the envelope, where $\sigma_{RS}$ is determined from the scatter around the best fit. The horizontal red line shows the upper limit in colour used for selection of AAOmega targets. 
\label{CMR_SDSS}}
\end{figure*}

The principal aim of SAMI-CRS was to gather as many cluster member redshifts as possible. With this aim in mind, we gathered pre-existing spectroscopy covering the cluster regions from the SDSS DR9 \citep{ahn2012}, 2dFGRS \citep{colless2001}, 6-degree Field Galaxy Survey \citep[6dFGS;][]{jones2009}, the Cluster and Infall Region Nearby Survey \citep[CAIRNS;][]{rines2003}, WIde-Field Nearby Galaxy cluster Survey \citep[WINGS;][]{cava2009}, NOAO Fundamental Plane Survey \citep[NFPS;][]{smith2004}, ESO Nearby Abell Cluster Survey \citep[ENACS;][]{katgert1996} and the A85 redshift catalogue of \citet{durret1998}. These data were used to determine obvious line-of-sight interlopers, which were subsequently removed, and to perform an initial allocation of spectroscopically confirmed members for the purpose of obtaining initial estimates of the velocity dispersions of the clusters, \rtwo, and the position of the cluster red sequence in $g-i$ colour. The $g-i$ versus $r_{petro}$ colour-magnitude diagrams for galaxies within a cluster-centric distance of 3\rtwo\, of clusters with VST/ATLAS and SDSS photometry are shown in Figures~\ref{CMR_VST} and \ref{CMR_SDSS}, respectively, where non-members are highlighted as blue diamonds and members as red diamonds. The member galaxies clearly show the presence of a red sequence. Only a very small fraction of member galaxies lie redward of the red sequence; this region is dominated by galaxies that, according to their redshifts, are background objects. We use this fact to remove objects beyond a limit in $g-i$ colour (shown as the horizontal red-dashed line in Figures~\ref{CMR_VST} and \ref{CMR_SDSS}) as likely background sources.  The $g-i$ cut is defined as follows. We fit the red sequence of a subset of the available confirmed members with $R<$\rtwo\, and $12 < r_{petro} < 18$ using an outlier-resistant linear fit\footnote{http://idlastro.gsfc.nasa.gov/ftp/pro/robust/robust\_linefit.pro}. An outlier-resistant dispersion around this best-fit line, $\sigma_{RS}$, was measured using the biweight estimator. The $g-i$ cut was defined as ${\rm BCG}_{col}+3\sigma_{RS}$ where ${\rm BCG}_{col}$ is the $g-i$ colour determined at the brightest cluster galaxy $r-$band magnitude using the linear fit to the red-sequence. For the clusters that have VST/ATLAS photometry, we did not apply this colour cut for galaxies brighter than $r_{petro}=16.5$ because the colours of these objects can be unreliable due to the aperture and background subtraction issues outlined in Section~\ref{old_vst}. Figures~\ref{CMR_VST} and \ref{CMR_SDSS} reveal that these cuts reject only a very small number (always less than 5 per cluster) of spectroscopically confirmed cluster members. Finally, we removed those galaxies that have $R>3$\rtwo\, and $r_{petro} > 19.5$.

We also selected a number of stellar objects for guiding and spectrophotometric calibration. Spectrophotometric standard stars were selected to have similar colours to the F-subdwarf BD+17 4708 in the same manner as described in \citet{bryant2015}. Guide stars were selected  to have magnitudes in the range $14<r<14.5$ and low proper motions. Blank sky regions for sky subtraction were selected by randomly sampling the region of sky covered by the input target catalogue. These sky regions were visually inspected to ensure that they are free of bright sources.

\subsection{AAOmega observations}\label{aaomega_obs}

The SAMI-CRS was conducted over seven nights using the 2dF/AAOmega multi-object spectrograph on the 3.9m Anglo-Australian Telescope. Three nights were allocated in Director's Discretionary time from 2013 September 10-13 (hereafter RUN1) and four nights from 2013 September 25-28 (hereafter RUN2) were awarded in addition to the SAMI-GS request. The 2dF instrument consists of 392 2\arcsec\ diameter fibres that can be allocated to objects over a two-degree field of view using a robotic positioner, as well as 8 fibres allocated to fiducial stars for guiding \citep{lewis2002}. The $\sim 40$m fibres feed light to the AAOmega dual-beam spectrograph \citep{saunders2004,smith2004, sharp2006}, which is bench-mounted in a stable, thermally controlled environment at the Coud\'{e} west room. For the SAMI-CRS we used the low resolution 580V and 385R gratings for the blue- and red-arms, respectively, where the light beam was split with a 5700\AA\, dichroic. This results in a wavelength coverage $3700-5850$\AA, ($5600-8850$\AA) at 3.53\AA\ (5.32\AA) FWHM resolution for the blue (red) arm. 

During the afternoon, the fibre configurations for the night were generated by a modified version of the {\sf TILER} code used by the GAMA survey and described in \citet{robotham2010}. Briefly, the code automatically determines the optimal centre for the field by attempting to maximise the spatial distribution of the spectroscopic completeness. The code then uses the {\sf CONFIGURE}\footnote{https://www.aao.gov.au/science/software/configure} software \citep{miszalski2006} to generate the night's fibre configurations. 
The {\sf CONFIGURE} software allows target prioritisation so that high priority targets are more likely to be allocated a fibre during the configuration process. We take advantage of this capability to maximise the number of cluster redshifts collected and the spectroscopic completeness within \rtwo, which is where the SAMI-GS will target. To that end, we set as highest priority (priority=9) those target galaxies within \rtwo\, and have no redshift information. At intermediate priorities (priority=8-6), we include galaxies with \rtwo $< R < 3$\rtwo and no redshift information, as well as those galaxies that have existing redshift information from the 2dFGRS or 6dFGS placing them near the cluster redshift ($|v_{pec}| < 4\sigma$). The lowest priorities (priority=5-1) are allocated to filler targets which have an existing SDSS redshift that places them close to the cluster redshift, with the priorities decreasing with radius in this low priority range. All objects having a redshift that places them well in the fore- or background of the cluster are excluded from the configurations, as are those targets that have colours indicating they are spurious detections, i.e., $r-i<-4$. In addition to these priorities, the {\sf TILER} code identifies objects in the input catalogues that are most likely to be impacted by limitations on the minimum allowable fibre separation ($\sim 40$\arcsec) due to the size of the fibre buttons. The objects most affected by collisions have their priorities increased, while the objects that are within 40\arcsec of these most clustered objects are removed from the input catalogue for the configuration of interest. By doing this, the most clustered targets are observed first, thereby lessening the impact of highly clustered objects on subsequent configurations and improving the survey efficiency \citep{robotham2010}. For each configuration, 25 fibres were positioned at blank sky regions for sky subtraction and 3 fibres were allocated to spectrophotometric standards.

Table~\ref{obs_table} summarises the number of fields and their respective exposure times. The observing sequence for each field observed in RUN1 and RUN2 generally included an arc exposure, two flat-field exposures (5s and 0.5s exposures for the blue- and red-arm, respectively) and three source exposures. During RUN1, we focussed on targets brighter than $r=19$ ($r=18.5$ for Abell~4038) and set the exposure time to 45min per field (taken as a set of $3\times$900s exposures), and only included the $19. < r <19.4$ targets as low-priority fillers. This magnitude limit was selected as a trade-off between the S/N ratio required to determine a redshift for a large fraction of the observed targets, and the minimum exposure time per field, which is limited by the re-configuration time of the 2dF robot (40-45 minutes). This strategy allowed us to maximise the number of fields and, therefore, the number of redshifts collected during RUN1. For RUN2, the fainter objects were increased in priority, and the majority of the tiles were targeted for 60min ($3\times$1200s exposures) so that the fainter objects with $19 < r < 19.4$ achieved sufficient S/N for redshift determination. During RUN2, objects with spectra too low in S/N to measure a reliable redshift during RUN1 were included in the input catalogues for re-observation. Those galaxies that had a lower-quality ($0.9 \leq z_{conf} < 0.95$; see Section~\ref{redshifting}) RUN1 redshift that placed them very close to the cluster redshift were added as filler targets. During RUN2, data were reduced and redshifted on the fly and any object for which a reliable redshift measurement was not possible was cycled back into the target catalogue for re-observation on subsequent nights.

In addition to the data collected in September 2013, we also included several sets of observations retrieved from the AAO archives (also listed in Table~\ref{obs_table}). For A85, there were two fields observed in 2006 and 4 fields in 2007, while Abell~168, Abell~3880 and Abell~2399 each had two extra fields observed as part of the OMEGAWINGS program \citep{Gullieuszik2015}. Except for the 2006 data \citep[see][]{boue2008}, the target selection for these archived datasets is not known. The data are processed in the same manner as the SAMI-CRS data, and are cross-matched with our input catalogues. Within the archived datasets, 1617 objects were not matched to objects in the SAMI-CRS input target catalogues. These non-matched objects were generally either fainter than the limiting magnitude of the SAMI-CRS input catalogue, or redder than the colour cut used for the particular cluster.

\begin{table*}
 \centering
  \caption{Summary of the SAMI-CRS and archival 2dF/AAOmega data.  \label{obs_table}}
  \begin{tabular}{@{}lcccccccc@{}}
  \hline
   Name     &     RUN1      & RUN2  & Archive & Seeing & ${\rm N_{field}}$ & ${\rm N_{spec}}$ & ${\rm N_z}$\\
\hline
APMCC 917/Abell 4038 & 5$\times$(45min) & 4$\times$(45min), 3$\times$(60min) & - & 1.6\arcsec - 4.2\arcsec & 14 & 5004 & 4424\\
                    &                          & 1$\times$(50min), 1$\times$(30min) &\\
Abell 3880 &3$\times$(45min) & 3$\times$(60min) & 1$\times$(60min), 1$\times$(120min) & 1.0\arcsec - 3.1\arcsec & 8 & 2522 & 2368\\

EDCC 442&2$\times$(45min) & 1$\times$(40min) & - & 1.4\arcsec - 2.0\arcsec  & 3 & 1019 & 840\\

\hline
Abell 168 &2$\times$(45min) & 4$\times$(60min) & 2$\times$(60min) & 1.4\arcsec - 3.9\arcsec & 8 & 2665 & 1960\\

Abell 2399 & 2$\times$(45min) & 4$\times$(60min) & 1$\times$(60min), 1$\times$(120min) & 1.4\arcsec - 2.9\arcsec & 8 & 2876 & 2480\\

Abell 119 &4$\times$(45min) & 5$\times$(60min) & - & 1.4\arcsec - 4.0\arcsec & 9 & 3224 & 2377\\

Abell 85&3$\times$(45min) & 5$\times$(60min) & 2$\times$(510min),1$\times$(250min) &1.3\arcsec - 2.9\arcsec & 14 & 4756 & 3966\\
                     &                          & &1$\times$(270min), 1$\times$(78min) &\\
                     &                          & &1$\times$(108min) &\\
\hline
\end{tabular}
\end{table*}

\subsection{Final Data reduction}

Following the two observing runs, the final data reduction was performed using a combination of the standard {\sf 2dFDR}\footnote{https://www.aao.gov.au/science/software/2dfdr} (version 6.28) software and a set of custom {\sf IDL} routines that offer several improvements over and above the standard {\sf 2dFDR} routines. The initial phases of the reductions are performed using {\sf 2dFDR} and include bias removal (using a fit to the overscan regions), tracking of the fibre position on the CCD using the flat-field exposures, cosmic ray identification and masking, and wavelength calibration using the arc frames. In the blue CCD, additional cosmetic structure was removed using master bias and dark frames which are the products of stacking 20-30 bias and dark frames.

\begin{figure*}
\includegraphics[width=.48\textwidth]{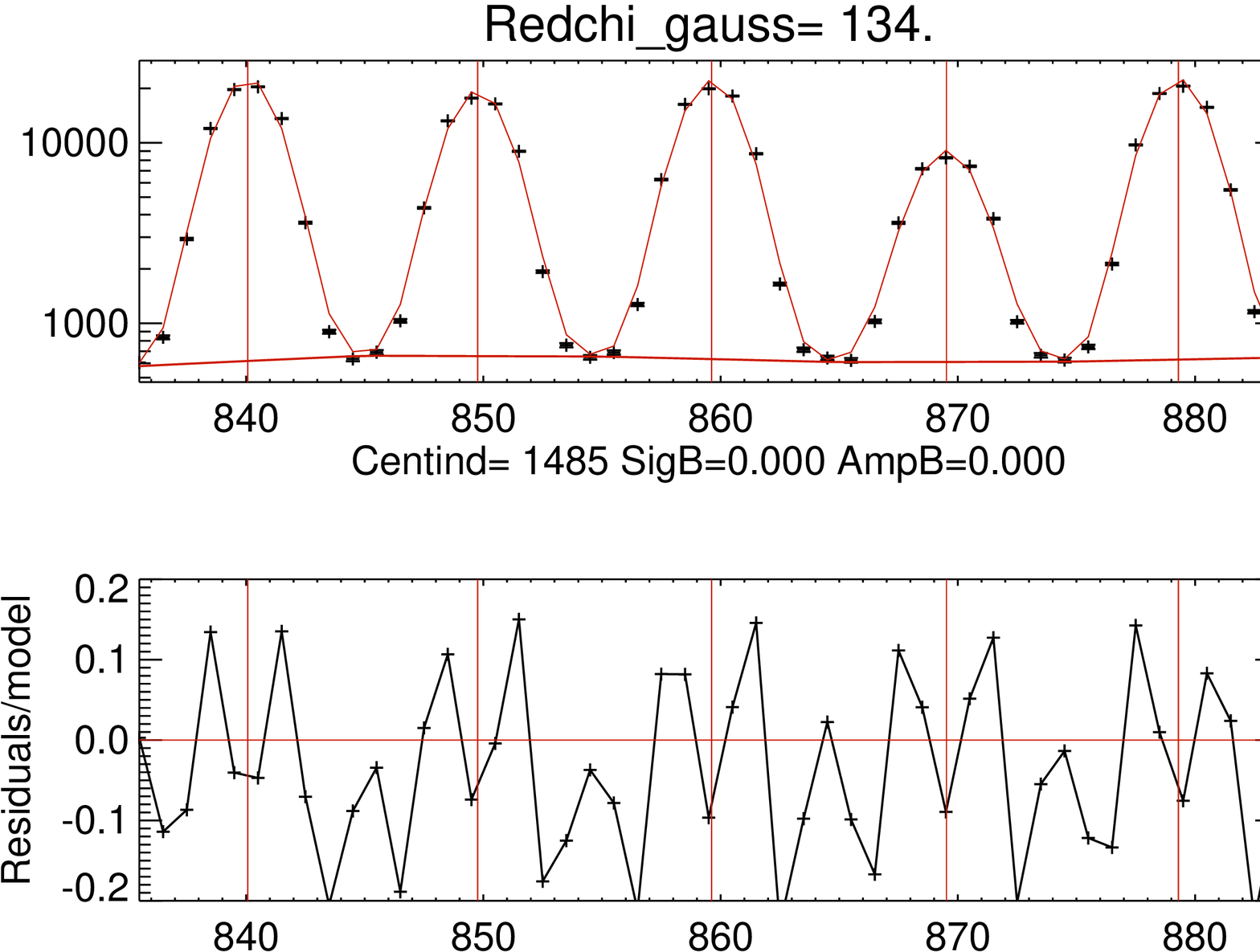}
\includegraphics[width=.48\textwidth]{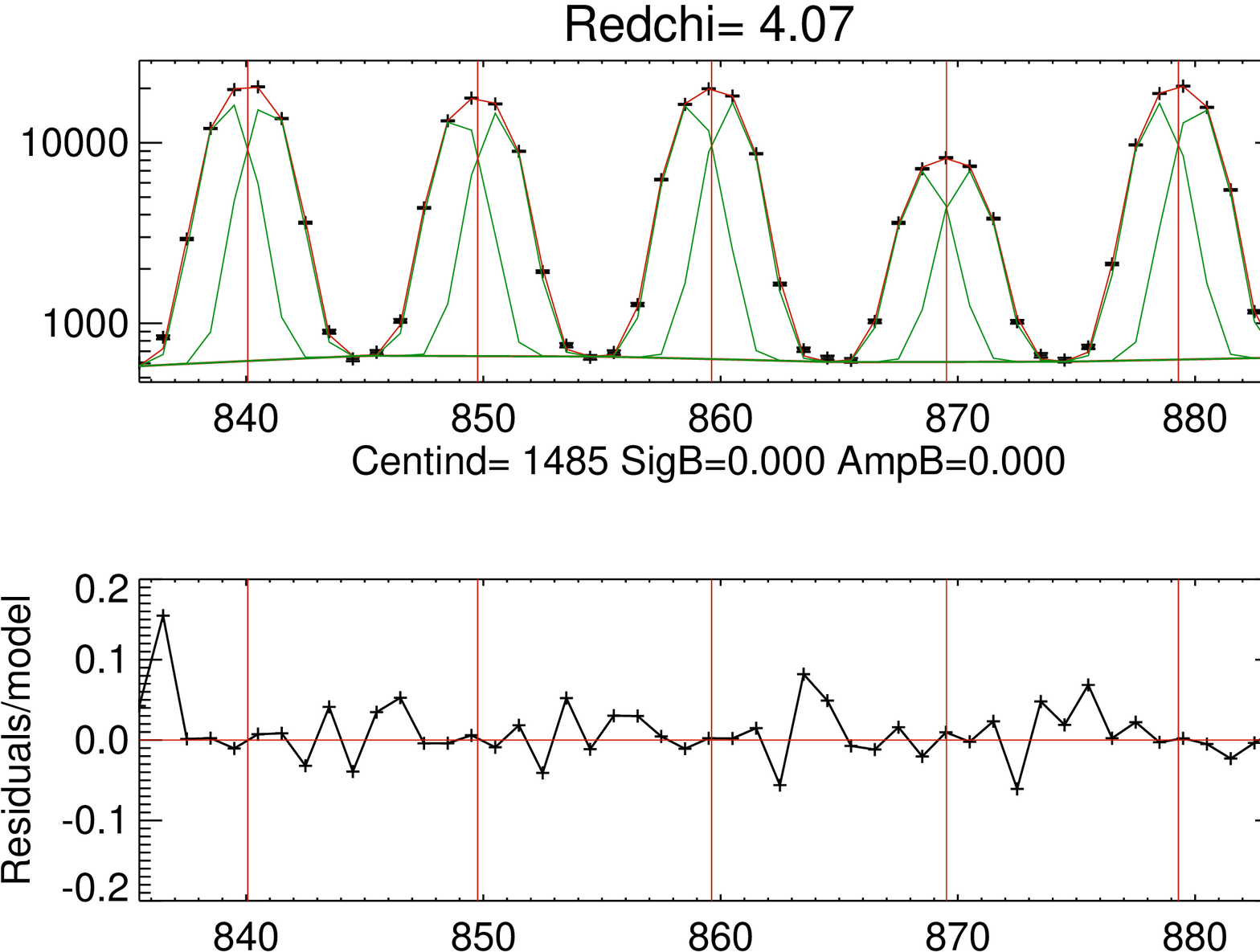}
\caption{{\it Upper left panel:} An example of a single Gaussian fit (red line) to the profile (black crosses) at column 1485, centred around row 860 for one of the flat-fields. {\it Upper right panel:} The double Gaussian fit (red line) to the profile. The green lines show the two Gaussian components, offset by $\pm \Delta$ around the {\sf 2dFDR}-defined fibre position. {\it Lower left and right panels:} Fractional residuals of the single and double Gaussian fits, respectively. The fibre position as defined by the {\sf 2dFDR} software is shown as a vertical red line for each fibre profile. The double Gaussian profile provides a significantly better fit, as indicated by the reduction in the reduced $\chi^2$ values (134 to 4.1) for only one extra degree of freedom, and also by the reduction in the residuals.
\label{fibre_profiles}}
\end{figure*}

Following these initial reduction steps, the custom {\sf IDL} routines were used to define the profiles of the fibres using the high S/N flat-field exposures. This step is vital for accurate extraction of flux for both the flat-field and object frames. The fibre profile for the 2dF/AAOmega combination is generally assumed to be well-described by a single Gaussian component \citep{sharp2010a}. However, we found that significant systematic residuals remain due to the more ``boxy'' nature of the fibre profile (Figure~\ref{fibre_profiles}) compared with a single Gaussian profile. This boxy profile structure is due to the convolution of the tophat fibre shape with the Gaussian PSF of the AAOmega spectrograph optics \citep{saunders2004,sharp2006}. The fit to the profile is vastly improved by using a double Gaussian profile where the amplitude, $A$, and dispersion, $\sigma$, of the two Gaussians are tied to the same value during the fitting. The positions of the two Gaussians are offset by an equal but opposite distance, $\Delta$, from the central position of the profile, $\overline{y}$, which is fixed to the value determined by the {\sf 2dFDR} tracking. The profile model for each fibre at column $x$ is defined as
\begin{equation}
P(y)=A(e^{{(y-(\overline{y} - \Delta))^2}\over{2\sigma^2}} + e^{{(y-(\overline{y} + \Delta))^2}\over{2\sigma^2}}),
\end{equation}
so that there is only one extra parameter over the single Gaussian case. The double Gaussian profile used is symmetric about $\overline{y}$ and provides an excellent description of the core of the fibre profile (see right panels of Figure~\ref{fibre_profiles}). The parameters $\Delta$ and $\sigma$ vary smoothly in the wavelength direction for each fibre. Therefore, only every 20th column is fitted and the results are interpolated onto the full 2048 resolution using a low-order polynomial fit. To account for the small contribution of flux to the fibre of interest due to crosstalk \citep{sharp2010a}, the four fibres surrounding the fibre of interest are fitted simultaneously (e.g., as shown in Figure~\ref{fibre_profiles}).

While the above procedure produces a very good description of the core of the fibre profile, there also exists a low-amplitude, very extended component to the profile that can be difficult to model accurately during the profile fitting. The cumulative effect of the broad component of the 400 fibres is a relatively smoothly varying (in the wavelength direction) background component that, in addition to the background produced by scattered light from reflections within the AAOmega spectrograph, can affect the accuracy of the flux extraction process if not removed. This background is subtracted prior to both the profile definition and flux extraction. The background component is determined for each fibre by selecting pixels near the midpoint between the fibres and fitting a B-spline model in the wavelength direction. The backgrounds are then interpolated onto the full $2048\times4098$ array using linear interpolation before being subtracted from the frame of interest. This background subtraction helps to minimise the impact of scattered light in flat-field frames, as well as scattered light due to bright stellar sources erroneously included in the input catalogues.

Having used the flat-fields to define the fibre profile shapes, and subtracted the background from the frame of interest, $\Delta$ and $\sigma$ are fixed and the flux is extracted by fitting the amplitude at each column for each fibre. Following the extraction, the relative chromatic response of each fibre is determined from the flat-fields by normalising them by the average flat-field spectrum, using the method described by \citet{stoughton2002}. The extracted object spectra are divided by the corresponding normalised flat-field spectrum. The wavelength solution determined by {\sf 2dFDR} using the arc frames is tweaked using the position of known skylines. The extracted spectra are then divided by their relative throughputs, determined using the flux measured in skylines. Sky subtraction is achieved in a similar manner to that described in \citet{stoughton2002}; a super-sampled sky is determined from the 25 sky fibres using a B-spline fit, which is then used to construct a sky spectrum sampled at the wavelength solution determined for each fibre and subsequently subtracted. The red-arm spectra are corrected for telluric absorption in a similar manner to that described in \citet{hopkins2013}. Briefly, a flux-weighted sum of object spectra (excluding very bright objects) is fitted with a polynomial after excluding regions affected by telluric absorption. The summed spectrum is then normalised by this polynomial, and regions not affected by telluric absorption are set to one, leaving only a template of the telluric absorption. Each spectrum is divided by this template, as are the associated variance vectors. Finally, the sky subtraction residuals near sky emission lines are removed using principal component analysis, as described by \citet{sharp2010c}. The separate frames are then combined by {\sf 2dFDR} using a weighted sum, which incorporates both a per-object variance weight and a weighting to account for varying sky conditions for each frame. The blue- and red-arm spectra are combined after being divided by an estimate of the throughput function for each arm. The red-arm is re-sampled from its native $\sim 1.5$\AA\, pixel scale to that of the blue-arm ($\sim 1.03$\AA) and the final reduced spectra cover a wavelength range $\sim 3730-8850$\AA.

\subsection{Redshift measurements, accuracy, precision and duplicate spectra}\label{redshifting}

The redshifting is performed by the {\sf IDL} task {\sf autoz},\footnote{http://www.astro.ljmu.ac.uk/\~ikb/research/autoz\_code/} described in detail in \citet[][]{baldry2014}. The code cross-correlates spectra with a set of templates where both the spectra and templates have been filtered to remove continuum and pixels with absolute values larger than 25 times the mean absolute deviation of the continuum-subtracted spectrum. This filtering helps to minimise the impact of spurious features associated with poor reduction, e.g., due to bad pixels, poor sky subtraction, etc.. As noted in \citet{baldry2014}, the clipping only removes real emission lines in high S/N cases where a redshift determination based on weaker features is possible. We use template IDs 2-14 and 40-49 \citep[see Table~1 in][]{baldry2014}, which corresponds to a subset of SDSS DR5 stellar templates\footnote{http://www.sdss.org/dr5/algorithms/spectemplates/} and a set of SDSS-BOSS galaxy eigenspectra \citep{bolton2012}. The redshift corresponding to the highest peak in the cross-correlation function, $r_x$, is selected and assigned a figure of merit, ${\rm cc_{FOM}}$ which is derived by comparing the $r_x$ value to the three next highest peaks, and adjusted based on the noise characteristics of the filtered spectrum, as outlined in \citet{baldry2014}. The $\rm cc_{\rm FOM}$ value is used to assign a redshift confidence, $z_{\rm conf}$, using the analytical function presented in Equation~8 of \citet{baldry2014} that has been calibrated using duplicated redshift measurements in the GAMA survey. The combination of the archived AAOmega data, as well as the strategy of reobserving many targets in the SAMI-CRS, meant that there were 7437 duplicate spectra for 3108 objects (after excluding stars). We used the duplicated spectra and their associated {\sf autoz} redshift and ${\rm cc_{FOM}}$ measurements to test the GAMA-based ${\rm cc_{FOM}}-z_{\rm conf}$ calibration. We do this by following the method of \citet{baldry2014} and compared the fraction of the duplicated redshifts that are discrepant (i.e., where $|\Delta cz| > 450$\kms) as a function of ${\rm cc_{FOM}}$. We confirm that the \citet{baldry2014} calibration is suitable for the SAMI-CRS data. Throughout the remaining analysis, only those redshifts with $z_{\rm conf} \geq 0.9$ are used.

\begin{figure*}
\includegraphics[angle=90,width=.3\textwidth]{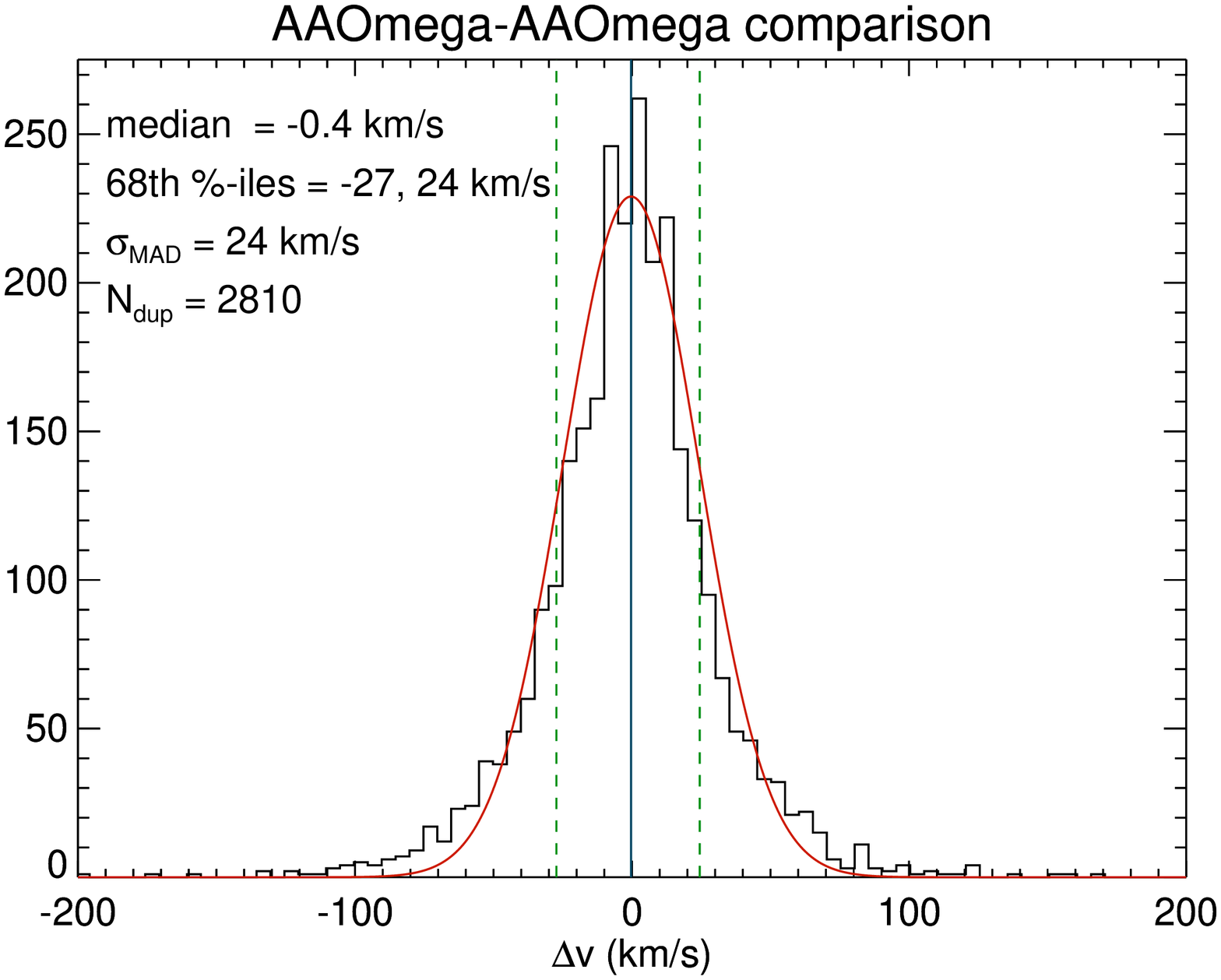}
\includegraphics[angle=90,width=.3\textwidth]{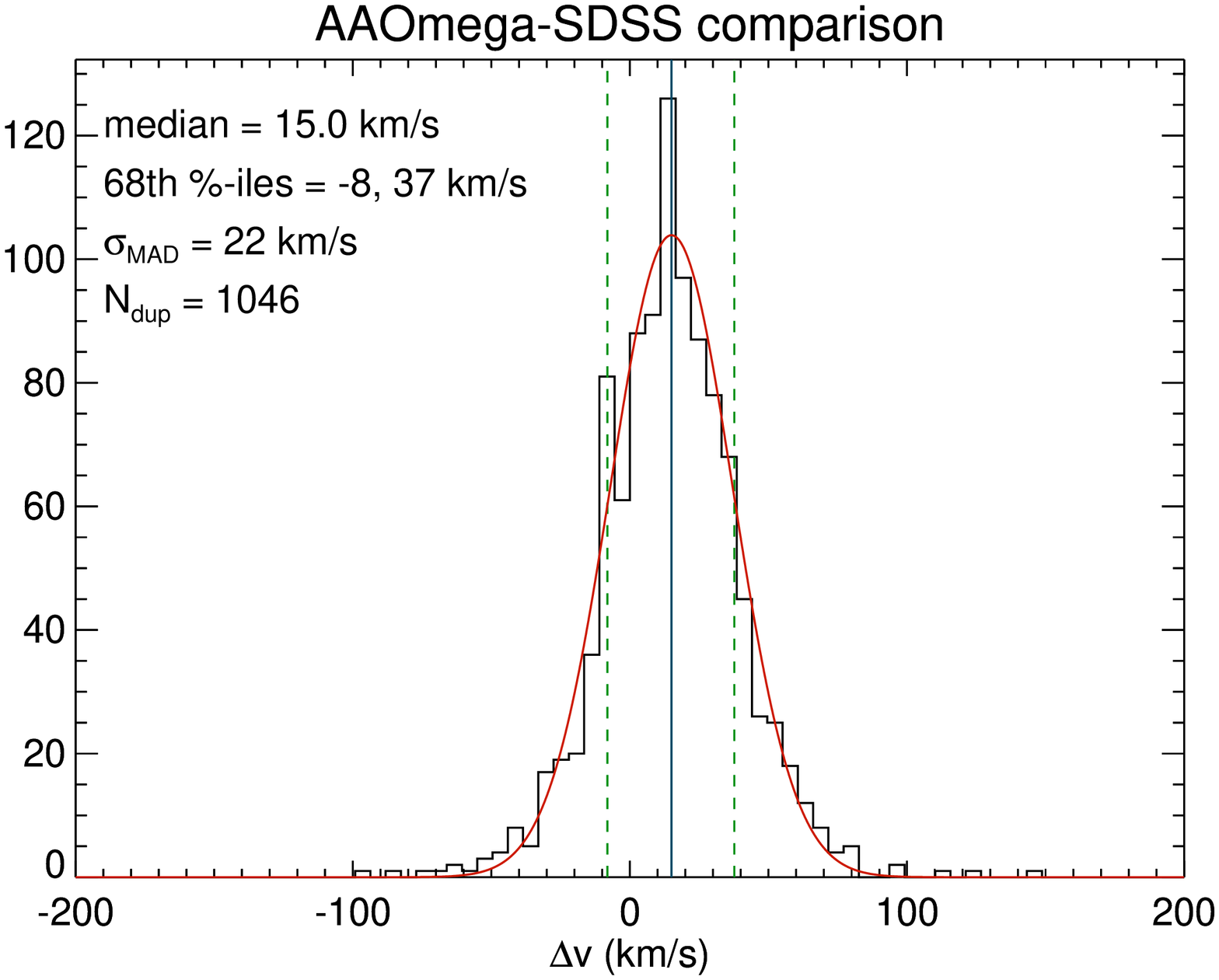}
\includegraphics[angle=90,width=.3\textwidth]{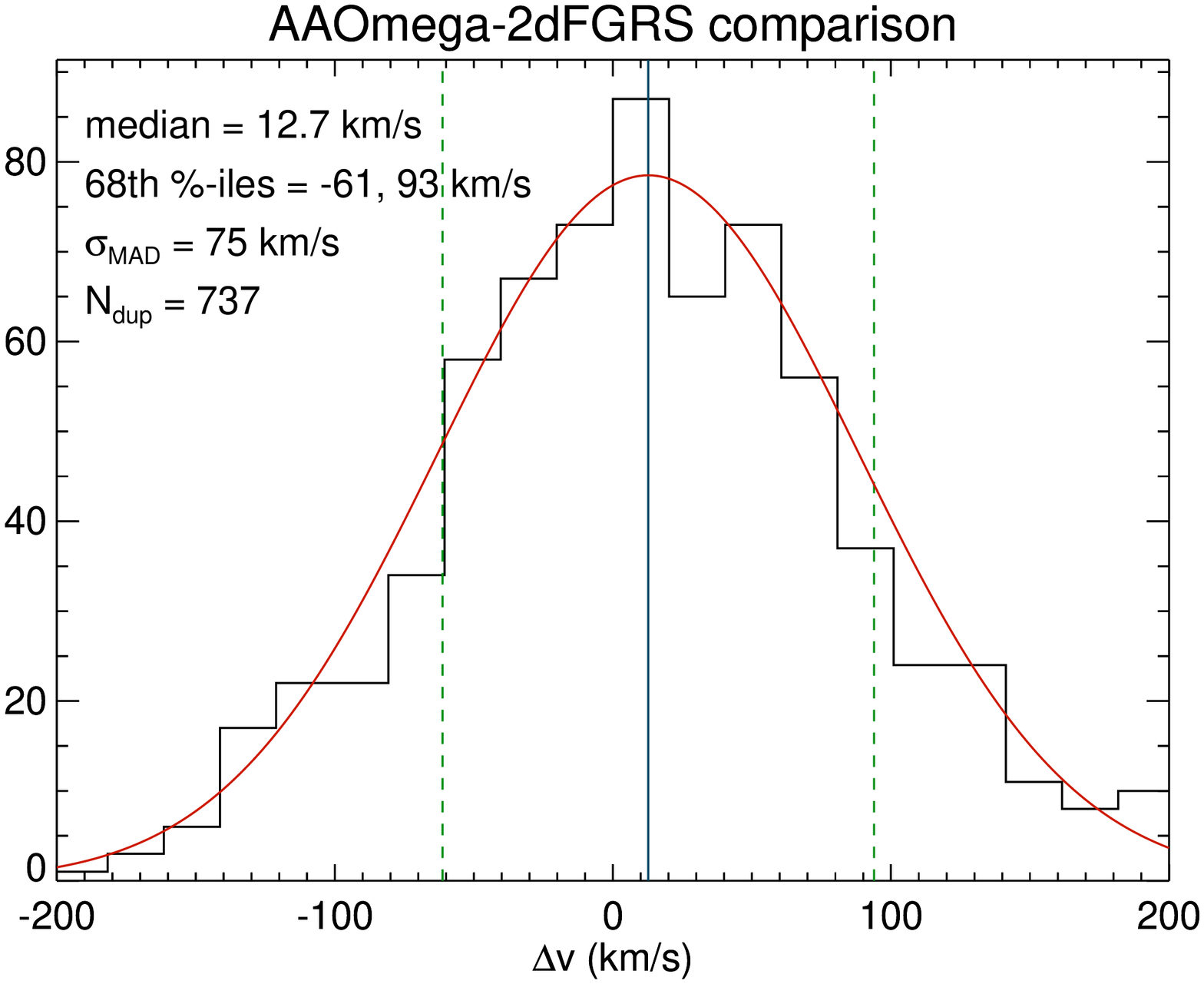}\\
\includegraphics[angle=90,width=.3\textwidth]{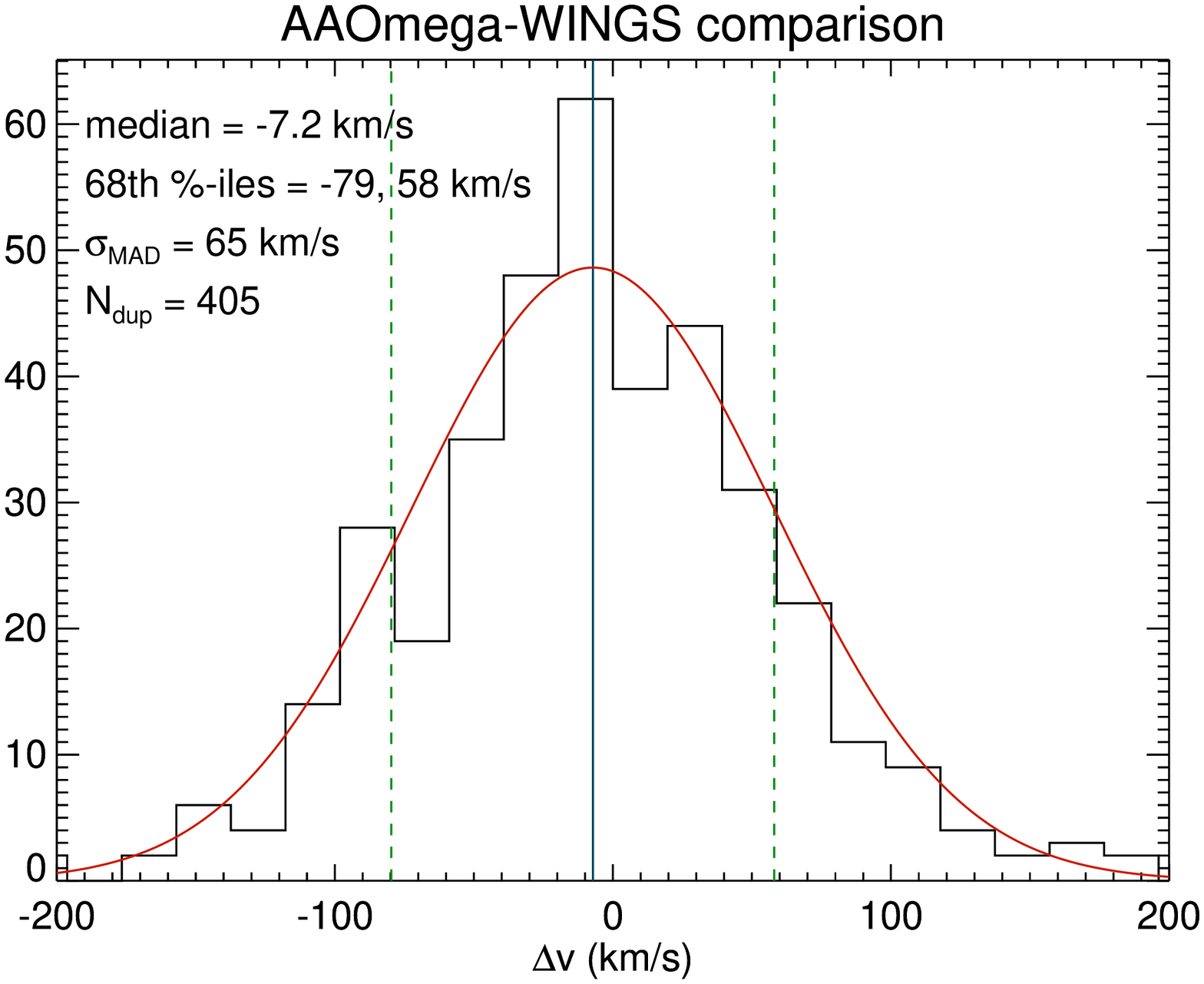}
\includegraphics[angle=90,width=.3\textwidth]{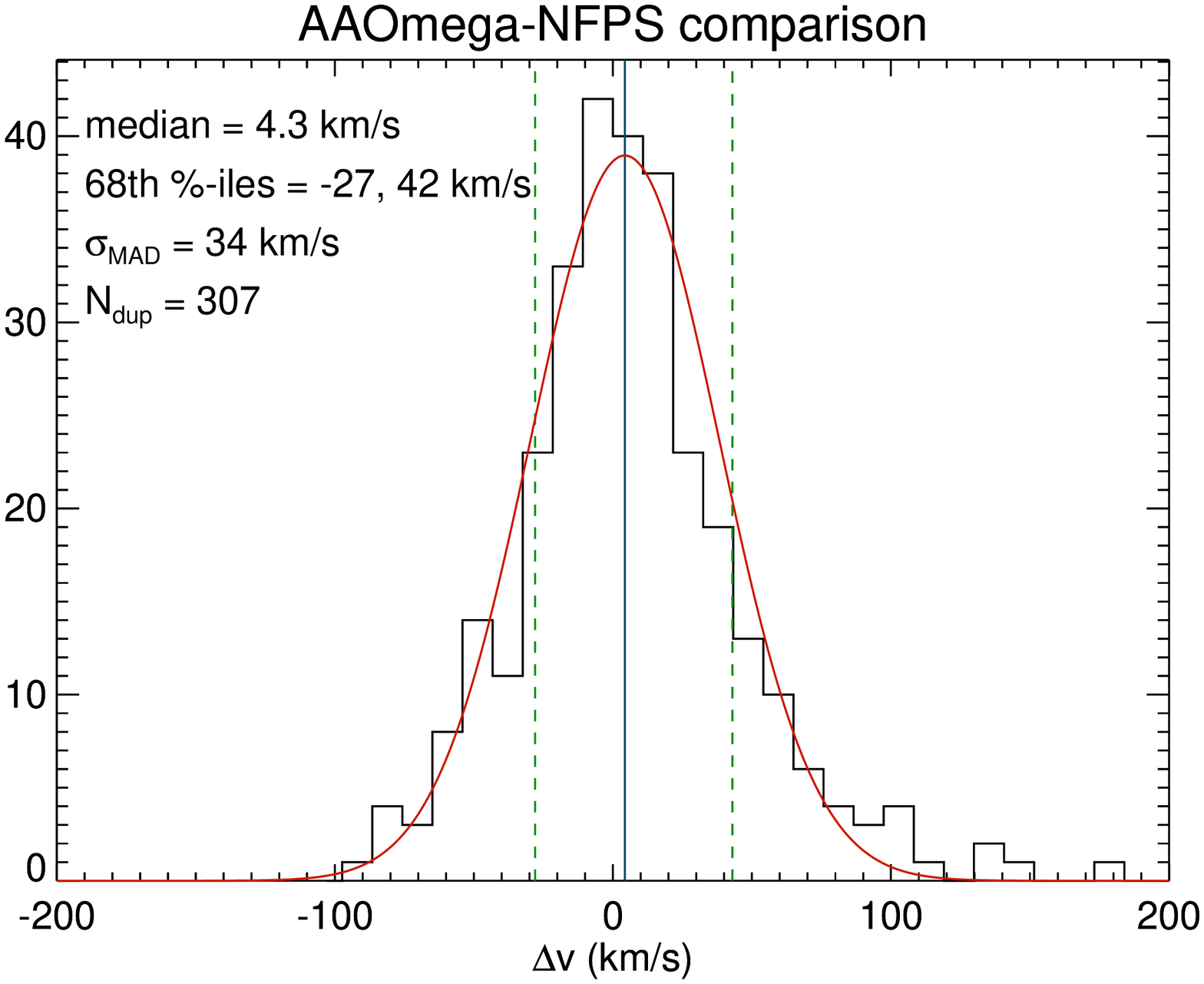}
\includegraphics[angle=90,width=.3\textwidth]{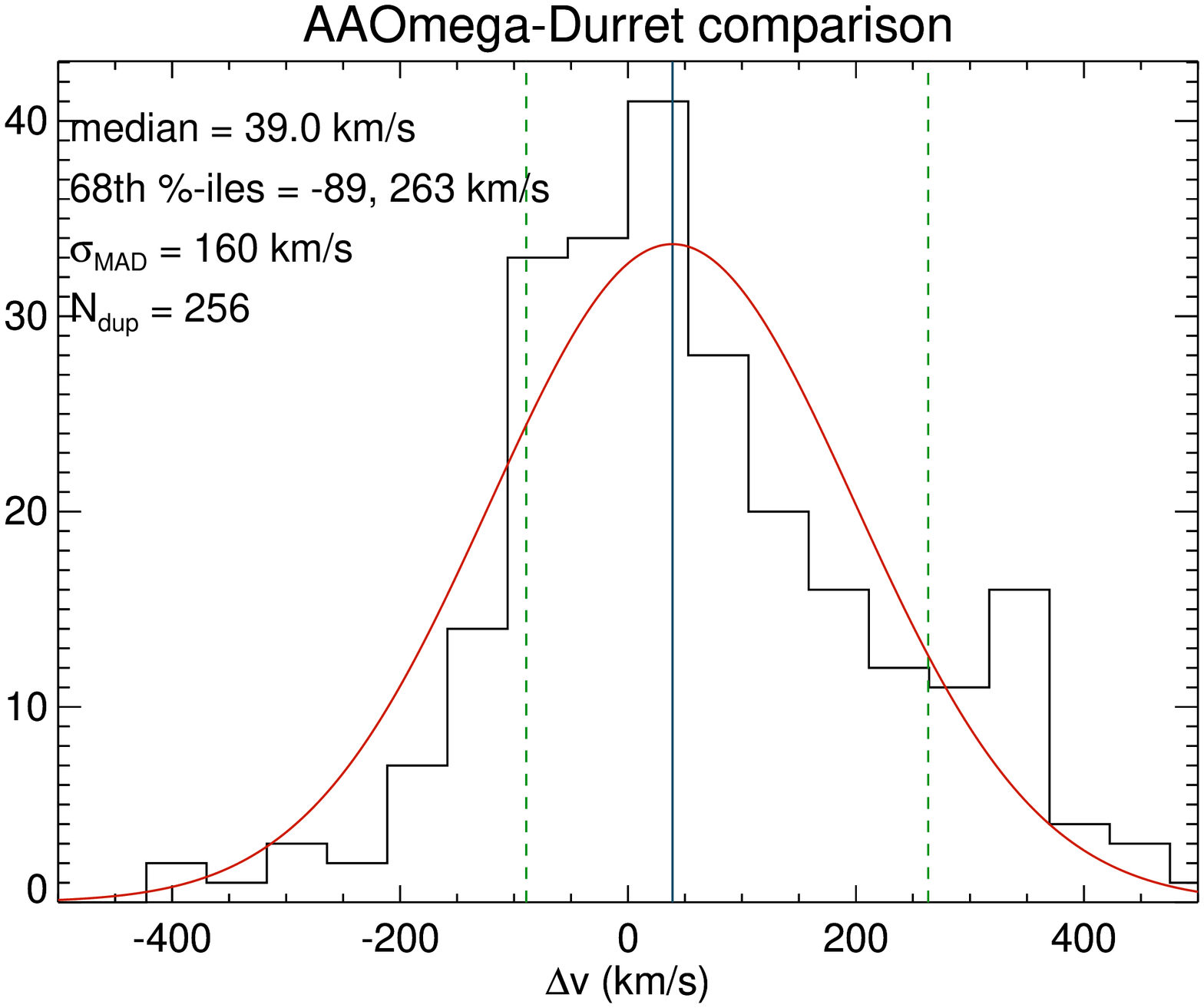}
\caption{The top left panel shows the distribution of $\Delta v$ for 2810 pairs of redshift from duplicate SAMI-CRS AAOmega observations of the same object. The solid red line shows a Gaussian distribution with $\mu=-0.4\,$\kms\, and $\sigma= 24\,$\kms. The top middle and right-most panels show the $\Delta v$ comparison for the SAMI-CRS-SDSS and SAMI-CRS-2dFGRS duplicate measurements, respectively. There is a small  shift in the sense that the SAMI-CRS measurements have redshifts that are systematically higher by $\sim 15$\kms. The bottom, left, middle and right panels show the $\Delta v$ distribution for the SAMI-CRS-WINGS, SAMI-CRS-NFPS and SAMI-CRS-Durret duplicate measurements, respectively. The SAMI-CRS-WINGS and SAMI-CRS-NFPS distributions show no systematic offset. The SAMI-CRS-Durret $\Delta v$ distribution shows a significant offset, with $\mu = 39$\kms. This offset is driven by an asymmetry in the distribution at $\sim 300$\kms. Also note that the scale of the $x$-axis ranges from $-500$ to $500$\kms\, for the SAMI-CRS-Durret comparison, whereas it is $-200$ to $200$\kms\, in the other panels. The origin of this asymmetry is unclear, although it is likely due to the \citet{durret1998} data given the other $\Delta v$ distributions are relatively symmetric.
\label{czdiffs}}
\end{figure*}

Within the sample of objects with duplicated measurements, there are 2047 extra-galactic objects that have 4448 spectra and 2810 redshift pairs where both redshift measurements have $z_{\rm conf} \geq 0.9$. These duplicates are used to determine the blunder rate and precision of the {\sf autoz} redshift measurements. The distribution of the pair $\Delta v = c({\rm ln}(1+z_1)-{\rm ln}(1+z_2))$ values is shown in the top left panel of Figure~\ref{czdiffs} where the difference is always in the sense that $cc_{\rm FOM, 1} > cc_{\rm FOM,2}$. The distribution is centred at $0\,$\kms\, with dispersion $\sigma_{\rm MAD}=\sim 24\,$\kms, which is consistent with the redshift precision measured for the GAMA survey \citep{liske2015}. The blunder rate is defined as the number of measurements where $|\Delta v| > 5\sigma_{\rm MAD}=120$\kms, and is $1.0$\% (N.B., using the blunder criterion defined in \citet{liske2015} of $|\Delta v| > 350$\kms\, returns a blunder rate of 0.6\%).  

We compare the SAMI-CRS redshifts to external survey measurements to determine the accuracy of the redshift measurements. The comparison with SDSS DR10, shown in the middle panel of the top row in Figure~\ref{czdiffs}, indicates that the SAMI-CRS redshifts are systematically higher than the SDSS ones by $\Delta v=15$\kms, similar to the offset found by \citet{baldry2014}. A similar offset is seen in the comparison with the 2dFGRS redshifts (top right panel, Figure~\ref{czdiffs}), although the scatter there is larger, and primarily driven by the larger uncertainties associated with the 2dFGRS redshifts \citep[mean redshift uncertainty $\sim 85$\kms;][]{colless2001}. There is good agreement between the SAMI-CRS and the WINGS and NFPS redshift measurements, although the \citet{durret1998} measurements appear to be asymmetric with a prominent excess at positive $\Delta v$ values, as indicated by the $68$th percentiles. The origin of this asymmetry is unclear, although given that the comparisons with other surveys show relatively symmetric distributions the cause likely lies with the \citet{durret1998} data.

In order to determine if the redshift uncertainties calculated by {\sf autoz} provide reasonable estimates of the true measurements uncertainty, and can therefore explain the spread in the $\Delta v$ values, we investigate the distribution of redshift differences normalised by the quadrature sum of the redshift uncertainties. The spread in the distribution of normalised redshift differences is $\sigma_{\rm MAD} = 0.65$, indicating that the redshift uncertainties can account for all of the scatter in the differences in the duplicated measurements and may be somewhat overestimated. We also compared the normalised redshift differences between the SAMI-CRS and external surveys. Again, there are significant differences that occur in the SAMI-CRS-Durret comparisons, that show an asymmetric distribution which favours positive offsets. In general, the external comparisons have $\sigma_{\rm MAD} < 1$ and confirm the results of the internal comparisons, i.e., the scatter in the repeat measurements is accounted for by the individual redshift uncertainties.

For many of the objects with duplicate spectra, a high-quality redshift could not be determined for any of the spectra due to their low S/N ratios. We attempt to recover these redshifts by combining the continuum-subtracted, high-pass filtered spectra as described in \citet{liske2015}. Prior to combination, the spectra are corrected for the shift due to the heliocentric velocity and interpolated onto a common wavelength grid. Following the combination, {\sf autoz} is used to determine the redshift and redshift confidence. This process produced an additional 319 reliable redshift measurements with $z_{\rm conf} > 0.9$.

\subsection{The combined redshift catalogue}\label{redshift_cat}

The SAMI-CRS redshift catalogue is combined with pre-existing redshifts from the other surveys mentioned in Section~\ref{redshifting} using the following selection rules. First, all duplicate redshift measurements from the SAMI-CRS are removed by selecting the redshift with the highest $z_{\rm conf}$ value. Second, the external  redshift catalogues are cross-matched with the input target catalogue using a matching radius of 3\arcsec. Where a target has both an external redshift measurement and a SAMI-CRS redshift with $z_{\rm conf} > 0.9$, the SAMI-CRS redshift is retained. If an object has no reliable SAMI-CRS redshift, but duplicated external redshift measurements, then the redshift with the lowest redshift uncertainty is selected. 

As noted in Section~\ref{aaomega_obs}, the archived AAOmega data targeted galaxies with fainter magnitudes \citep[in particular the][observations]{boue2008} and, therefore, have no existing object in the SAMI-CRS input catalogue. Of these additional objects, 1277 had reliable redshift measurements (out of a total 1617 additional spectra). Similarly, a handful of objects (less than one percent of the total number of redshifts) from the external catalogues have no match in the input target catalogue. The majority of these unmatched objects occur in the clusters covered by the VST/ATLAS photometry and are due to objects that fall within small holes in coverage in at least one of the $g-$, $r-$ or $i-$band images \citep[][]{shanks2015}. There were several duplicate redshifts due to shredded galaxies, along with mis-classified stars, that were removed from the catalogues after visual inspection. The remaining unmatched objects form a separate catalogue and are included in the determination of the cluster properties in Section~\ref{cluster_properties}, but not in the selection of SAMI-GS targets in Section~\ref{sami_targets}. 

The final catalogue contains 11855 reliable redshift measurements with ${\rm R} < 2{\rm R}_{200}$ that are matched to the SAMI-CRS input catalogue. Of these measurements, 9278 come from the SAMI-CRS, 1213 from the SDSS, 1123 from the 2dFGRS, 106 from the 6dFGS, 56 from the Durret catalogue, 29 from the CAIRNS, 26 from the WINGS and 24 from the NFPS.

\subsection{Spectroscopic completeness}\label{spec_comp}

Our goal for the redshift survey was to reach a similar magnitude limit and spectroscopic completeness level (within 1\rtwo\, for each cluster) to that obtained by the GAMA-I survey \citep{driver2011}, which is the survey from which the primary SAMI-GS targets are selected \citep{bryant2015}. The overall spectroscopic completeness at our nominal magnitude limit ($r_{petro} =19.4$) for each cluster is listed in Table~\ref{clus_table} and is greater than 90 per cent within \rtwo\, for all but the cluster Abell~119, where it is 89 per cent. The majority of the targets that do not have a reliable redshift have been observed, but the spectrum was of too low S/N to produce a reliable redshift. We note that while the original input catalogues for the SAMI-CRS were selected based on the photometry described in Section~\ref{input_cat}, throughout this section we have updated the photometry for the objects in the input catalogues with the latest measurements described in Section~\ref{new_photo}. The impact of this change mainly affects the clusters with VST/ATLAS photometry at close to the limiting magnitude of our spectroscopy, where the $r_{petro} =19.4$ magnitude limit becomes less well-defined. However, the use of the updated photometry allows for a consistent check of how the spectroscopic completeness in the redshift survey affects the selection of targets for the SAMI-GS described in Section~\ref{SAMI_TS}.

\begin{figure*}
\includegraphics[angle=0,width=.95\textwidth]{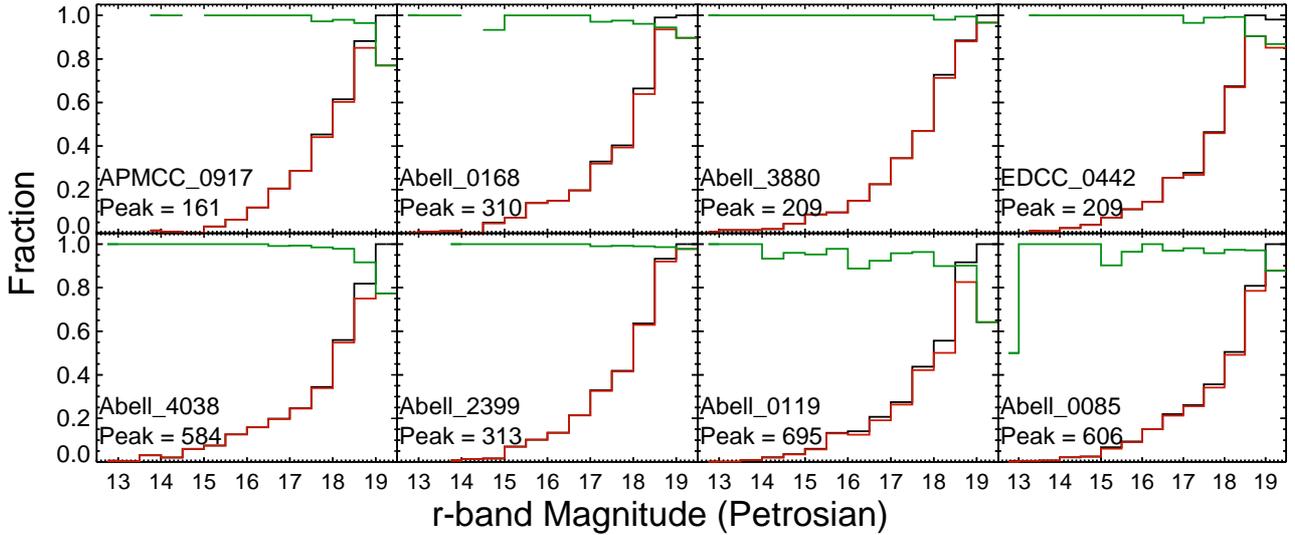}
\caption{Fractional galaxy numbers and completeness as a function of $r$-band Petrosian magnitude for galaxies within 2\rtwo. Black histogram shows the input target catalogue, the red histogram shows the number of galaxies with reliable redshift measurements, and the green line shows the spectroscopic completeness as a function of $r$-band magnitude.
\label{spectro_completeness_mag}}
\end{figure*}

\begin{figure*}
\includegraphics[angle=0,width=.95\textwidth]{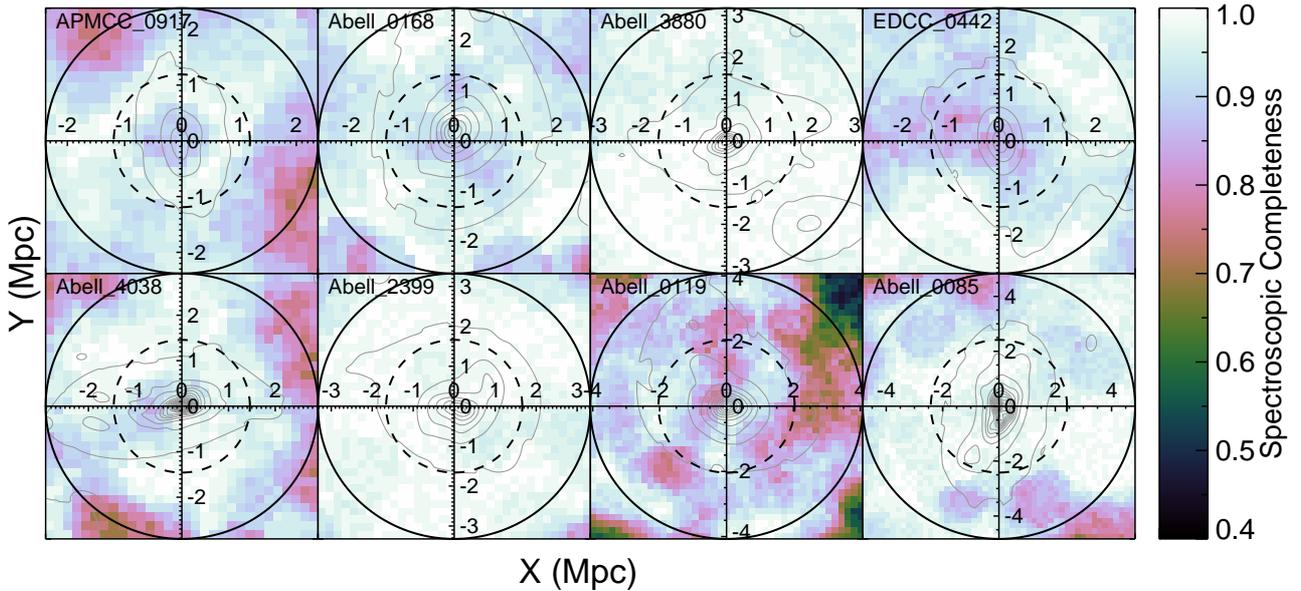}
\caption{The spatial distribution of the spectroscopic completeness. For each $150\times150\,{\rm kpc^2}$ pixel, a radius, $R$, was defined which contains 50 galaxies from the input catalogue. The spectroscopic completeness at that pixel is defined as $N(<R)_z$/50. The black dashed circle shows the \rtwo\, radius, and the black solid circle shows the 2\rtwo\, radius. The grey contours show the member galaxy isopleths as shown in Figure~\ref{gal_dist}.
\label{spectro_completeness_spat}}
\end{figure*}

To determine if the completeness is homogeneous across both the magnitude range and the spatial extent on the sky, we investigated the spectroscopic completeness as a function of $r$-band Petrosian magnitude in Figure~\ref{spectro_completeness_mag} and also as a function of position on the sky in Figure~\ref{spectro_completeness_spat}. The spectroscopic completeness per magnitude bin (green histograms in Figure~\ref{spectro_completeness_mag}) is calculated as the ratio of the number of galaxies for which a reliable redshift measurement exists (red histograms in Figure~\ref{spectro_completeness_mag}) to the number of galaxies in the input catalogue (black histograms in Figure~\ref{spectro_completeness_mag}). The spatial distribution of the completeness (Figure~\ref{spectro_completeness_spat}) is calculated at each pixel by determining the radius to the 50th nearest target galaxy to the pixel of interest. The number of targets with a reliable redshift measurement, $N_z$, within that radius is then determined, with the completeness computed as $N_z$/50. 

\begin{figure*}
\includegraphics[angle=0,width=.95\textwidth]{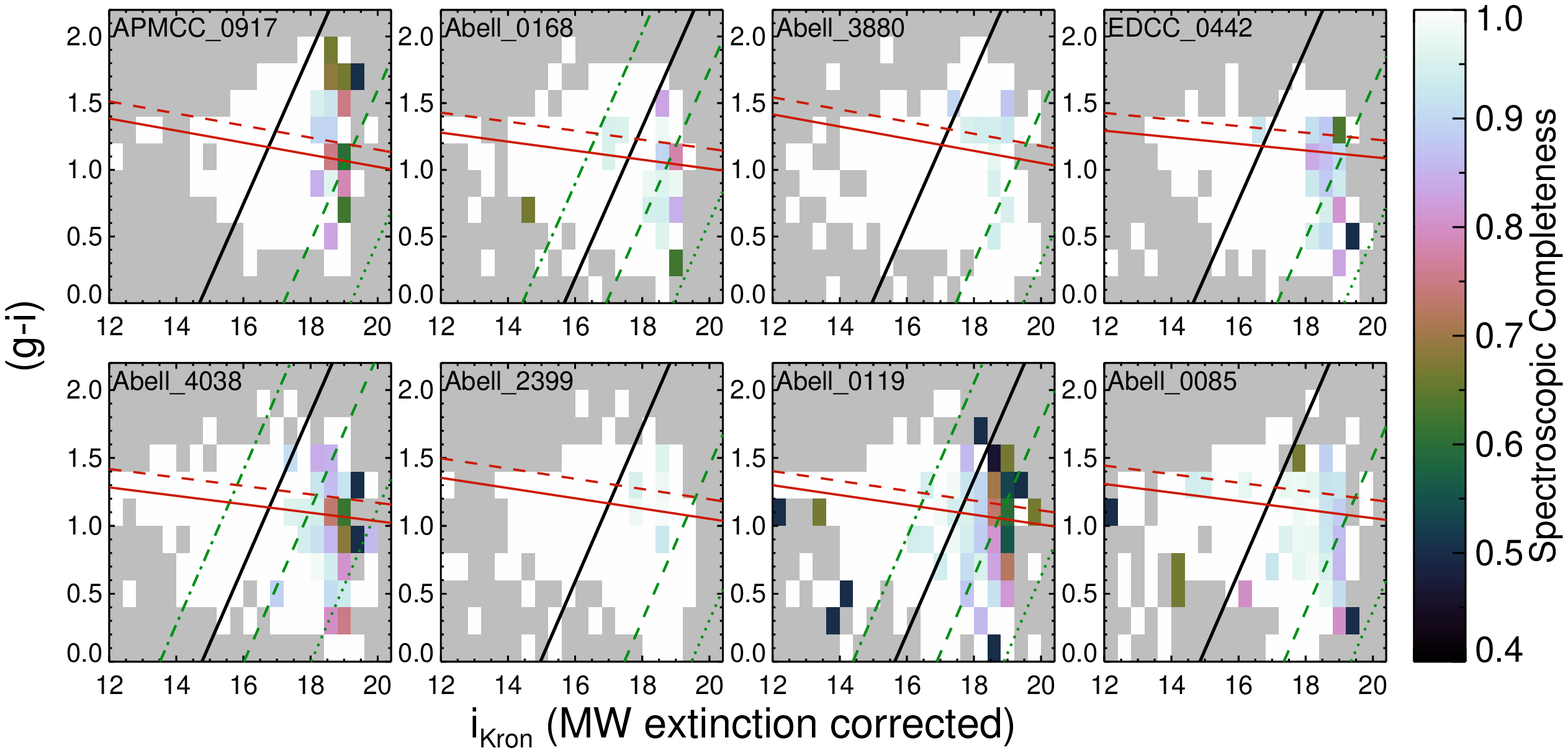}
\caption{The spectroscopic completeness as a function of position in $(g-i)$ vs $i$ for galaxies with $R<2$\rtwo. The solid red line shows the position of the cluster red sequence and the dashed red line shows the upper $2\sigma$ scatter around the red-sequence. The green lines show the $i$-band magnitude as a function of $(g-i)$ colour for stellar mass limits used in the main portion of the SAMI-GS, i.e., ${\rm log}_{10} (M^*_{\rm approx}$/\msolar$) = 8.2, 9.0,\, {\rm and}\, 10$ (dotted, dashed and dot-dashed lines, respectively). These trends are determined using Equation~\ref{mass_proxy}. The solid black line shows the trend for the stellar mass limit of the cluster of interest; either ${\rm log}_{10} (M^*_{\rm approx}/$\msolar$) = 9.5\, {\rm or}\, 10$, depending on $z_{\rm clus}$ (see Section~\ref{sami_targets}). 
\label{spectro_completeness_gmi}}
\end{figure*}

Clearly, we do not reach the high level of completeness achieved in the GAMA-I survey ($\sim 98\%$) for all of the clusters. In particular, for the clusters A119, APMCC0917 and A4038 the completeness drops below $80\%$ for galaxies fainter than $r_{\rm petro} = 19$. To determine if the lower spectroscopic completeness at fainter magnitudes will impact the selection of SAMI-GS targets described in Section~\ref{sami_targets}, we investigate the spectroscopic completeness as a function of position in the colour-magnitude diagram. Since the $g$- and $i$-band magnitudes are used to determine the stellar mass proxy for SAMI-GS target selection (see Equation~\ref{mass_proxy}), we plot the completeness in $(g-i)$ versus $i$ in Figure~\ref{spectro_completeness_gmi}. Overplotted are lines showing how the $i$-band magnitude varies with $g-i$ colour for the stellar mass limits ${\rm log}_{10} (M^*_{\rm approx}$/\msolar$)=8.2,\, 9.0$ and $10.0$. These are the stellar mass limits used for the main SAMI-GS primary target selection for galaxies in the redshift range probed by the clusters \citep{bryant2015}. The $(g-i)$-$i$ trends are determined from Equation~\ref{mass_proxy} and using the cluster redshift, $z_{\rm clus}$. The two clusters A4038 and A119 have low spectroscopic completeness ($<60\%$) at the main SAMI-GS limits for their redshifts (${\rm log}_{10} (M^*_{\rm approx}$/\msolar$)=8.2,\, 9.0$, respectively), particularly for galaxies on the cluster red-sequence (shown as red line in each panel of Figure~\ref{spectro_completeness_gmi}). However, for the reasons outlined in Section~\ref{sami_targets}, we set a lower limit of ${\rm log}_{10} (M^*_{\rm approx}$/\msolar$)=9.5$ for the primary cluster targets when $z_{\rm clus} < 0.045$. The black lines in Figure~\ref{spectro_completeness_gmi} show how the $i$-band magnitude varies with $g-i$ colour for the stellar mass limits determined for the clusters in Section~\ref{sami_targets}. At these stellar mass limits, the spectroscopic completeness is $>95$ per cent for all clusters. Moreover, the depth of the survey (at least 3 magnitudes fainter than the knee in the cluster luminosity function) allows for the collection of a large number of spectroscopically confirmed members even at the relatively poorer completeness levels reached for A119. The large number of cluster member redshifts will enable the robust characterisation of the dynamical properties of the clusters. We therefore conclude that, despite not quite achieving our initial goals, the SAMI-CRS is sufficient for our purposes. Importantly, Figure~\ref{spectro_completeness_gmi} shows that at the stellar mass limits used to define primary targets for the  SAMI-GS in Section~\ref{SAMI_TS}, the spectroscopic completeness is very high and will not impact the target selection for the SAMI-GS.

\section{Cluster membership and global parameters}\label{cluster_properties}
In this section, we describe the selection of spectroscopically confirmed cluster members and parameters derived from the member redshifts such as the cluster redshift, velocity dispersion and mass. These parameters are listed for each cluster in Table~\ref{clus_table}.
\subsection{Determination of cluster membership, velocity dispersion and \rtwo}\label{subsect:memsel}

The allocation of cluster membership is a multi-step process. First, obvious interlopers are rejected as non-members if they lie further than a projected distance of $R=6\,$Mpc from the cluster centre and have peculiar velocity $|v_{\rm pec}| \ge 5000\,$\kms\, where $v_{\rm pec}=c(z-z_{\rm CCG})/(1+z_{\rm CCG})$, $z_{\rm CCG}$ is the bright central cluster galaxy (CCG) redshift, which is a good initial approximation of the cluster redshift. The projected distances are measured from the R.A. and decl. of the cluster centres listed in Table~\ref{clus_table}. In the majority of cases, the selection of the cluster centre is obvious; there is a single bright, CCG for A3880, EDCC~442, A119 and A85 which marks the cluster centre. However, for APMCC0917, A4038, A168 and A2399 there are one or more candidates for a CCG. In those cases, the coordinates of the brightest CCG closest to the peak in galaxy surface density (see Section~\ref{cluster_structure}) was used for the cluster centre. For the clusters A168, APMCC0917 and A2399, the CCG closest to the peak in the galaxy density distribution was not the brightest galaxy in the cluster. In fact, for these three clusters the brightest cluster galaxies were located as far as $800$\,kpc from the defined cluster centre. As we will show in Section~\ref{cluster_structure}, A168 and A2399 host substructures associated with these bright galaxies. We note that the centres for A168 and A2399 differ from those listed in \citet{bryant2015}.

\begin{figure*}
\includegraphics[angle=0,width=.42\textwidth]{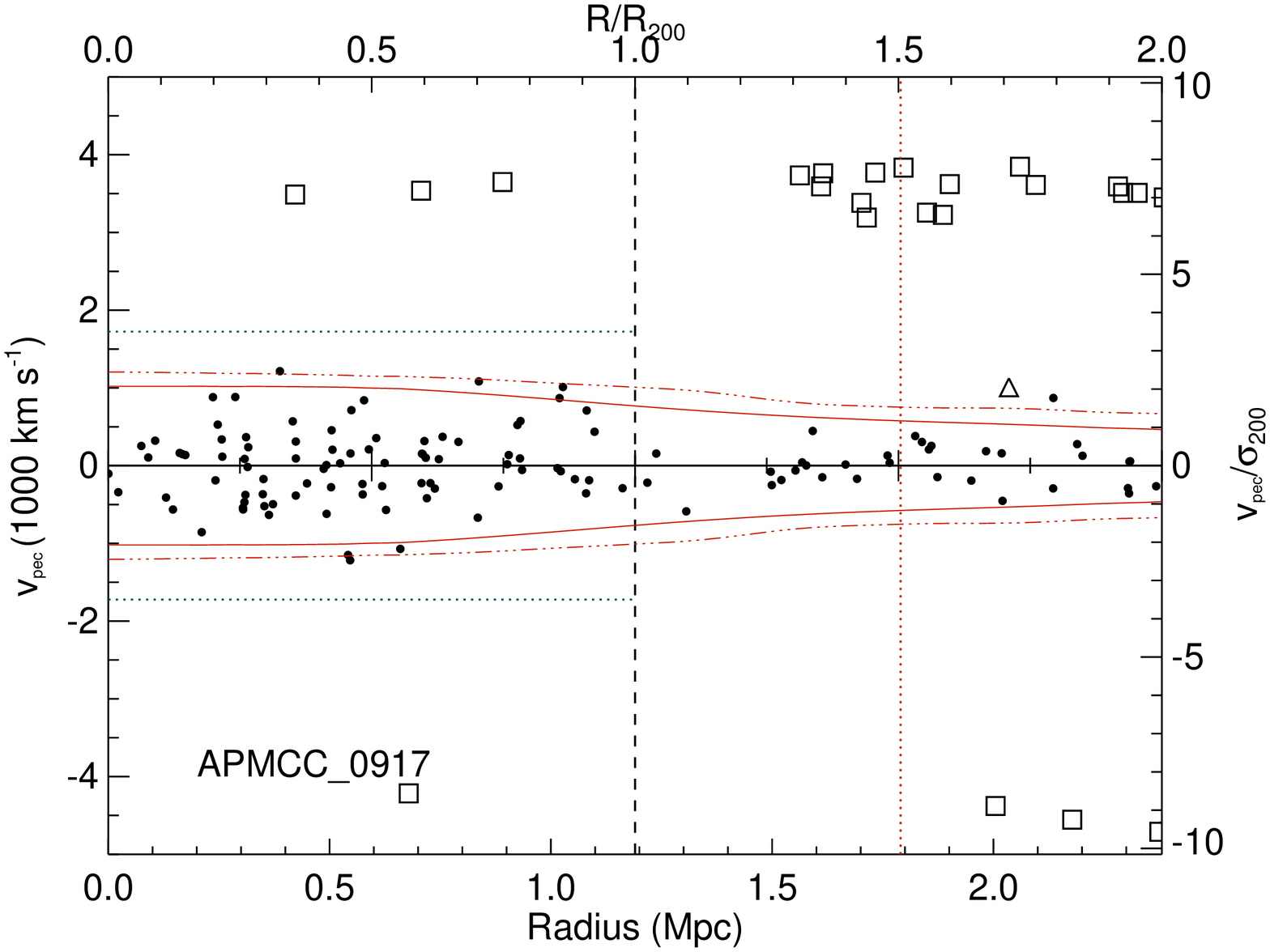}
\includegraphics[angle=0,width=.42\textwidth]{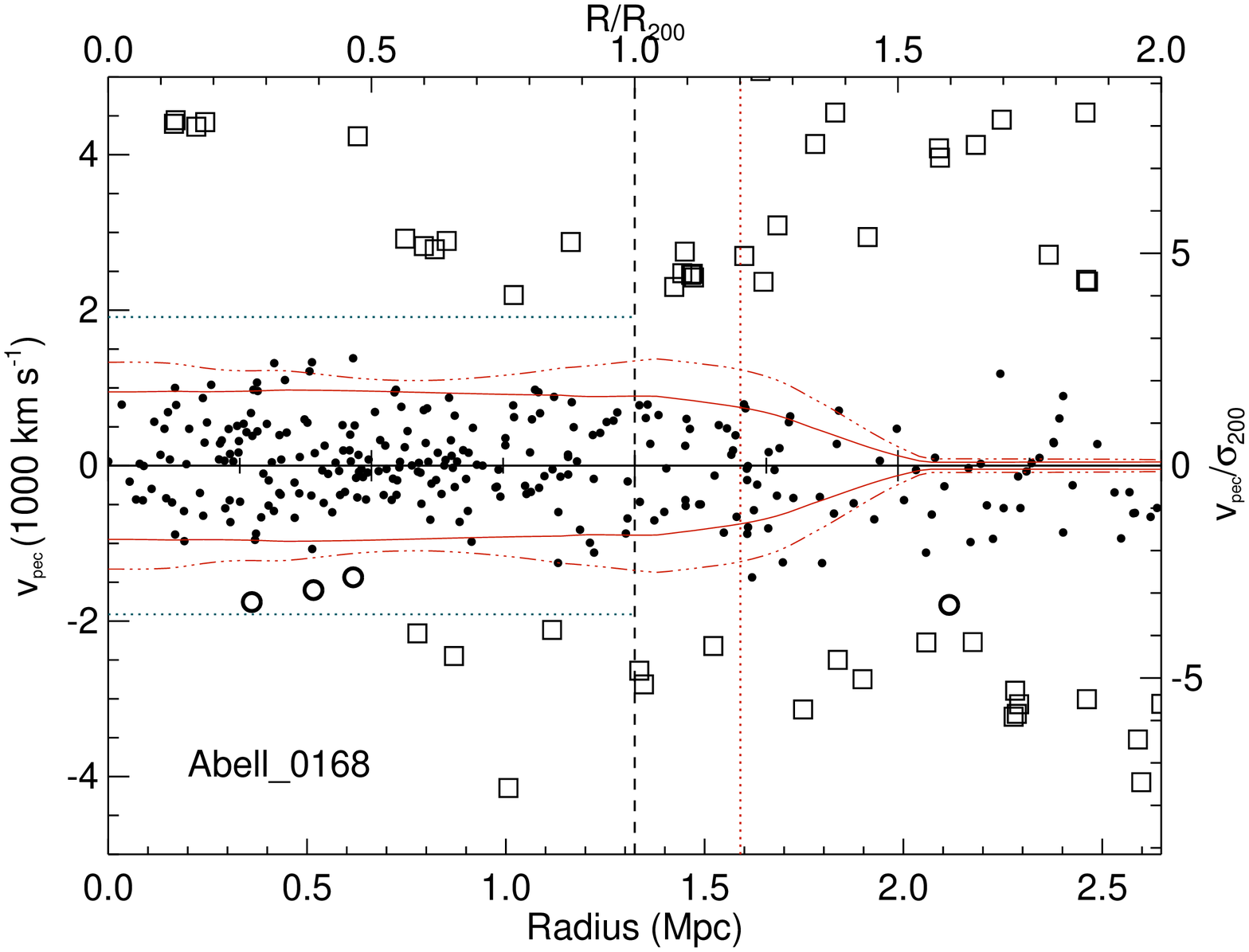}\\
\includegraphics[angle=0,width=.42\textwidth]{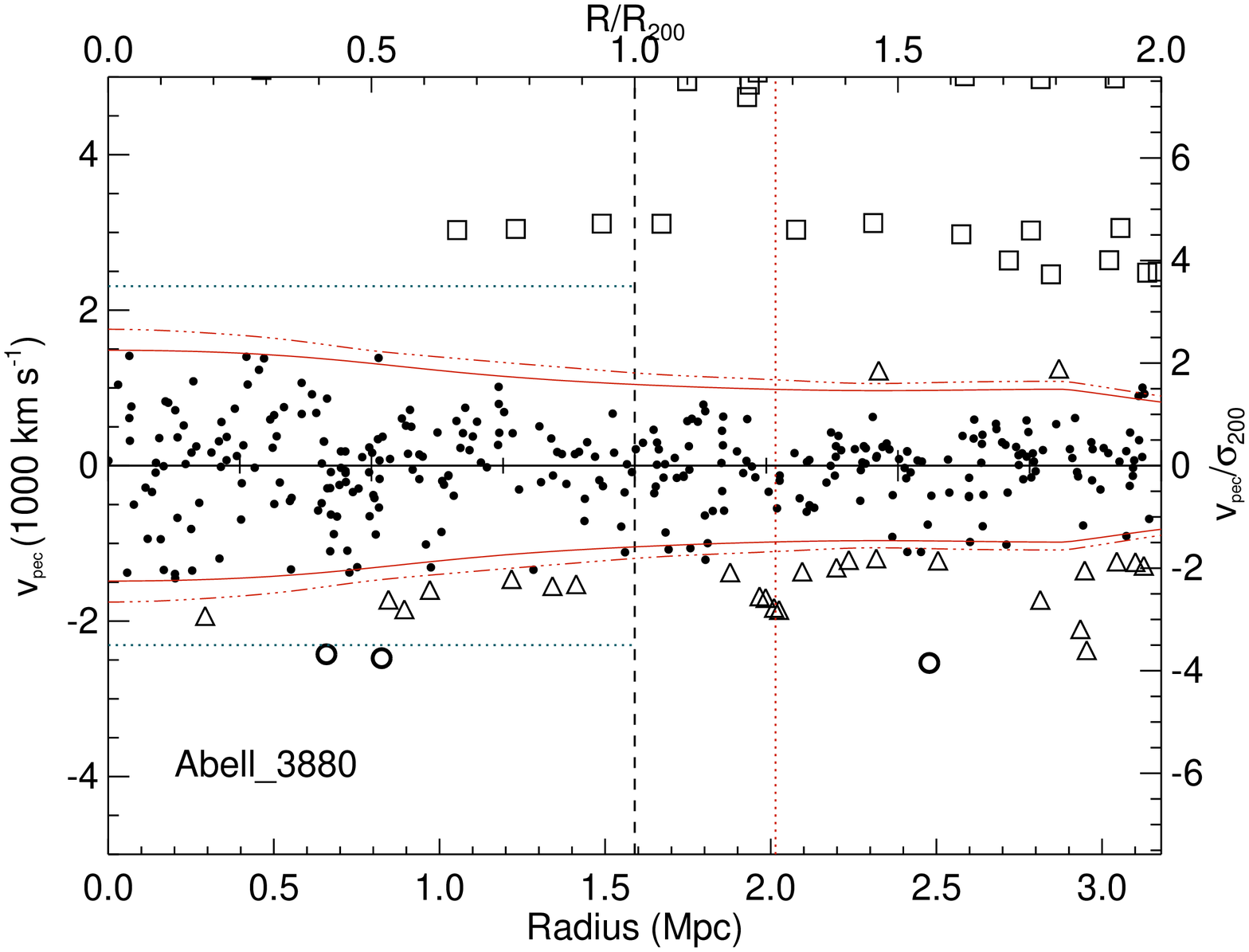}
\includegraphics[angle=0,width=.42\textwidth]{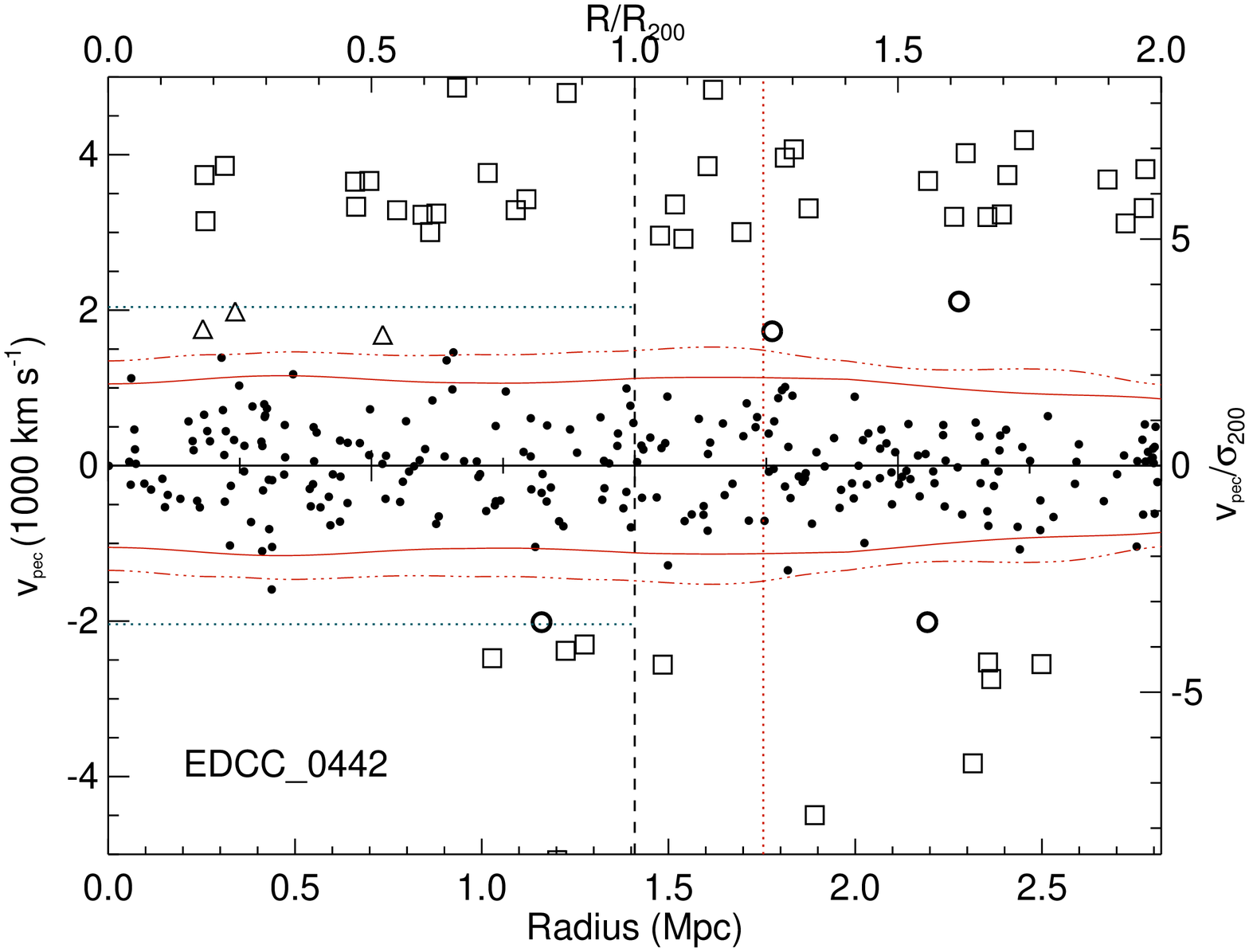}\\
\includegraphics[angle=0,width=.42\textwidth]{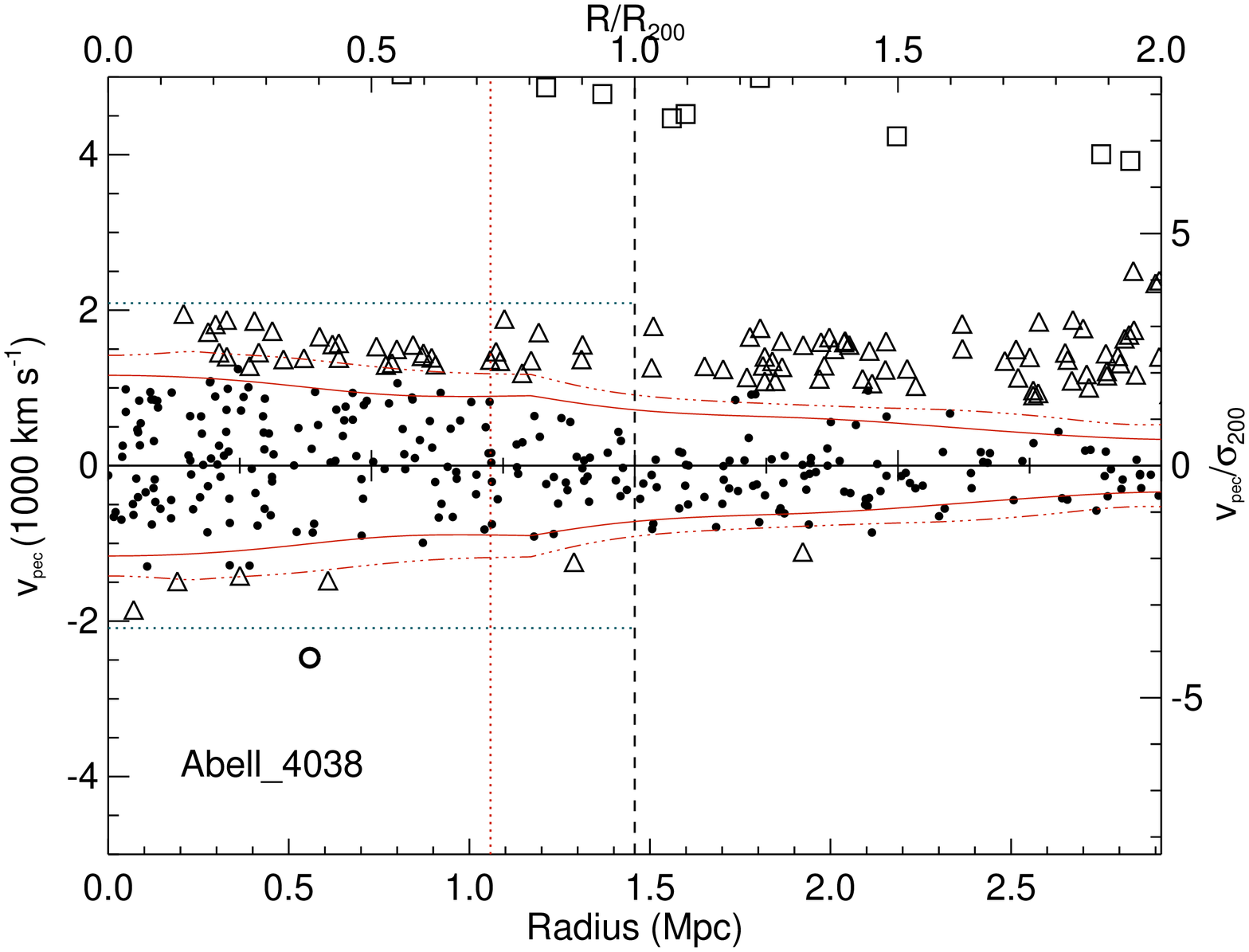}
\includegraphics[angle=0,width=.42\textwidth]{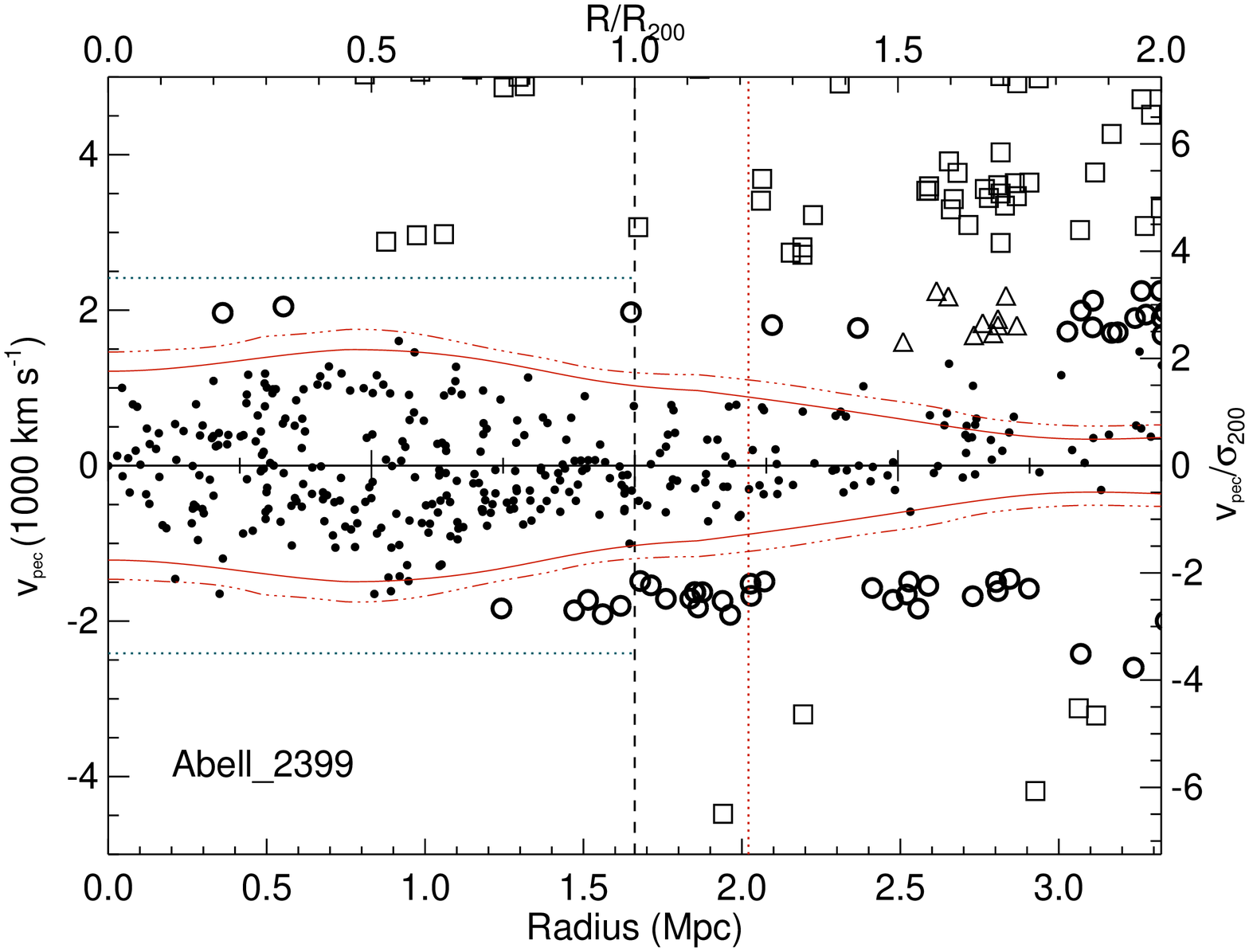}\\
\includegraphics[angle=0,width=.42\textwidth]{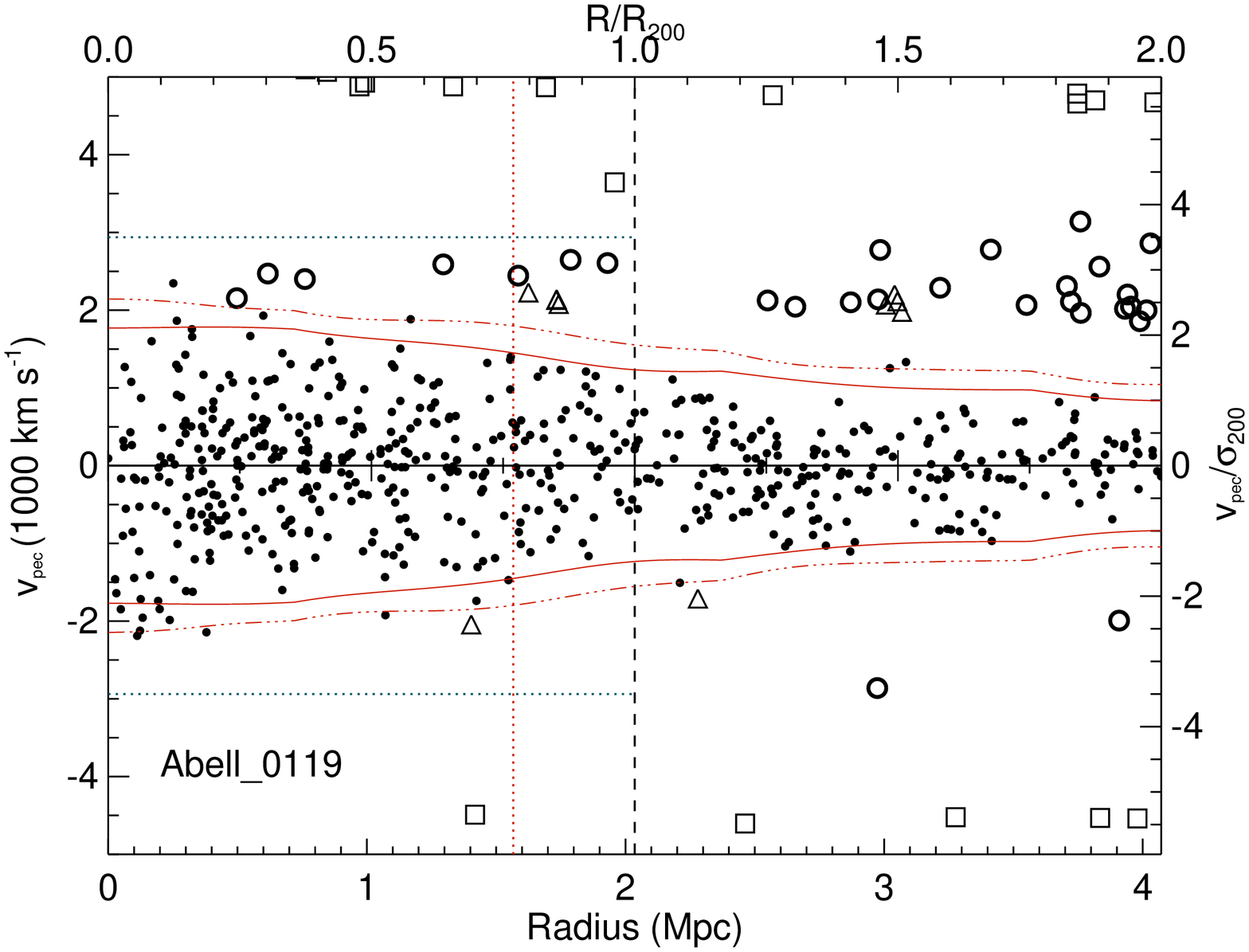}
\includegraphics[angle=0,width=.42\textwidth]{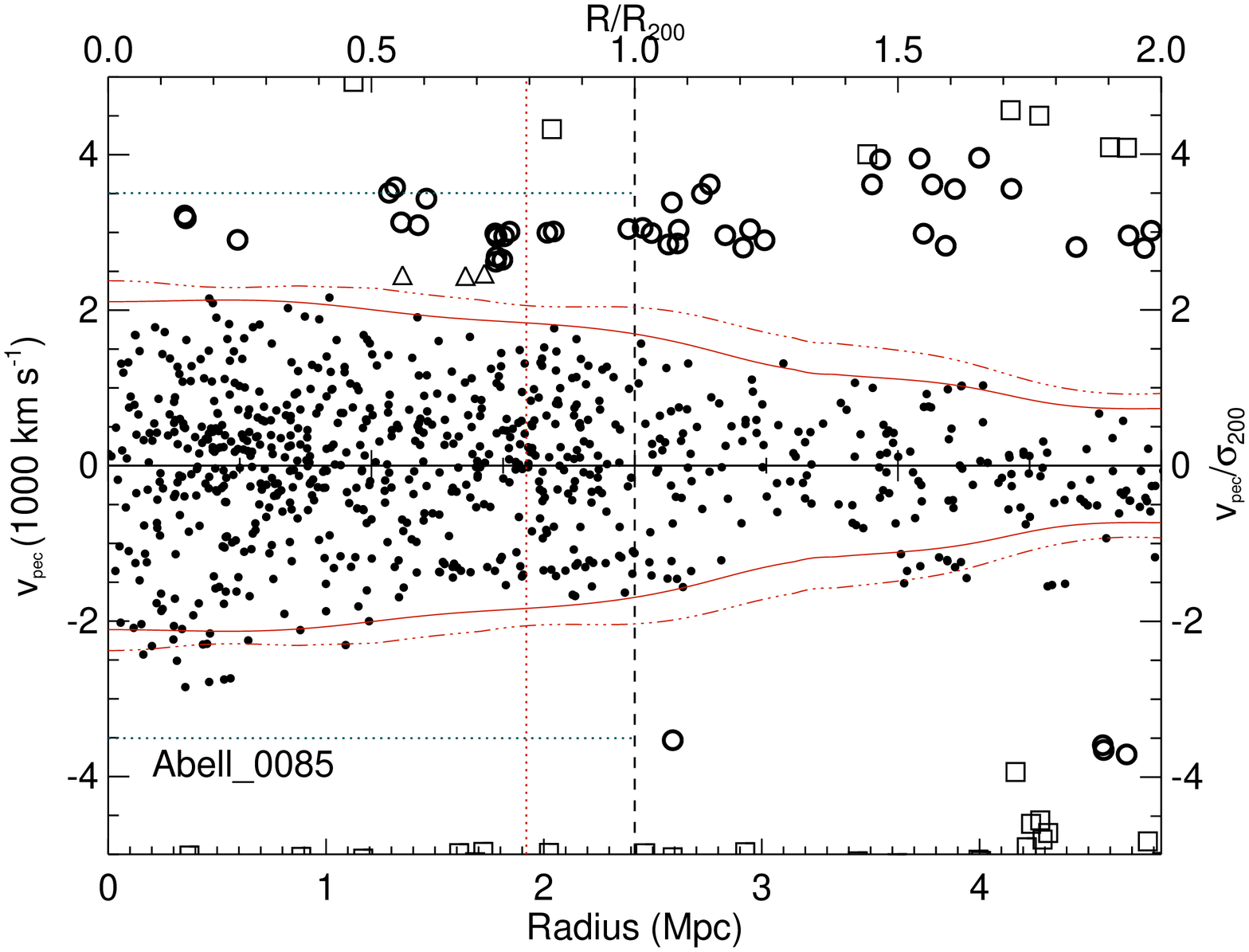}\\
\caption{These figures show the phase-space distribution of galaxies within $c|(z-z_{clus})/(1+z_{clus})| < 5000\,$\kms and $R < 2$\rtwo. The galaxies defined as cluster members are shown as filled black circles. The caustics, which define the $v_{esc}$ profile based on the cluster members, are shown as solid red lines. Non-members have shapes that reflect the step at which they were rejected; open squares show galaxies rejected because $|v_{\rm pec}| > 3.5\sigma_{200}$, open circles show galaxies rejected by the shift-gapper, and open triangles show galaxies rejected by the caustics. The vertical dashed lines show the $r_{200}$ radius and the vertical dotted red line shows the radial limit of the SAMI FOV (0.5 degree radius).  The blue horizontal dotted lines show the $3,5\sigma_{200}$ limits used for the selection of SAMI targets. This selection is allowed to be looser than the caustics selection which may change with more data.
\label{mem_allocation}}
\end{figure*}

Following this cut in peculiar velocity and clustercentric distance, the remaining galaxies are used to obtain an estimate of the cluster velocity dispersion, $\sigma_{200}$, using the biweight scale estimator, which is a robust estimator of scale in the presence of outliers \citep{beers1990}. The value of $\sigma_{200}$ is determined from those galaxies within the virial radius, which is estimated as $R_{200}=0.17\sigma_{200}/{\rm H(z)}$\,Mpc \footnote{where $R_{200}$ is the cluster radius within which the mean density is 200 times the critical density, where the cluster density distribution is assumed to follow that of a single isothermal sphere \citep{carlberg1997}}. Since $R_{200} \propto \sigma_{200}$, the process is iterated until the values of $R_{200}$ and $\sigma_{200}$ are stable. A second cut in peculiar velocity is then applied such that those galaxies with $|v_{\rm pec}| > 3.5\sigma_{200}$ are removed from the member sample. The galaxies that are removed by the $3.5\sigma_{200}$ cuts are shown as black open squares in Figure~\ref{mem_allocation}. 

The above method of using only velocity information is sufficient for the identification of obvious non-members, however, it is not a completely rigorous approach to interloper rejection \citep{denhartog1996,vanhaarlem1997, wojtak2007a,wojtak2007b}. More robust techniques for identifying line-of-sight interlopers utilize the peculiar velocity as a function of cluster-centric-distance. Here, for the second step in selecting cluster members we use a slightly modified version of the ``shifting-gapper'' technique \citep{fadda1996} which has the advantages of being a fast, model-independent method of interloper rejection. The ``shifting-gapper'' is applied as follows. Centred at the radius of each potential cluster member, an adaptive annular bin containing at least $N=50$ other potential cluster members is generated. Within this bin, the galaxies are sorted in order of increasing $v_{\rm pec}$. The velocity difference between successive galaxies is determined as $v_{\rm gap} = v_{\rm i+1} - v_{\rm i}$. Any galaxy that is separated by a $v_{\rm gap} > \sigma_{200}$ from the adjacent galaxy is rejected as a non-member, as are all galaxies with $v_{\rm pec}$ larger than (or smaller than for negative $v_{\rm pec}$) the newly defined non-member. Galaxies identified as non-members using this method are shown in Figure~\ref{mem_allocation} as black open circles. We note that the choice of $\sigma_{200}$ as the maximum allowed gap is somewhat arbitrary, although it was found to produce good results here (see Figure~\ref{mem_allocation}), and in other clusters \citep[e.g.,][]{zabludoff1990, owers2009a, owers2009b, owers2011a, nascimento2016}.


The next step in the procedure involves using the adaptively-smoothed distribution of galaxies in $v_{\rm pec}$-radius space to locate the cluster caustics \citep{diaferio1999}. The caustics trace the escape velocity of the cluster as a function of cluster-centric radius and, therefore, robustly identify the boundary in $v_{\rm pec}-$radius space between bona-fide cluster members and line-of-sight interlopers \citep[e.g.,][]{serra2013, owers2013, owers2014}. Identifying the location of the caustics in the projected-phase-space (PPS) diagram requires determining an adaptive smoothing kernel that minimises statistical fluctuations without over-smoothing real structure in dense regions. 

Our procedure for determining such an adaptive kernel follows the general procedure outlined by \citet[][]{silverman1986} (see also \citet{pisani1996} and \citet{diaferio1999}). Briefly, an initial pilot estimation of the density distribution in PPS is determined by smoothing the PPS distribution with a kernel of fixed width. The width of this kernel, $\sigma_{\rm smth}$, is determined by the Silverman's rule of thumb estimate $\sigma_{\rm smth} = A \sigma_{\rm dist} N^{-1/6}$ where $N$ is the number of data points, $A=0.8$ \citep[which is 25 percent below the optimal value for a Gaussian kernel, as recommended by][ to avoid over-smoothing in the presence of multi-modality]{silverman1986},  and $\sigma_{\rm dist}$ is an estimate of the standard deviation of the distribution. The final value for $\sigma_{\rm dist}$ is taken to be the minimum of a number of estimators including the standard deviation, median-absolute deviation, the interquartile range, sigma-clipped, biweight and the standard deviation estimated when including higher order Gauss-Hermite polynomials \citep[as described in][]{owers2009a,zabludoff1993}. The estimate for $\sigma_{\rm smth}$ is determined separately for the distributions in the $v_{\rm pec}$ and radial direction; the $\sigma_{\rm dist}$ is determined from the distribution of galaxies with cluster-centric distances less than \rtwo. 

The pilot estimate of the density distribution is used to define the local kernel widths $\sigma_{\rm R,v_{\rm pec}} = h_{\rm R,v_{\rm pec}} (\gamma/f_P({\rm R,v_{pec}}))^{1/2}$, where $f_P({\rm R,v_{pec}})$ is the pilot density at the point of interest, ${\rm log}(\gamma) = {\overline{{\rm log}(f_P({\rm R,v_{pec}}))}}$, and the $h_{\rm R,v_{pec}}$ values control the amount of smoothing in the x- and y-directions. The $h_{\rm R,v_{pec}}$ values are determined iteratively by using least-squares cross validation as described elsewhere \citep{silverman1986, diaferio1999}. The locally adaptive smoothing kernels are used to produce the final estimate of the density in PPS, $f({\rm R,v_{pec}})$.

Having adaptively smoothed the PPS distribution, the location of the caustics need to be determined. This is achieved by determining the value $f({\rm R,v_{pec}})=\kappa$ that minimises $(\langle {\rm v_{\rm esc}}({\rm R}_{200})^2 \rangle - 4 \sigma_{200}^2)^2$ where $\sigma_{200}$ is the velocity dispersion determined within $R_{200}$ using the biweight estimate, 
\begin{equation}
\langle {\rm v_{\rm esc}(R}_{200})^2 \rangle=\int_{0}^{R_{200}} A^2_{\kappa}(R)\varphi(R) dR/\int_0^{R_{200}} \varphi(R) dR
\end{equation}
with $\varphi(R)=\int f({\rm R,v}) dv$ \citep{diaferio1999}. The value of $A_{\kappa}(R)$ is the location of the caustic amplitude that traces the escape velocity as a function of radius for a given $\kappa$\, value. As described in \citet{serra2013}, due to asymmetries in the velocity component of the $f({\rm R,v_{pec}})$ distribution, a single $\kappa$\, value  results in two distinct velocity choices for $A_{\kappa}({\rm R})$, ${\rm v_{pos}(R)}$ and ${\rm v_{neg}(R)}$, where in general $|{\rm v_{neg}(R)}| \ne {\rm v_{pos}(R)}$. For the purpose of membership selection, the choice of $A_{\kappa}({\rm R})$ is somewhat subjective, but in general the chosen $A_{\kappa}({\rm R})$ is the one that falls on the cleanest side of the ${\rm v_{\rm pec}}$ distribution. For example, for A85 and A4038 the separation between the main cluster and the line-of-sight interlopers is far cleaner on the $v_{neg}(R)$ side of the PPS distribution, and so we set $A_{\kappa}(R) =|v_{neg}(R)|$.
Uncertainties on the values of $A(R)$ are estimated as described in \citet{diaferio1999}, i.e., $\delta A(R)/A(R) \simeq \kappa/{\rm max}(f({\rm R,v_{pec}}))$. For member selection, we reject any galaxy at radius $R$ that has $|v_{pec}| > A(R) +\delta A(R)$ as interlopers (shown as open triangles in Figure~\ref{mem_allocation}), while any galaxy that was initially rejected as a non-member during the shifting-gapper selection, but has $|v_{pec}| \leq A(R)$ is reinstated as a cluster member. This process of shift-gapper plus caustics member allocation is iterated until the number of members within 2\rtwo\, becomes stable. At each iteration the cluster redshift, the $\sigma_{200}$ and the \rtwo\, are remeasured. The number of spectroscopically confirmed cluster members within \rtwo\, and 2\rtwo\, is shown in Table~\ref{clus_table}. In total, there are 1935 and 2899 confirmed members within \rtwo\, and 2\rtwo, respectively.

\subsection{Cluster mass measurements}\label{cluster_masses}

For each cluster the final catalogue of cluster members is used to determine the cluster redshift, velocity dispersion, \rtwo\, and $M_{200}$ listed in Table~\ref{clus_table}. The virial radius estimate, \rtwo, and velocity dispersion, $\sigma_{200}$, are determined iteratively as in Section~\ref{subsect:memsel}. The cluster redshift is determined using the biweight estimator of location \citep{beers1990} for galaxies within 2\rtwo. The mass, $M_{200}$, is determined within \rtwo\, using both the virial and caustic estimators. The virial mass determination follows the same prescription described elsewhere \citep{girardi1998,owers2009a, owers2013}. Briefly, the corrected virial mass is
\begin{equation}\label{mvir}
M (R < R_{200}) = M_{vir} - C = {3 \pi \over 2} {\sigma_v^2 R_{PV} \over G} -C
\end{equation}
where $C\approx 0.19M_{vir}$ is an approximation to the surface pressure term correcting for cluster mass external to \rtwo\, \citep{girardi1998} and the projected virial radius is
\begin{equation}\label{rpv}
R_{PV} = {N_{200}(N_{200}-1) \over \sum_{i=j+1}^{N_{200}} \sum_{j=1}^{i-1} R^{-1}_{ij}}
\end{equation}
where $R^{-1}_{ij}$ is the projected separation between the $i$th and $j$th galaxies and $N_{200}$ is the number of galaxies within \rtwo. The uncertainties provided in Table~\ref{clus_table} are estimated by propagating the uncertainties on $\sigma_{200}$ and $R_{PV}$, which are estimated by using jackknife resampling.

The caustic masses are determined from the caustic amplitudes, $A(R)$, which are estimated from the adaptively smoothed PPS described in Section~\ref{subsect:memsel}. However, for the mass measurement we use the more conservative criterion $A(R) = {\rm min}((|v_{neg}(R)|, v_{pos}(R))$ so as to minimise the impact of asymmetry in the velocity distribution due to, e.g., substructure and interlopers \citep{diaferio1999}. The caustic mass estimator within a radius $R$ is given in Equation~13 of \citet{diaferio1999} and is
\begin{equation}\label{Mcaust}
	M(<R) = {F_{\beta} \over G} \int_0^R A^2(R) dR
\end{equation}
where the $F_{\beta}$ term is a calibration parameter that accounts for the combined effect of the radial dependence of the gravitational potential, the mass density and orbital anisotropy profiles. \citet{diaferio1999} argued that the combined effect of these profiles is a parameter that is a very slowly varying function of radius, and can be estimated as a constant. Here, we set $F_{\beta}=0.7$, which is the value suggested by \citet{serra2011} based on numerical simulations, although we note that other authors have suggested lower values are more appropriate \citep[][]{diaferio1999,gifford2013a,svensmark2015}. This parametrisation, along with the assumption of spherical symmetry, are significant sources of systematic uncertainty in the caustic mass measurements \citep{gifford2013b,svensmark2015}. The statistical uncertainties measured on the caustic masses are determined from the uncertainty estimate on the position of the caustic following the method outlined in \citet{diaferio1999}. The virial masses are systematically larger than the caustic masses, although the $1\sigma$ uncertainties generally overlap for any given cluster. Regardless of this offset, the updated mass measurements presented in Table~\ref{clus_table} show that the clusters selected sample the full mass range expected for rich clusters.

It is important to note that the mass determinations outlined above differ from those used to measure halo masses for the GAMA portion of the survey \citep[see][for details]{robotham2011}. The halo mass estimates of \citet{robotham2011} are virial-like and are calibrated using halos from simulated mock catalogues. Future analyses will use halo mass as an environment metric, so it is important to  determine the scaling between the SAMI-CRS and GAMA halo masses due to the different mass estimators.  To determine this, we select groups from the GAMA group catalogue (version 8; G3CV08) that have $z < 0.2$, $\sigma > 300\,$\kms\, and more than 30 members as defined by the Friends-of-Friends algorithm in \citet{robotham2011}, returning 56 halos. We use the iteratively defined centroids, the redshifts and the $\sigma$ values from the G3CV08 as initial inputs, and then determine membership and mass as outlined above for galaxies within $6\,$Mpc and $v_{\rm pec} \pm 3.5\sigma$ of the group. We compare the virial and caustic masses to those determined by \citet{robotham2011} after rescaling the GAMA masses by the suggested calibration factor \citep[A=10;][]{robotham2011}, and also to our assumed cosmology. We find that the GAMA-determined group masses are a factor of 1.37 (1.25) and 1.66 (1.46) larger than the virial and caustic mass estimates where the numbers in brackets are the ratios determined for GAMA groups with $N_{200} > 50$. This difference is unsurprising since the GAMA masses are calibrated to simulated halo masses that can differ from the $M_{200}$ estimates used here \citep{jiang2014}. We also compare the $\sigma$ measured in G3CV08 to that determined using the method presented above, and find that the GAMA $\sigma$ values are lower by a factor of 0.96. It is recommended that any future analyses using halo masses from both the cluster and GAMA samples should implement these factors, i.e., in addition to scaling to match cosmology, the cluster virial masses should be multiplied by 1.25 in order to match GAMA halo masses.

\subsection{Cluster structure}\label{cluster_structure}
The cluster mass measurements derived in Section~\ref{cluster_masses} may be impacted by the presence of cluster-cluster merger activity. The virial mass measurements assume a fully virialised, spherically symmetric cluster and the placement of the caustic location, $\kappa$, that traces the escape velocity profile as described in Section~\ref{subsect:memsel} assumes that the mass within \rtwo\, is distributed as an isothermal sphere. These assumptions are violated where there exists significant dynamical substructure due to cluster merger activity. Moreover, the dynamical growth of clusters may produce additional environmental influences on galaxies over and above those encountered in a relaxed system. We therefore investigate the structure of the clusters within \rtwo\, using three indicators of dynamical substructure: a one-dimensional test for structure on the velocity distribution, a two-dimensional indicator based on galaxy positions, and a three-dimensional indicator which uses the combination of position and velocity. 

\begin{figure*}
\includegraphics[angle=0,width=.4\textwidth]{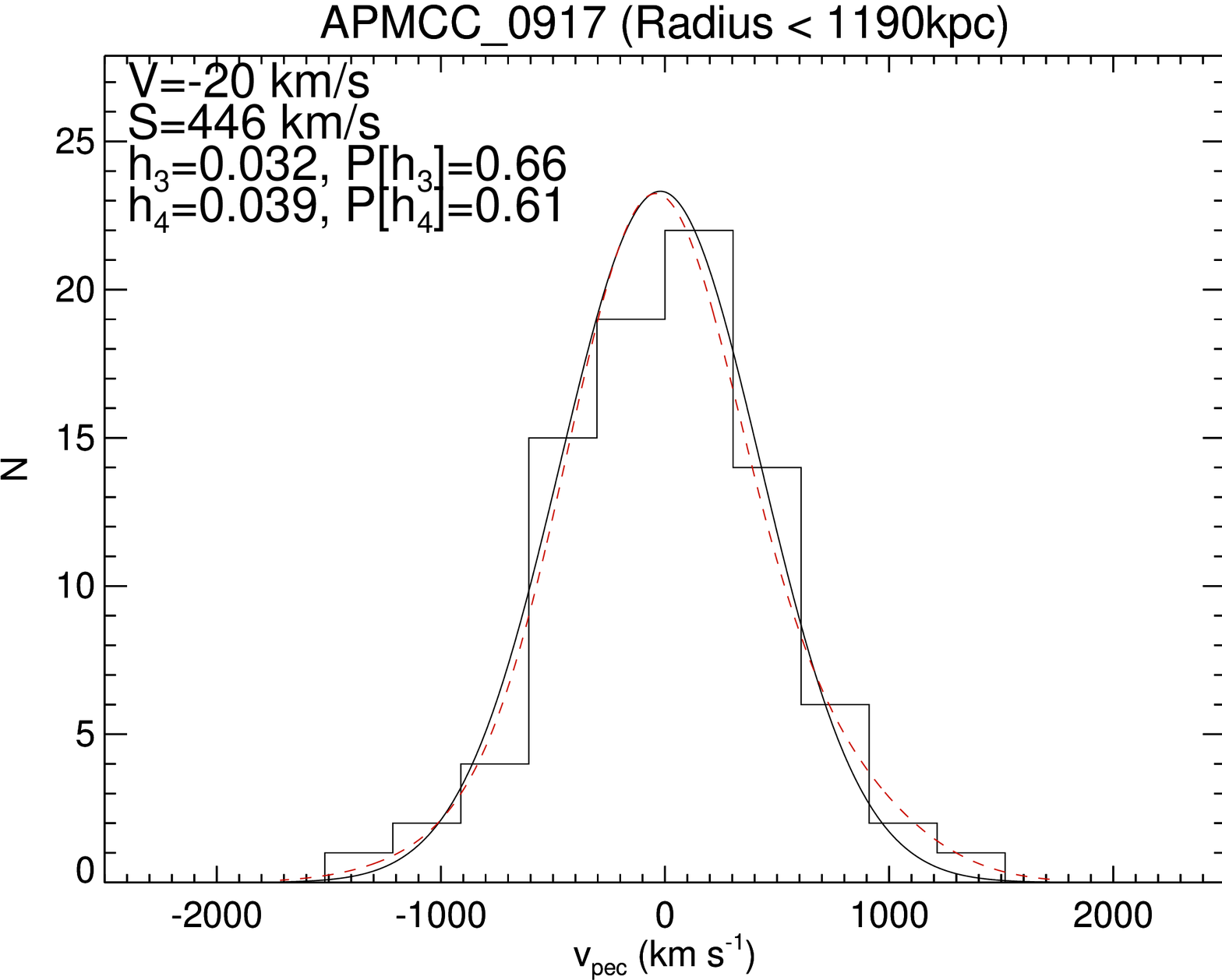}
\includegraphics[angle=0,width=.4\textwidth]{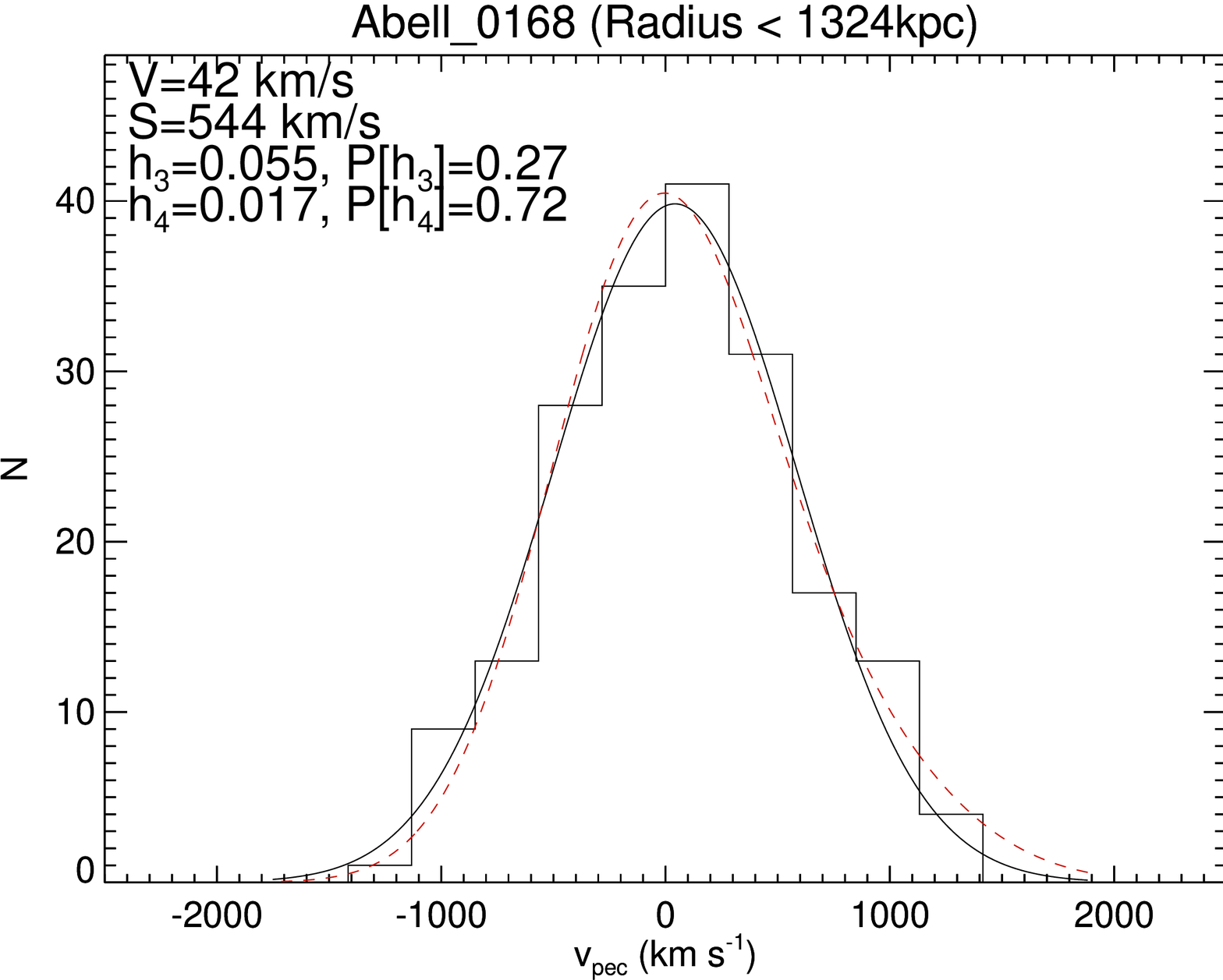}\\
\includegraphics[angle=0,width=.4\textwidth]{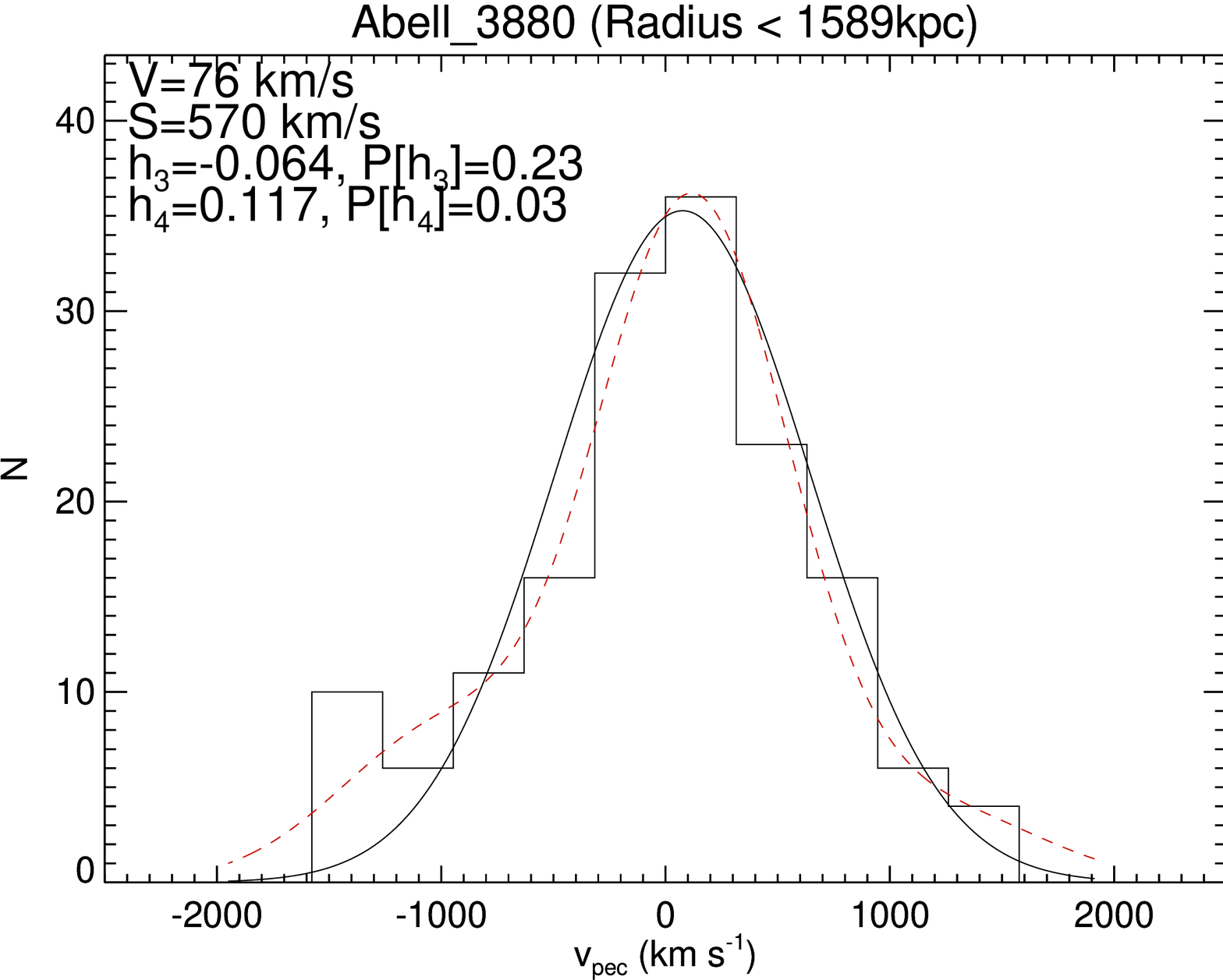}
\includegraphics[angle=0,width=.4\textwidth]{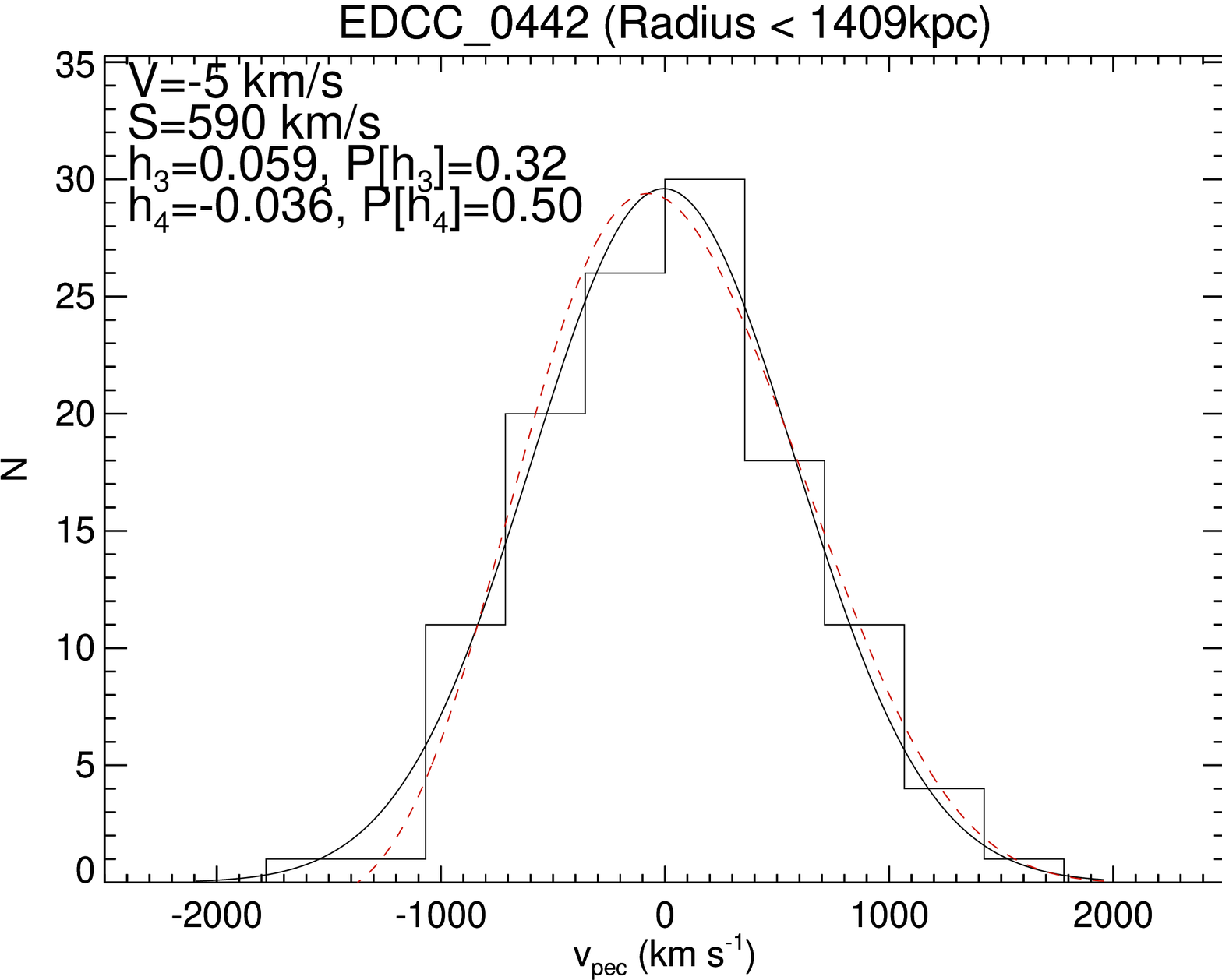}\\
\includegraphics[angle=0,width=.4\textwidth]{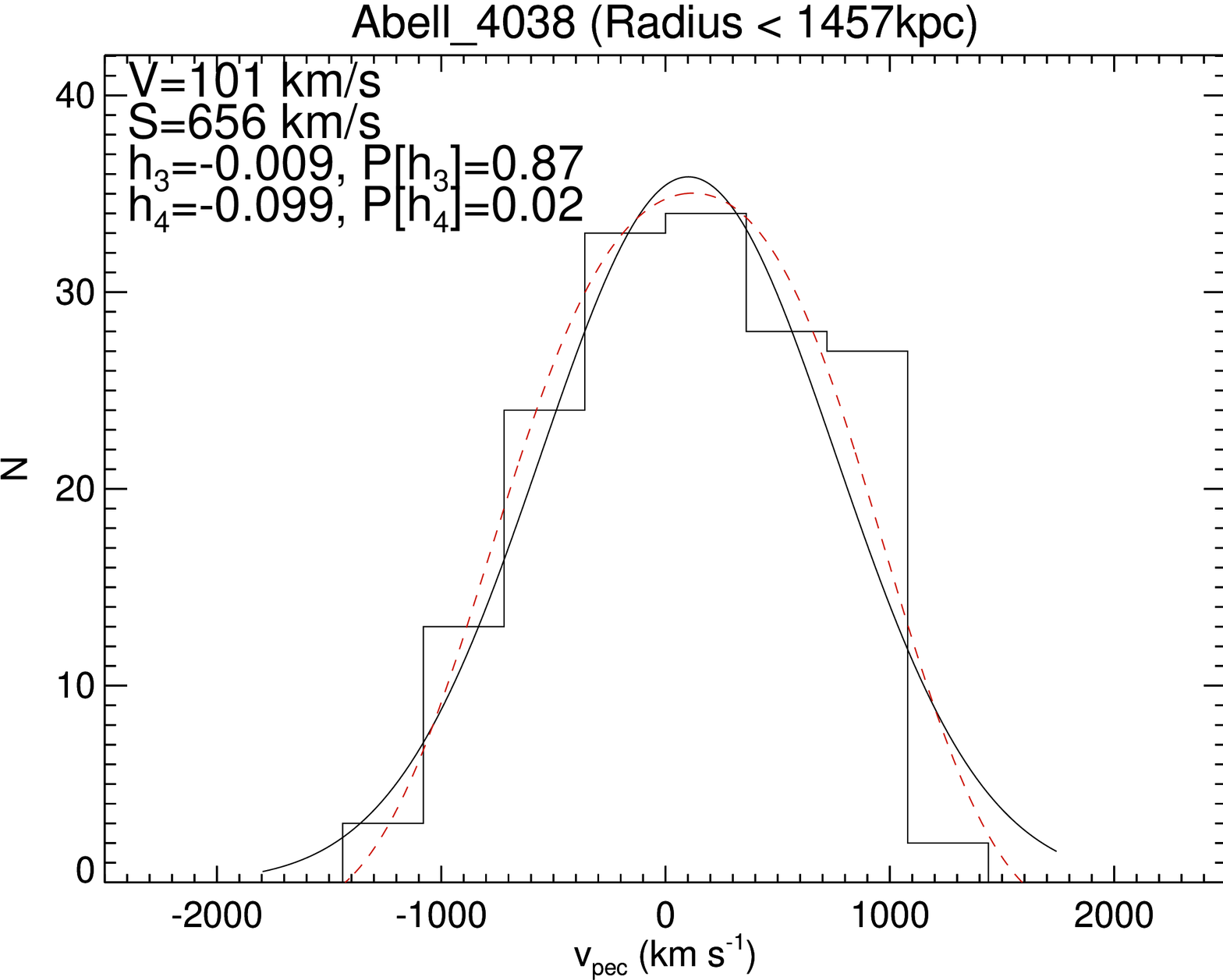}
\includegraphics[angle=0,width=.4\textwidth]{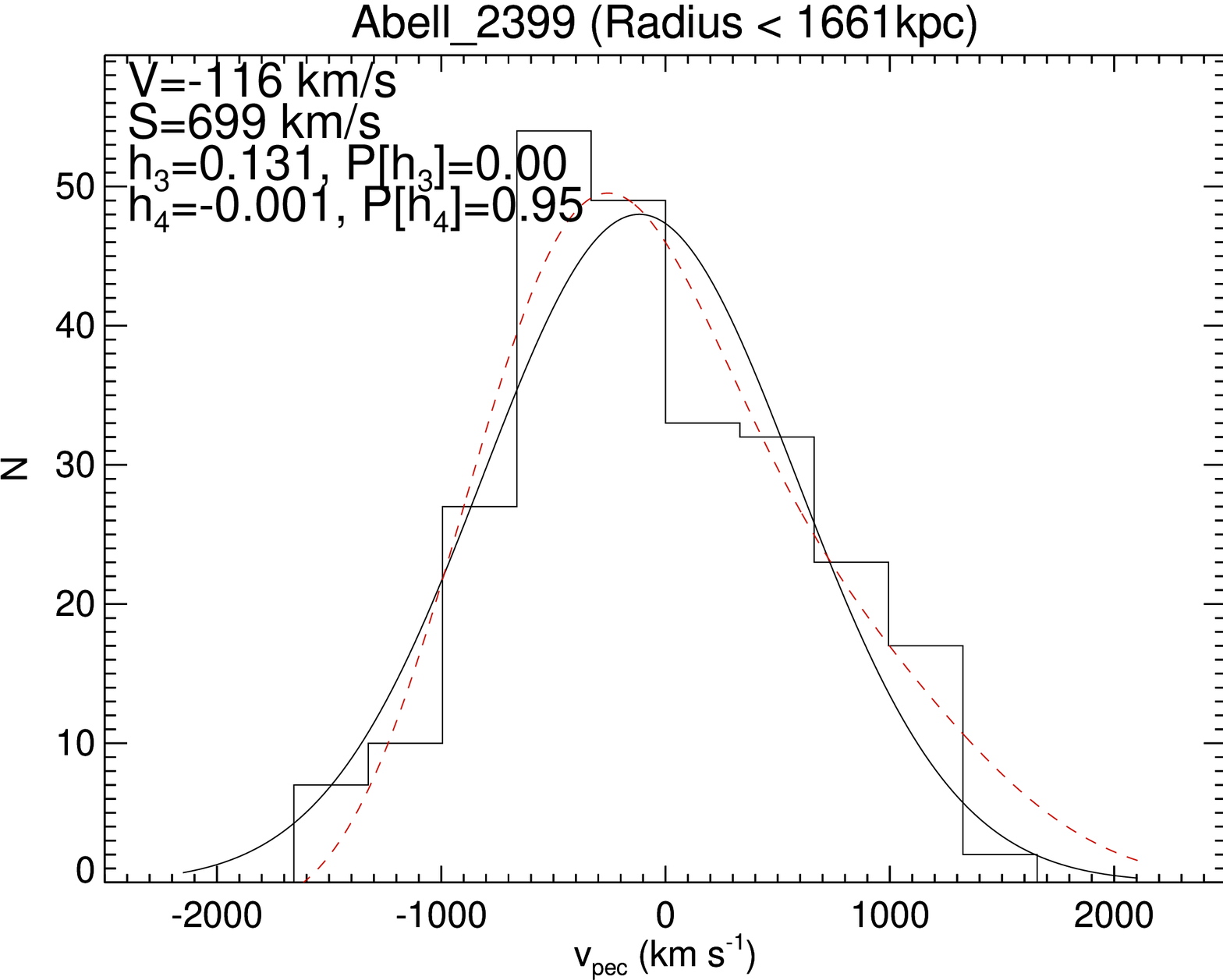}\\
\includegraphics[angle=0,width=.4\textwidth]{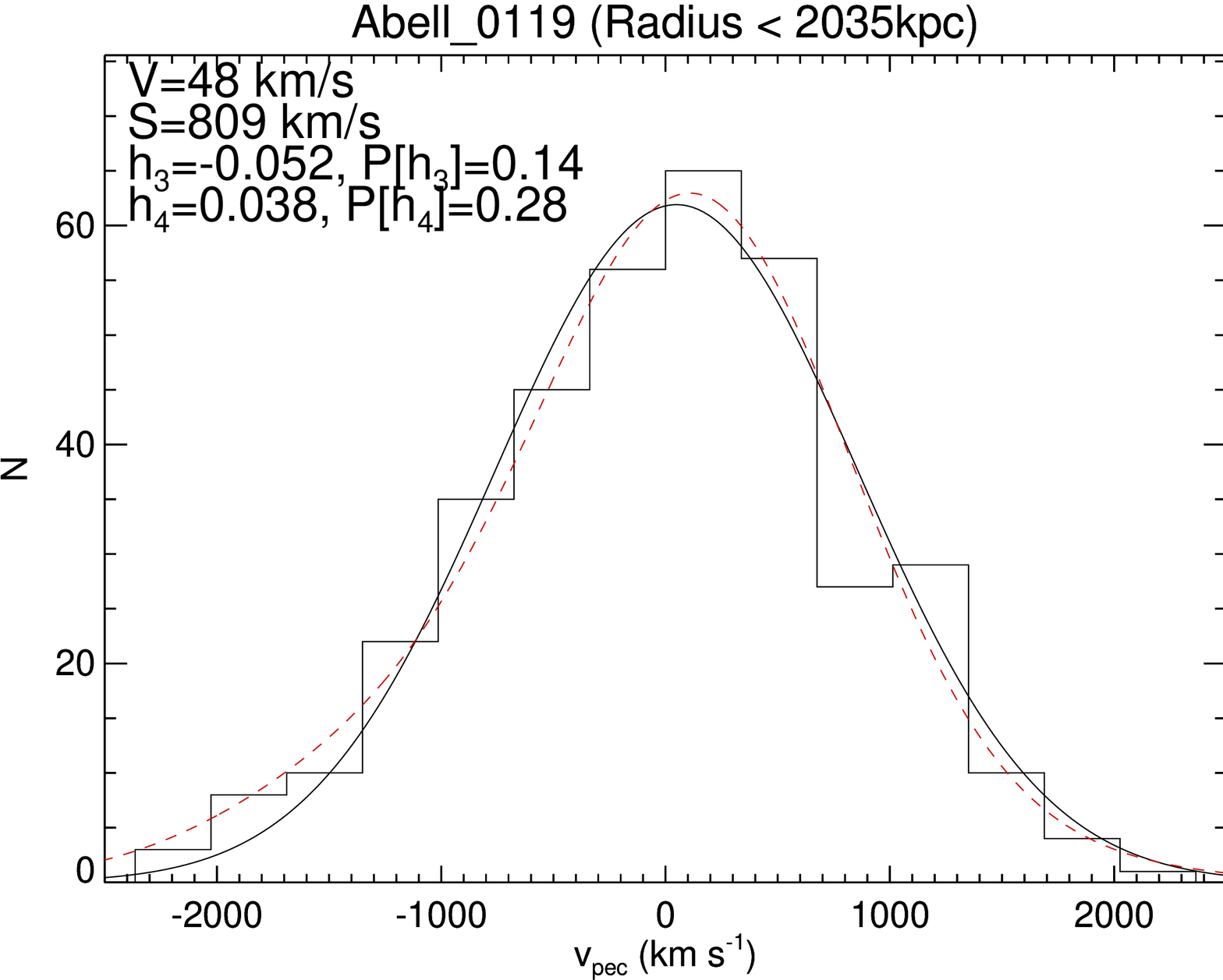}
\includegraphics[angle=0,width=.4\textwidth]{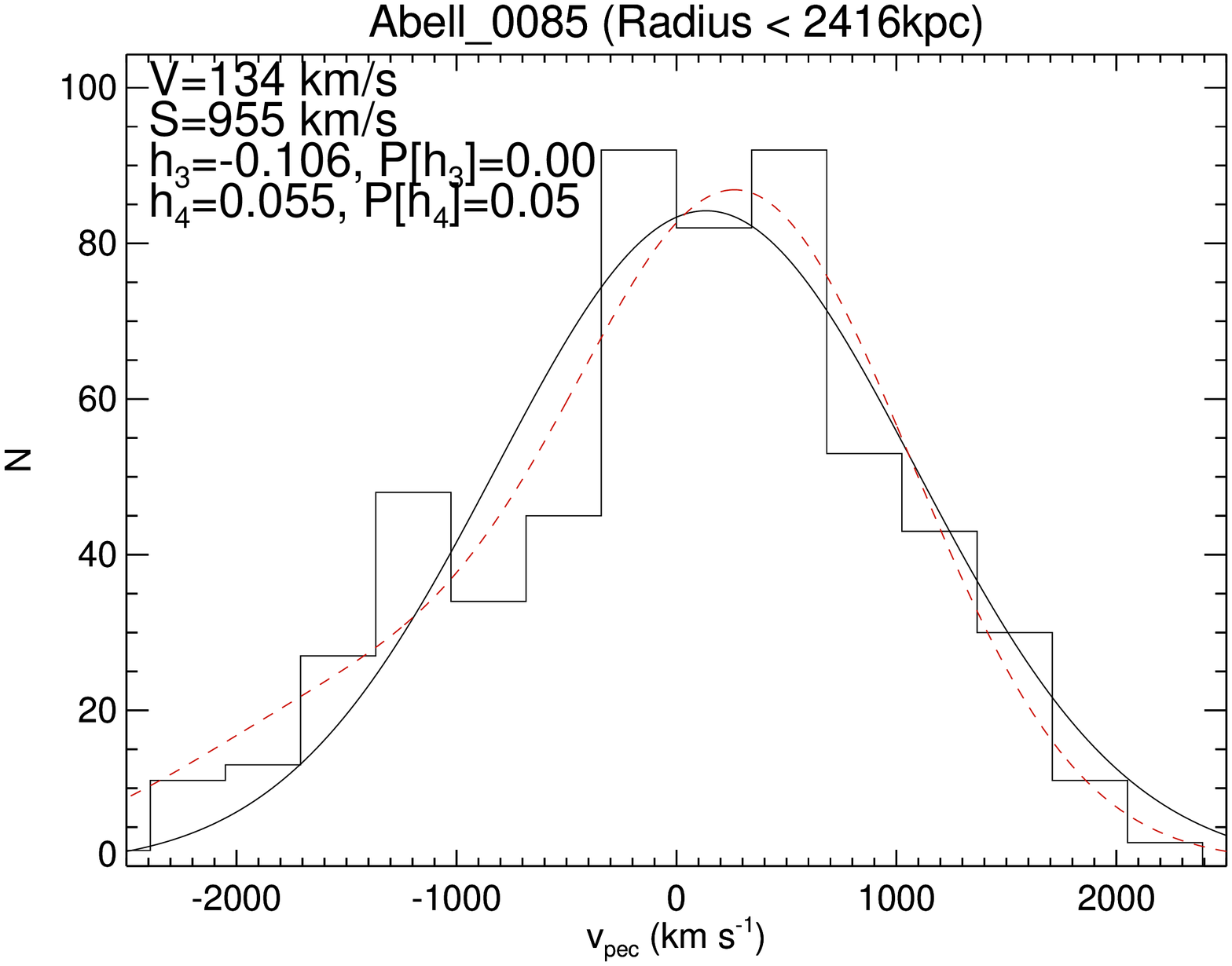}\\
\caption{The histograms show the velocity distributions for cluster members within \rtwo. The bin size is determined as $3S/(N)^{(1/3)}$ where $N$ is the number of galaxies within \rtwo. The solid black line shows the best-fitting Gaussian, while the dashed red line shows the Gauss-Hermite reconstruction of the velocity distribution. In each panel, the best-fitting Gaussian parameters, the Gauss-Hermite terms $h_3$ and $h_4$, representing the asymmetric and symmetric deviations from a Gaussian shape, and their associated level of significance are given in the upper left corner. 
\label{vel_dist}}
\end{figure*}

The one-dimensional indicator uses a Gauss-Hermite decomposition of the velocity distribution to test for departures from the Gaussian shape expected of fully virialised clusters. The method for this Gauss-Hermite decomposition is described in detail elsewhere \citep{zabludoff1993, owers2009a}. This method is sensitive to mergers that are occurring on an axis that is highly inclined to the plane of the sky, which can cause significant skewness or kurtosis in the velocity distribution \citep[e.g., as seen in Abell~2744][]{owers2011a}. The skewness (the $h_3$ term in the Gauss-Hermite series) and kurtosis (the $h_4$ term in the Gauss-Hermite series)  are measures of asymmetric and symmetric deviations from a Gaussian shape. The binned velocity distributions (black histograms), along with the Gauss-Hermite reconstructions (red dashed curves), and the best-fitting Gaussian (black curves) for each cluster are shown in Figure~\ref{vel_dist}. The probability that the observed $h_3$ and $h_4$ terms are simply due to random fluctuations in the distribution, $\rm{P[h_{3, 4}]}$, is determined as described in \citet[][]{zabludoff1993}. Briefly, we measure the $h_3$ and $h_4$ terms for 10000 random Gaussian distributions with the same sample size as the cluster of interest and using the best-fit mean and dispersion (listed in the top left of each panel in Figure~\ref{vel_dist}). The comparison of the distributions of the simulated $h_{3, 4}$ terms with the observed values shows the frequency with which random fluctuations produce $h_{3, 4}$ values more extreme than those measured for the clusters. 

In cases where mergers occur close to the plane of the sky, the velocity distribution may not show significant departures from a Gaussian shape due to the small line-of-sight velocity difference between the systems \citep[e.g., as seen in Abell~3667;][]{owers2009b}. In these cases, merging substructures may be identified as localised enhancements in the projected galaxy density distribution \citep{geller1982}. These enhancements are best revealed in the adaptively smoothed isopleths shown as black contours in Figure~\ref{gal_dist}. The adaptive smoothing is performed using the two-step procedure described in Section~\ref{subsect:memsel}.

\begin{figure*}	
\includegraphics[angle=0,width=.45\textwidth]{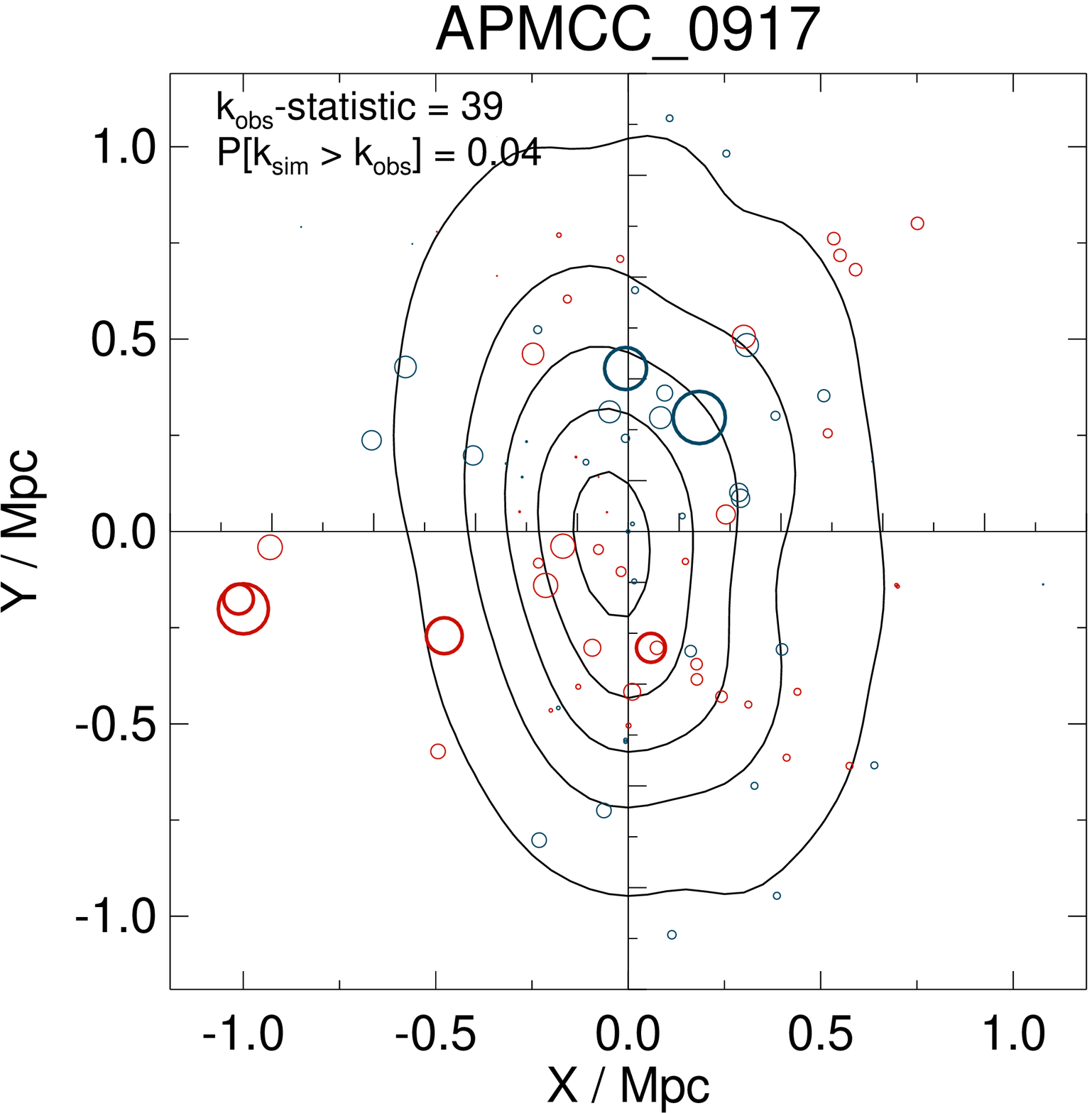}
\includegraphics[angle=0,width=.45\textwidth]{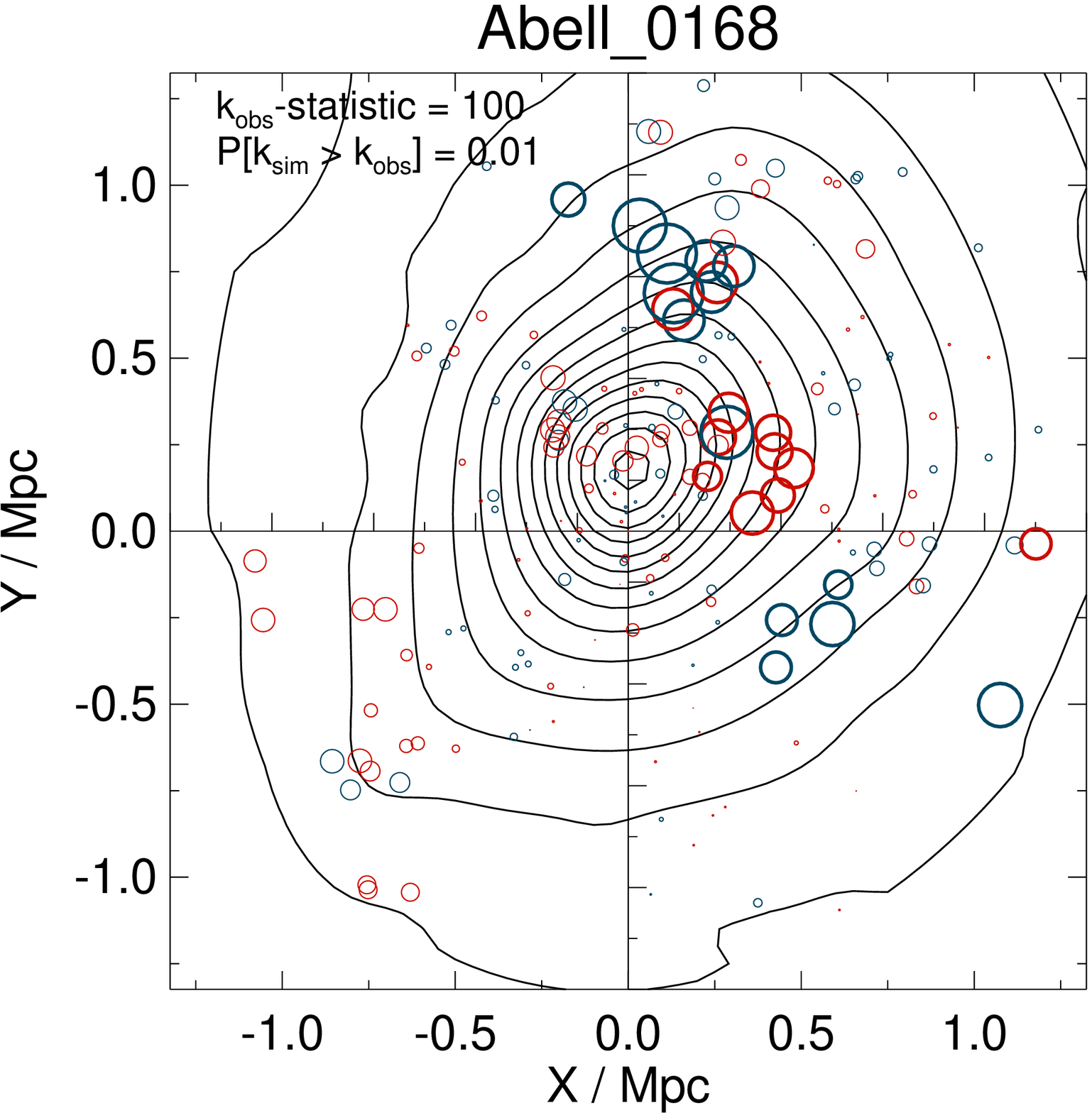}\\
\includegraphics[angle=0,width=.45\textwidth]{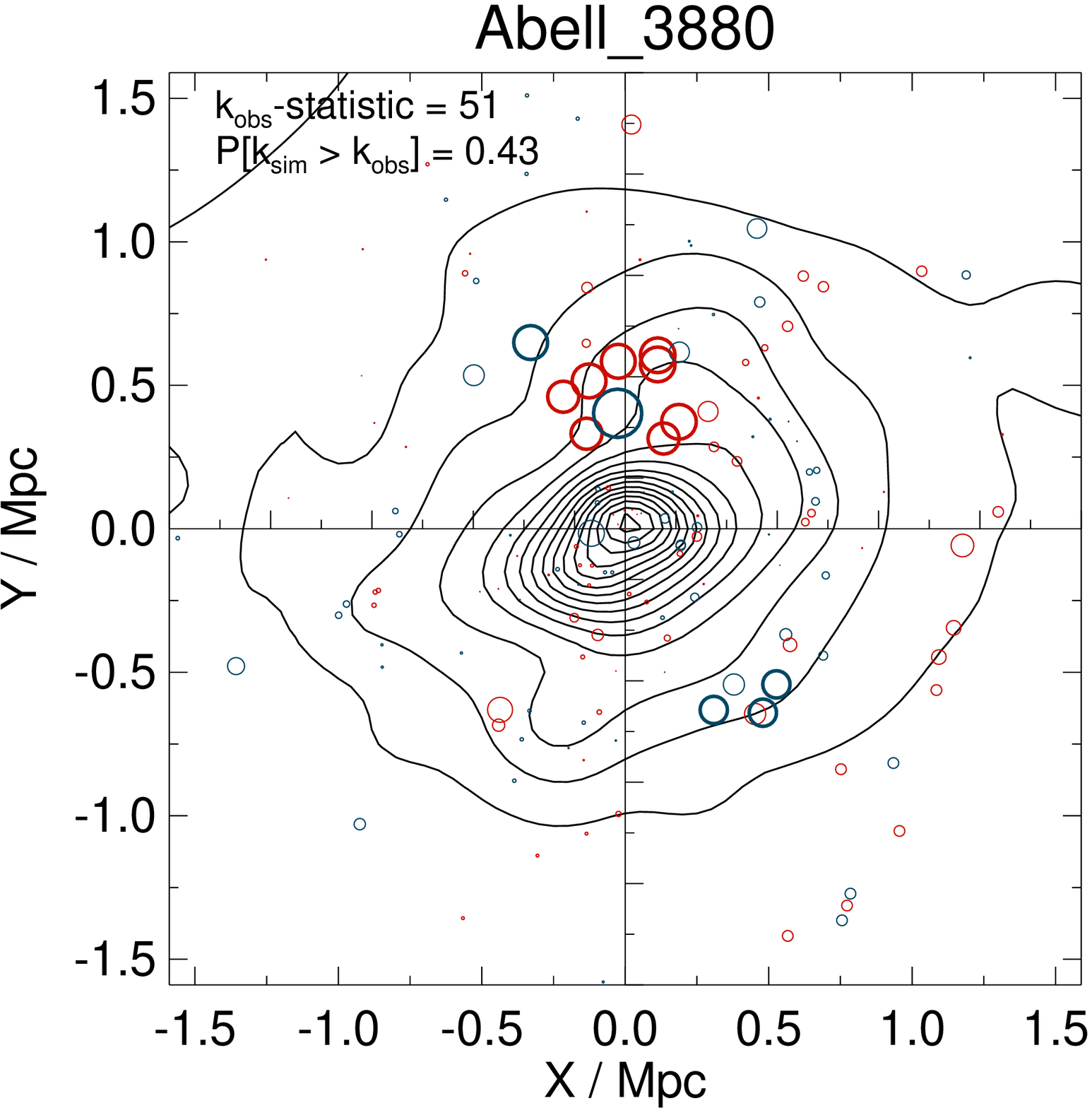}
\includegraphics[angle=0,width=.45\textwidth]{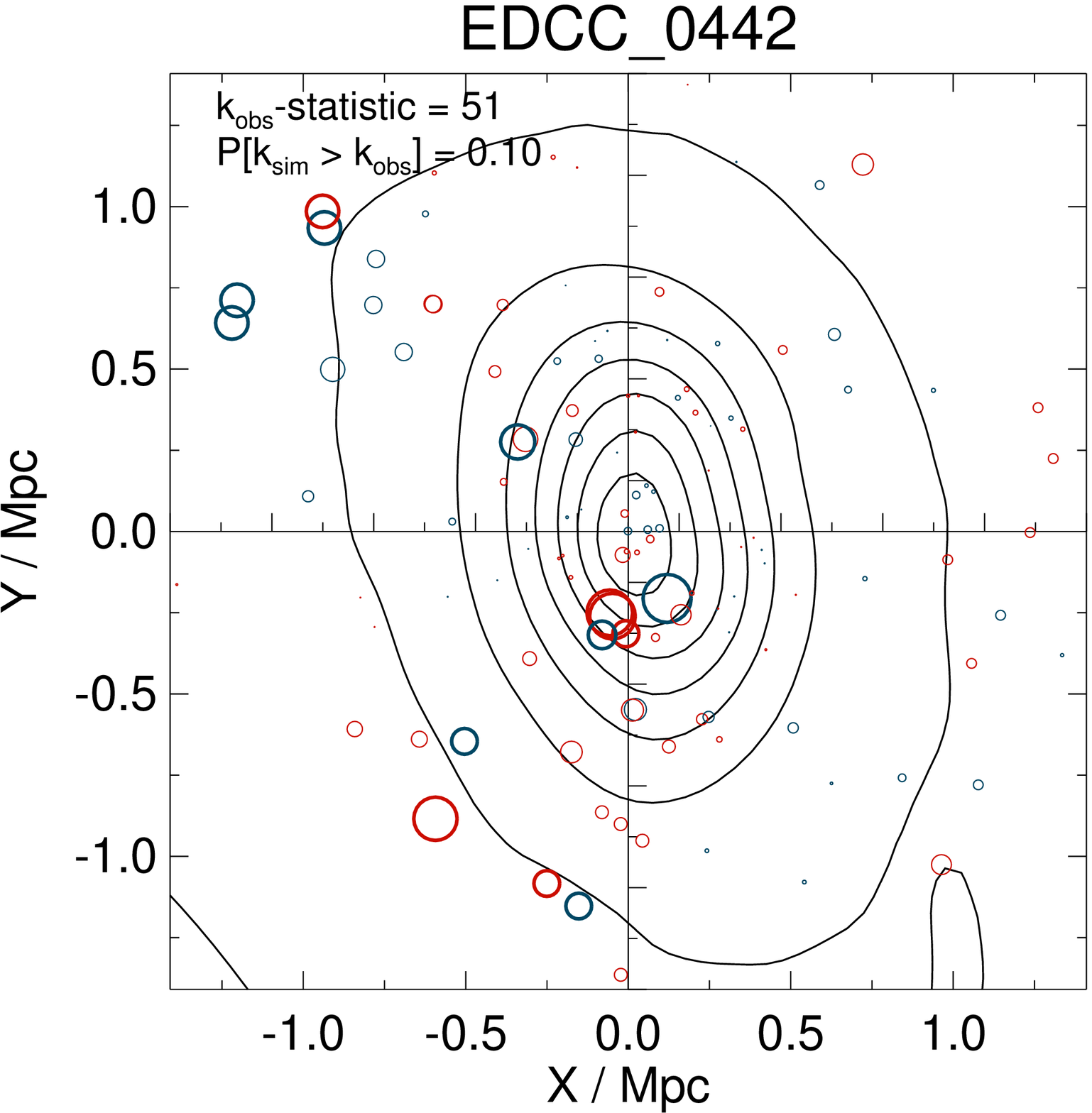}\\
\caption{In each panel, the black contours show the galaxy isopleths derived from the adaptively smoothed galaxy distribution. The circles centred at the position of each member galaxy show the results of the $\kappa$-test. The radius of the each circle is proportional to $-{\rm log10} P_{KS}$, where $P_{KS}$ is the probability that the local and global velocity distributions are drawn from the same parent distribution. The circles are coloured blue or red for galaxies with positive and negative $v_{\rm pec}$, respectively. Clusterings of large circles indicate that the local velocity distribution is different from the global. Bold circles highlight those galaxies where only $5\%$ of simulated $-{\rm log10} P_{KS}$ values are larger than the observed values. The observed $\kappa$-values, and the probability that a value larger than this one occurs in the randomly shuffled simulations are given in the top left of each panel.
\label{gal_dist}}
\end{figure*}

\begin{figure*}	
\includegraphics[angle=0,width=.45\textwidth]{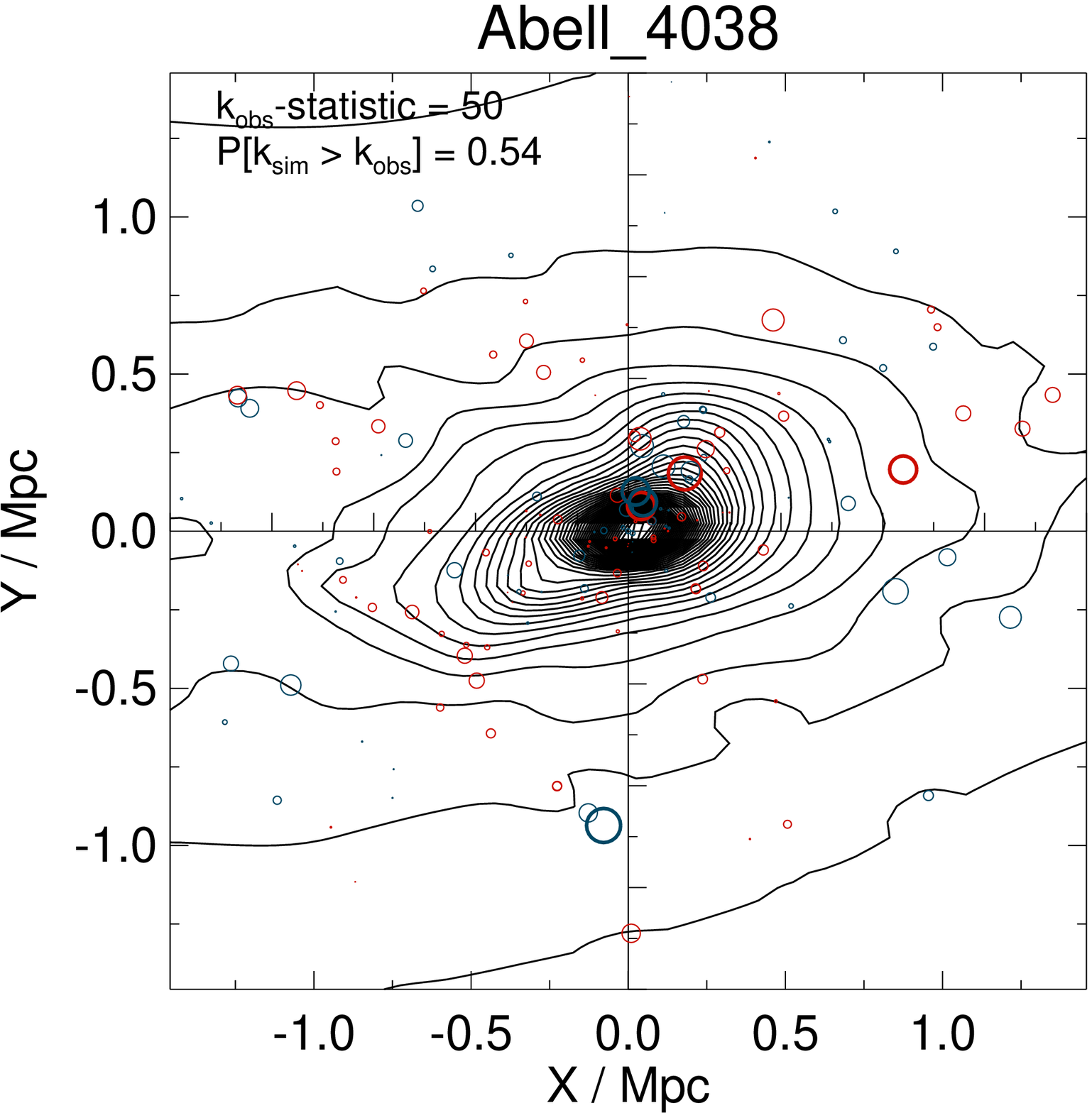}
\includegraphics[angle=0,width=.45\textwidth]{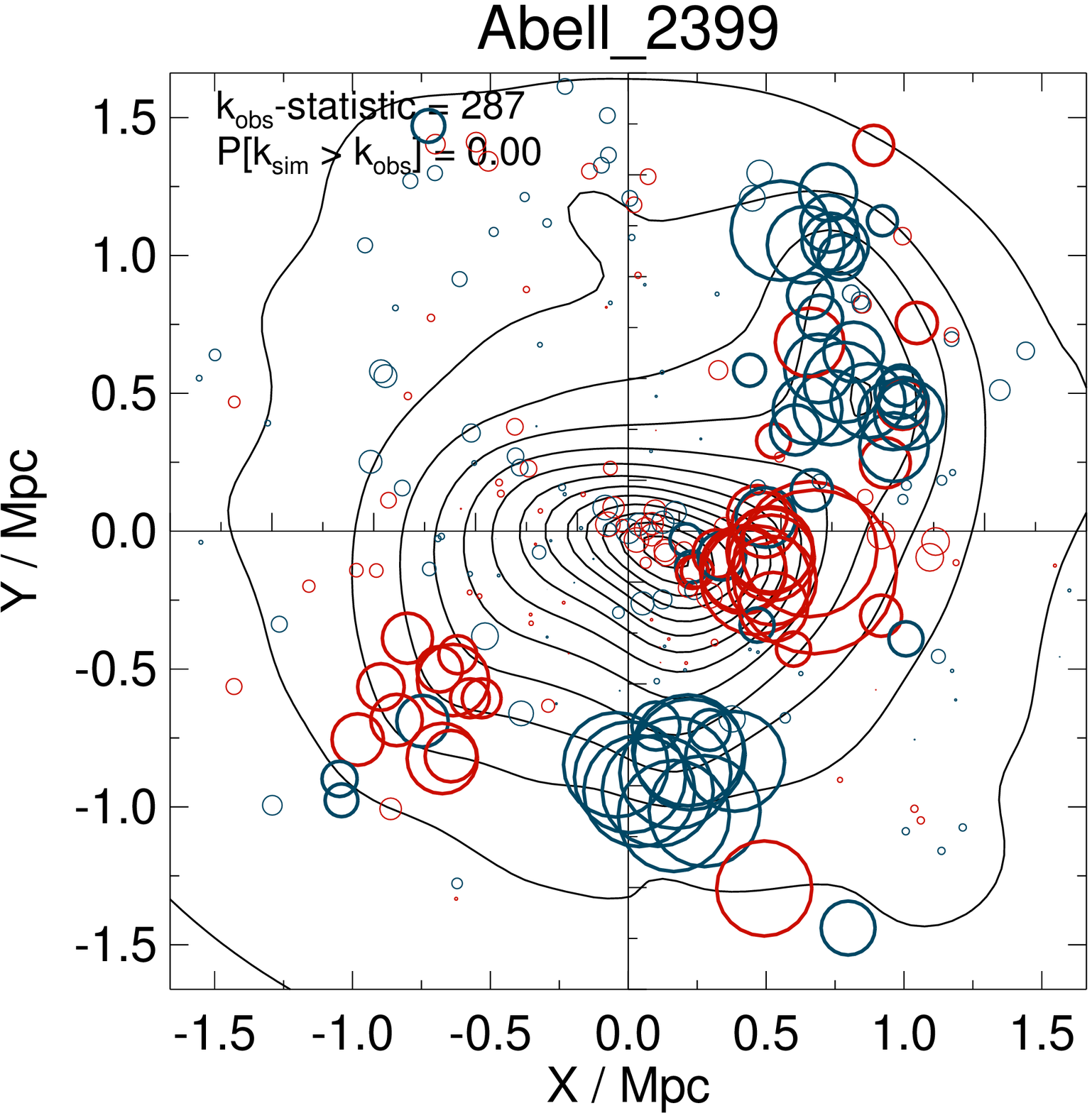}\\
\includegraphics[angle=0,width=.45\textwidth]{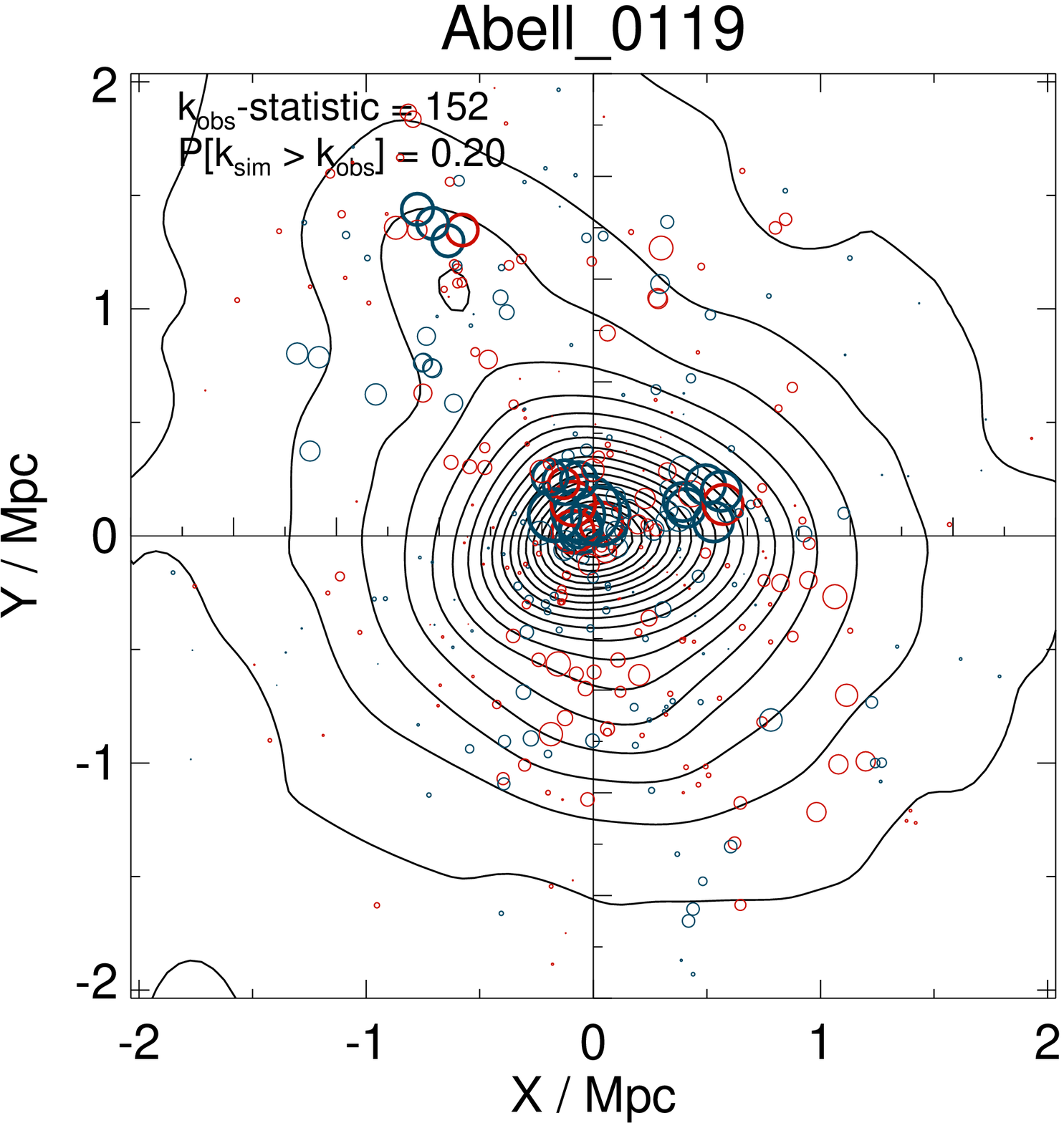}
\includegraphics[angle=0,width=.45\textwidth]{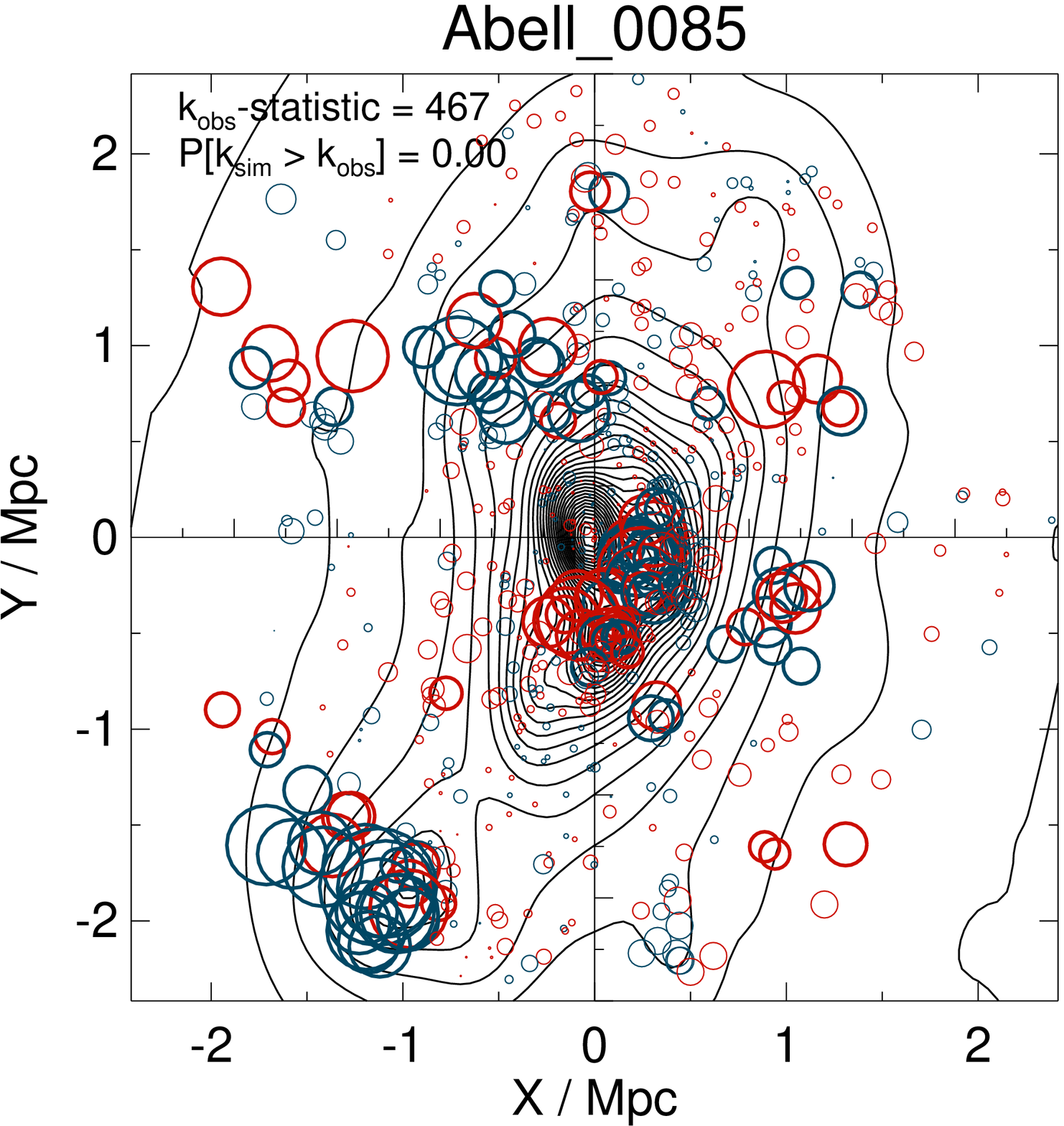}\\
\contcaption{}
\end{figure*}

Tests that use the combination of the position and velocity information are the most reliable in detecting substructure \citep{pinkney1996}. Here, we implement the $\kappa$-test which uses the Kolmogorov-Smirnov (KS) statistic to determine whether the local velocity distribution to a galaxy of interest is likely to be drawn from the same parent distribution as the global cluster velocity distribution \citep{colless1996}.  By doing so, the test is able to identify local departures from the global cluster kinematics, which indicate the presence of dynamical substructure. Here, local is defined to be the $\sqrt{N}$ nearest neighbours, where $N$ is the number of cluster members within \rtwo. The results of this test are presented in Figure~\ref{gal_dist} where the size of the circles are proportional to $-{\rm log_{10}} P_{KS}$, where $P_{KS}$ is the KS-probability that the local and global velocity distributions are drawn from the same parent distribution. That is, regions where there are groupings of large circles indicate that the kinematics there are different from the global cluster kinematics. The circles are colour-coded according to the velocity of the galaxy of interest: red and blue for positive and negative $v_{\rm pec}$ with respect to the cluster, respectively. The overall level of substructuring according to the $\kappa$-test is calculated by determining $\kappa= \sum_{i=1}^{N} -{\rm log_{10}} P_{KS}$. The significance of $\kappa$ is determined by remeasuring $\kappa$ 10,000 times and with each iteration removing any correlation between position and velocity by randomly reassigning the galaxy velocities, while maintaining positions. The value of $P[\kappa_{sim} > \kappa_{obs}]$ listed at the top left of each panel in Figure~\ref{gal_dist} gives an indication of the probability that a $\kappa_{obs}$ value as high as the measured one could occur in the simulations due to random fluctuations. Furthermore, circles with large $-{\rm log_{10}} P_{KS}$ values that rarely occur in the simulations (less than $5\%$ of the time) are highlighted in bold.

Combining the results of the tests, it is clear that the clusters A2399 and A85 show significant evidence for the presence of substructure in all three indicators. The existence of substructure within A85 is well-known from both X-ray observations \citep{kempner2002, durret2005, fogarty2014, ichinohe2015} and previous optical observations \citep{durret1998, bravo2009}. For A2399, the clustering of large, preferentially red, bubbles $\sim 560$\,kpc just south of due west of the cluster centre is coincident with the brightest cluster galaxy which has $v_{pec} \sim 1200\,$\kms. The X-ray observations of \citet{bohringer2007} indicated strong bimodality in A2399, while the X-ray contours shown in Figure~12 of \citet[][]{fogarty2014} show a plume of X-ray emission extending to the north associated with this structure. These pieces of evidence strongly indicate ongoing merger activity in A2399. There is a marginal ($P[\kappa_{sim}>\kappa_{obs}] \sim 0.01$) indication of kinematical substructure present in A168, which is a well-known post-core-passage cluster merger occurring with an axis aligned close to the plane of the sky based on its X-ray observations \citep{hallman2004, fogarty2014}. Inspection of the $\kappa$-test results for A168 in Figure~\ref{gal_dist} indicate the presence of a kinematical substructure $\sim 800$\,kpc just East of North that is coincident with the brightest cluster galaxy, as well as one of the X-ray substructures detected in \citet{hallman2004} and \citet[][]{fogarty2014}. There is a small substructure in A119 that is apparent as a local excess in the projected galaxy density $\sim 1$\,Mpc to the northeast of the cluster centre. This substructure is associated with a bright cluster galaxy and has previously been detected using optical data \citep{kriessler1997}. The remaining clusters, APMCC0917, A3880, EDCC0442 and A4038 show no strong evidence for substructure.

\citet{pinkney1996} showed that virial mass estimators can overestimate the total cluster mass by as much as a factor of 2 during cluster mergers. They found that the impact on the measured mass is strongest in the case of low mass-ratio mergers that are very close to pericentric passage and where the merger axis is within $30\deg$ of the line of sight. For those mergers, the high relative velocity of the merging subclusters causes strong bi-modality in the velocity distribution, leading to an overestimation of the velocity dispersion and therefore the mass. For less pathological merger scenarios, they found that masses may be overestimated by $\sim 50\%$, while for mergers with an axis close to the plane of the sky (e.g., as in A168), there is minimal impact on the measured masses. None of the velocity distributions shown in Figure~\ref{vel_dist} exhibit strong bi-modality, although the clusters A85 and A2399 have $h_3 \sim 0.1$, indicating asymmetric departures from a single Gaussian shape at the 10\% level. However, the velocity dispersions determined for A85 and A2399 do not appear to be significantly affected by this asymmetry; compare the dispersion measurements in Table~\ref{clus_table} with those in the upper left corner of Figure~\ref{vel_dist}. In principle, the caustic mass measurements should be less-affected by substructure (with the caveat that the velocity dispersion is used in determining where to place the caustics) so it is reassuring that the caustics and virial mass estimators for A85 and A2399 are not significantly different. We conclude that the clusters with substructure, particularly A2399 and A85, may have their masses overestimated by up to $\sim 50\%$, although they are unlikely to be overestimated by the factor of 2 seen in more extreme situations.

\section{SAMI target selection}\label{sami_targets}

Having described the SAMI-CRS and the global cluster properties, we now use those results as input for selecting the cluster targets for the SAMI-GS. In this section, we first describe our improved procedure for measuring aperture- and PSF-matched photometry, and verify the photometric precision and accuracy using internal and external measurements. We then describe the selection of cluster targets for the SAMI-GS.

\subsection{Updated Photometry}\label{new_photo}

The SAMI-GS target selection is outlined in \citet{bryant2015}. Briefly, the selection used the combination of redshift and a stellar mass proxy that is derived from the empirical relationship between the stellar mass-to-$i$-band ratio and $(g-i)$ colour defined by \citet[][]{taylor2011}. The stellar mass proxy requires an accurate determination of both $(g-i)$ colour and total $i$-band magnitude. To that end, we produce new photometric catalogues for the SAMI clusters where a more careful matching of aperture and image-PSF homogenisation has been achieved. The process is outlined below. 

\subsubsection{SDSS clusters}\label{new_sdss_phot}

For the four clusters that lie within the SDSS footprint (Abell 85, Abell 2399, Abell 168 and Abell 119) the photometry has been re-measured using the {\sf IOTA} software, which was used to measure aperture-matched photometry for the GAMA survey \citep[]{hill2011,driver2016}. Prior to running the {\sf IOTA} software, all image frames in the $ugriz$ bands and within a $4\deg \times 4\deg$ region surrounding the cluster centres were retrieved from the SDSS DR10 database. Each frame was convolved to a common PSF (FWHM=2\arcsec) and renormalised to a common zero-point (ZP) \citep[][]{hill2011,driver2016}. For each of the $ugriz$ bands, the convolved, renormalised frames were then combined using the {\sf SWarp} software \citep{bertin2002, bertin2010SWARP}. The photometry for each of the combined $ugriz$ images was measured using the {\sf SExtractor} software \citep{bertin1996,bertin2010SEX} in dual-image mode with the $r$-band image used for detection. This method is consistent with that used to measure the photometry in the GAMA portion of the SAMI-GS.

\subsubsection{VST/ATLAS clusters}

The images produced by CASU for the VST/ATLAS survey are not suitable for measuring accurate photometry. The primary reason behind this is that the illumination correction is not applied to the images themselves, but rather to the photometric catalogues in a post-processing step \citep{shanks2015}. This illumination correction accounts for the $\sim 0.2$mag centre-to-edge gradient due to the impact of scattered light on the flat-fielding. To overcome this issue, raw VST/ATLAS data in the $gri$-bands were retrieved from the ESO archive and reduced using the {\sf Astro-WISE} optical image reduction pipeline \citep{mcfarland2013} for the clusters Abell~3880, EDCC~0442, the Abell~4038/APMCC0917 region and also for the partially-covered Abell~85 region (which will be used for quality control tests in Section~\ref{sdss_vst_comp}). The reduction follows the general procedure outlined for the Kilo-Degree Survey data releases described in \citet{dejong2015}, with several modifications that are required to account for the two-pointing dither pattern (as opposed to the KiDs 5-pointing dither pattern) and the brighter magnitude limits of the VST/ATLAS survey (which affects the astrometric calibration). The reduction produces calibrated coadded images that have the illumination correction applied and have photometric zero-points tied to the ESO nightly standards. 

Because there can be significant variation in the night-to-night photometric zero-points, we measure the aperture- and PSF-matched photometry separately for each $1\deg \times 1\deg$ $gri$ coadded image and correct the photometry on a tile-by-tile basis as follows. On each full resolution $gri$ coadded image, {\sf SExtractor} is used to detect sources and extract photometric parameters. Stellar magnitudes are measured using 6.4\arcsec\, diameter apertures and are corrected for PSF-dependent aperture losses using a curve-of-growth (COG) correction \citep{howell1989}. The COG correction is determined from $\sim 1000$ stars in each tile with magnitudes in the range $15 < gri < 20$. For each of these stars, the differential magnitude profile is measured through a series of concentric apertures. At the radius of each aperture, the median value of the $\sim 1000$ differential magnitude values is used to determine the median differential magnitude profile. The aperture correction is determined by integrating the median differential profile out to a 15\arcsec\, diameter aperture. Following the method described in Section~\ref{old_vst}, for each tile the bright, non-saturated stars are cross-matched with APASS stars. The median difference between the APASS and COG-corrected total magnitudes are used to derive a correction which ties each $gri$-band tile to the APASS photometry. The corrections are generally of the order 0.05\,mag, although in some cases were as high as 0.15\,mag. A further small correction is then applied to place the $gri$ photometry onto the SDSS system using the colour transforms derived by \citet{shanks2015}. 

The aperture- and PSF-matched photometry is measured in a similar manner to that described for the SDSS data (Section ~\ref{new_sdss_phot}). Since the VST/ATLAS imaging has higher image quality than the SDSS \citep{shanks2015}, we convolved each $gri$-band tile to a common $1.5\arcsec$ FWHM. For the convolution, it is assumed that the seeing measurement provided within the tile header, ${\rm FWHM_{seeing}}$, is accurate and that the PSF has a Gaussian shape. The amount by which the image is convolved is ${\rm FWHM_{convolve}} = \sqrt{1.5^2 - {\rm FWHM_{seeing}}^2}$. We run {\sf SExtractor} in dual-image mode using the $r$-band image as the detection image. For each tile the separate photometric catalogues for the full resolution and convolved $gri$-bands are matched to the full-resolution $r$-band catalogue. Finally, the photometric catalogues for the separate tiles are concatenated. Duplicated measurements due to the $2\arcmin$ overlap between adjacent tiles are accounted for by keeping the photometric measurement with the highest quality as determined based on the {\sf SExtractor} flags. We make use of these duplicated measurements of $15 < r < 20$ stars in the overlap regions to test the tile-to-tile variations in the photometric calibration. For each cluster, we find that the median difference in the distribution of $gri$-magnitudes is always smaller than $\pm 0.016\,$mag and the dispersion in the distribution is smaller than 0.05\,mag. This internal comparison indicates that the VST/ATLAS photometry has good precision, with low tile-to-tile variation.

\subsubsection{Comparison of SDSS and VST/ATLAS photometry}\label{sdss_vst_comp}

\begin{figure*}
\includegraphics[angle=0,width=.45\textwidth]{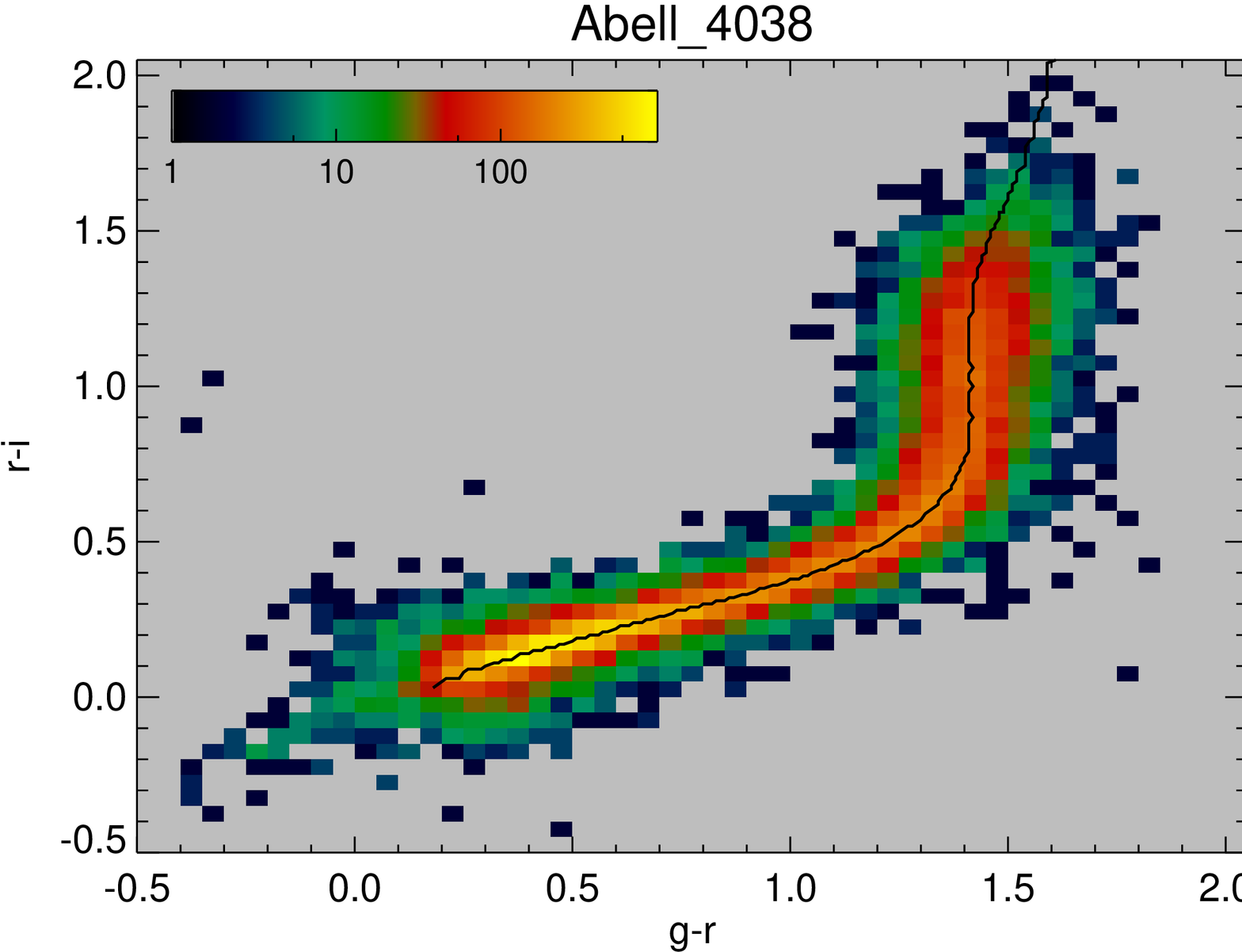}
\includegraphics[angle=0,width=.45\textwidth]{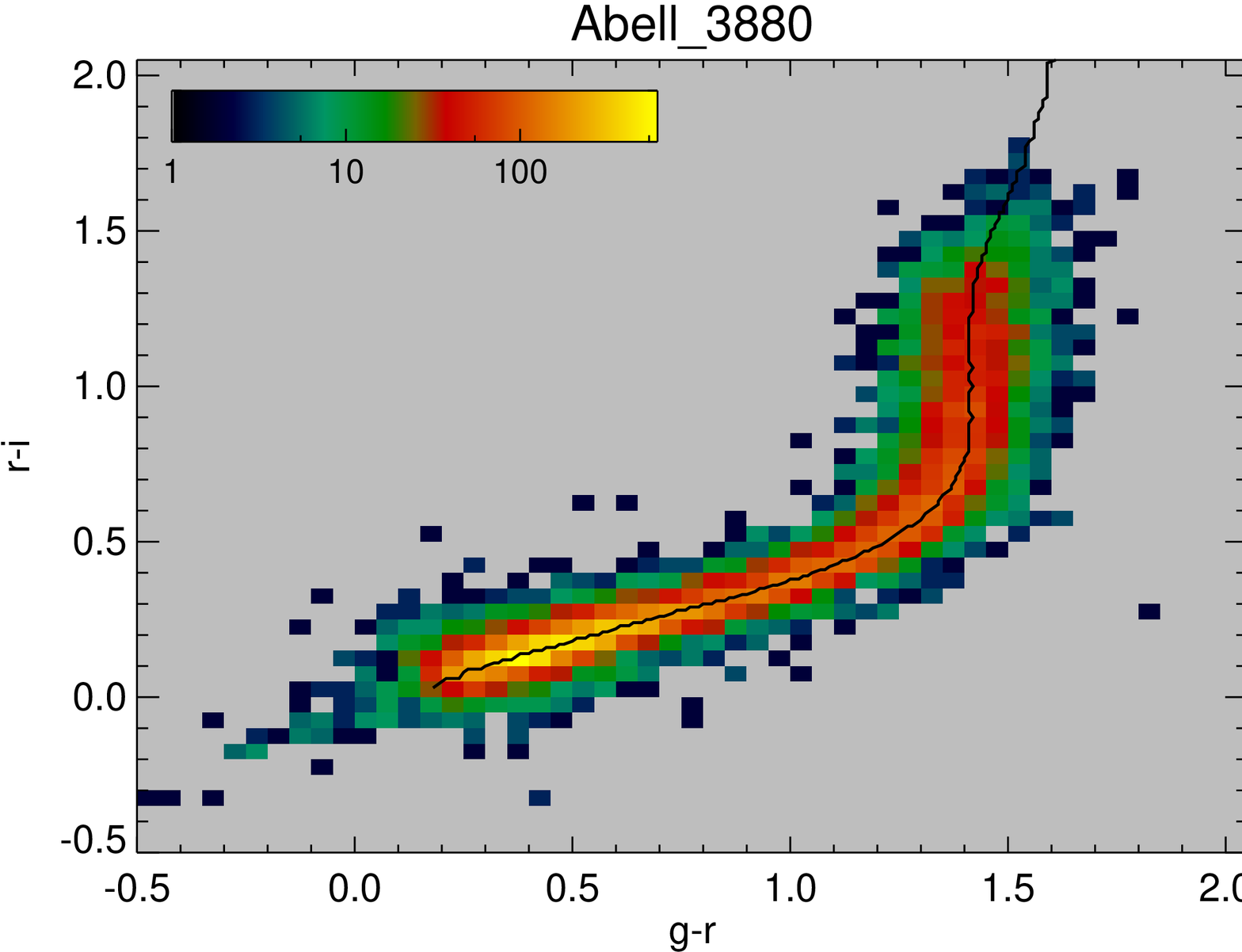}\\
\includegraphics[angle=0,width=.45\textwidth]{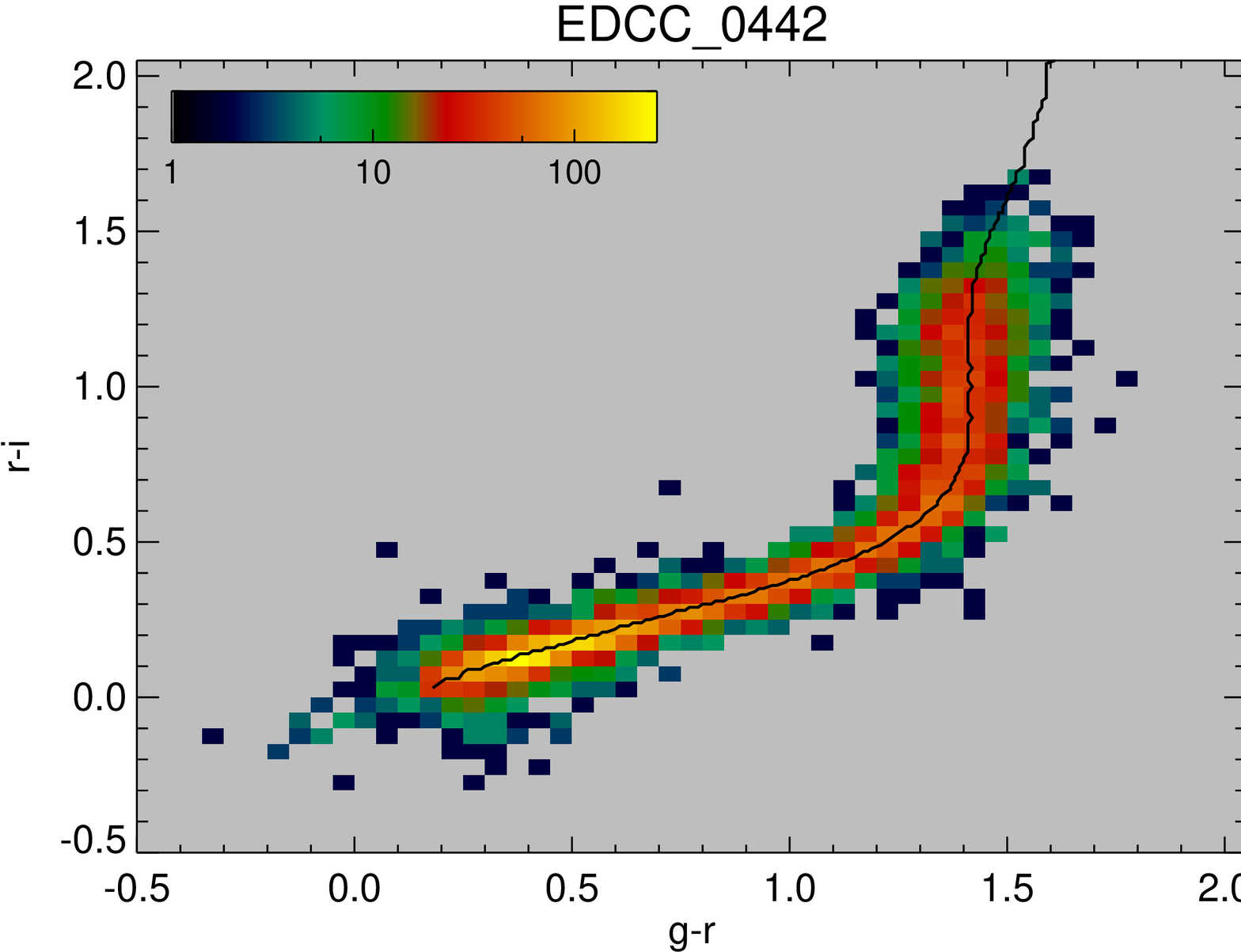}
\includegraphics[angle=0,width=.45\textwidth]{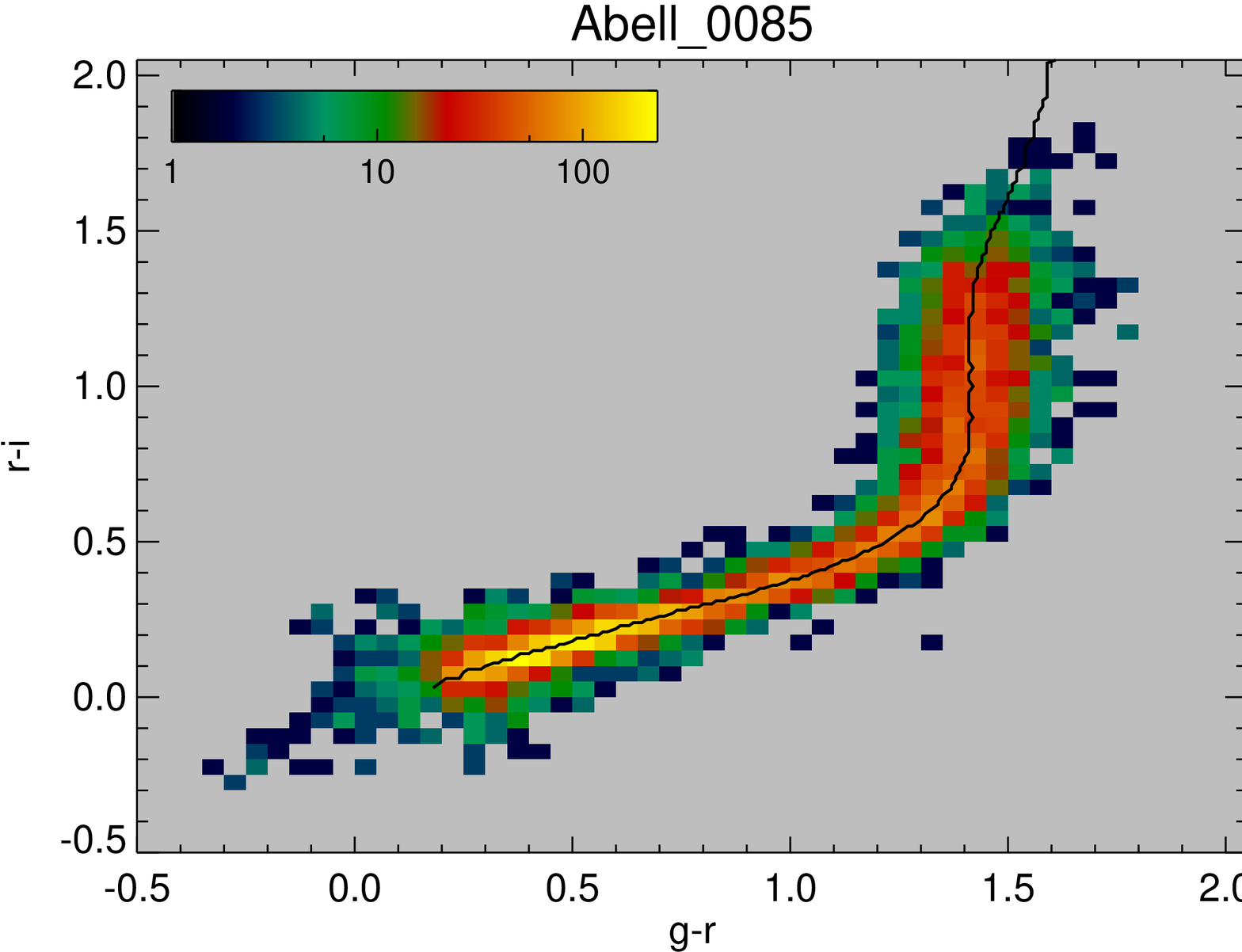}
\caption{Extinction-corrected $(g − r)$ vs $(r − i)$ colour-colour diagrams for stars selected from VST/ATLAS photometry for the clusters A4038, A3880, A85 and EDCC0442.  The stars have COG-corrected aperture magnitudes in the range 15 < r < 20. The colour scale gives the number of stars in each bin, as indicated by the colour bar shown in the upper left corner of each panel. in the upper left corner show the  The black line overlaid on these diagrams shows the median stellar locus of high-quality SDSS stars from \citet{covey2007}. These plots confirm that there are no large systematic offsets in colours measured with the VST/ATLAS photometry when compared with the SDSS. 
\label{gmr_rmi_cols}}
\end{figure*}

\begin{figure*}
\includegraphics[angle=0,width=.45\textwidth]{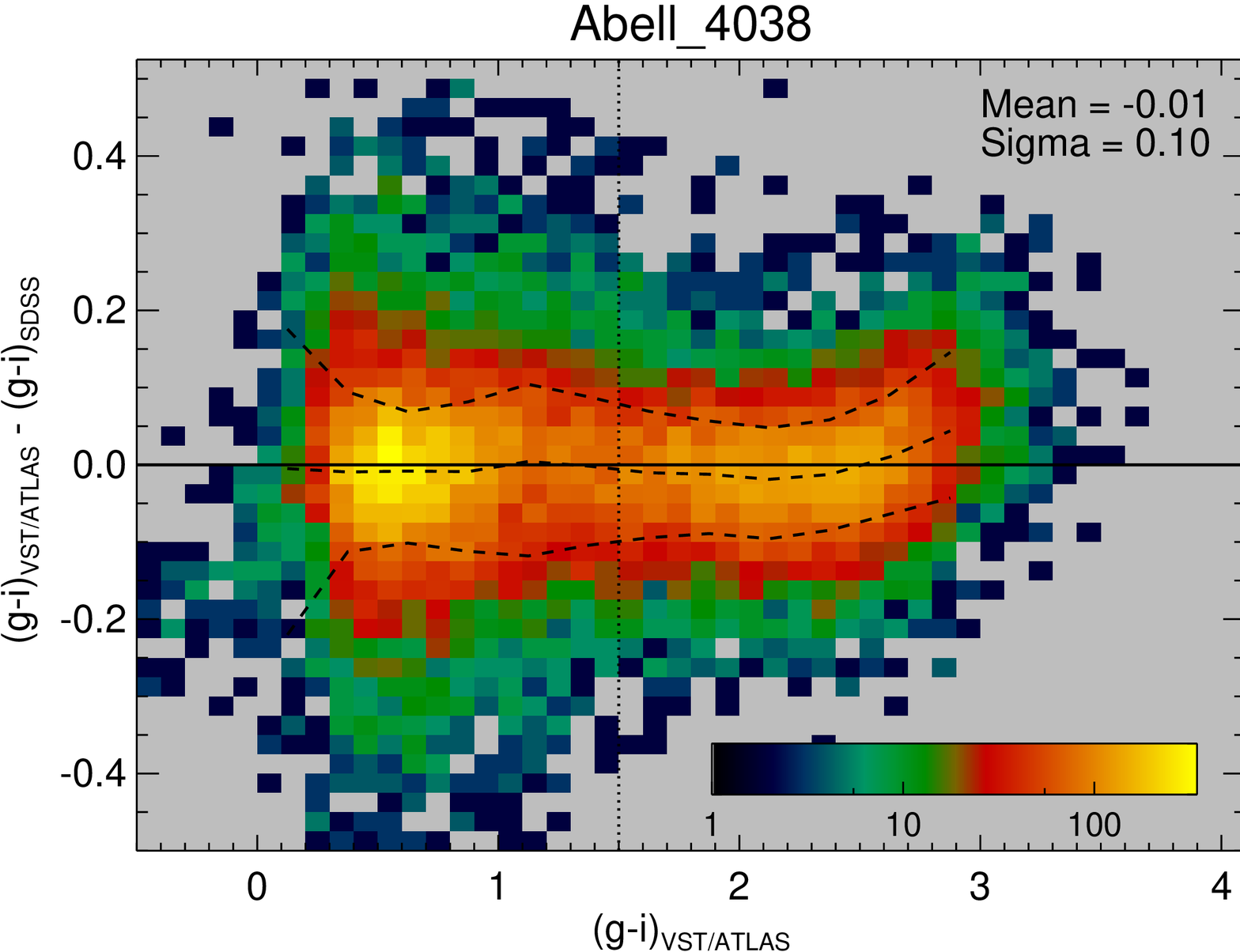}
\includegraphics[angle=0,width=.45\textwidth]{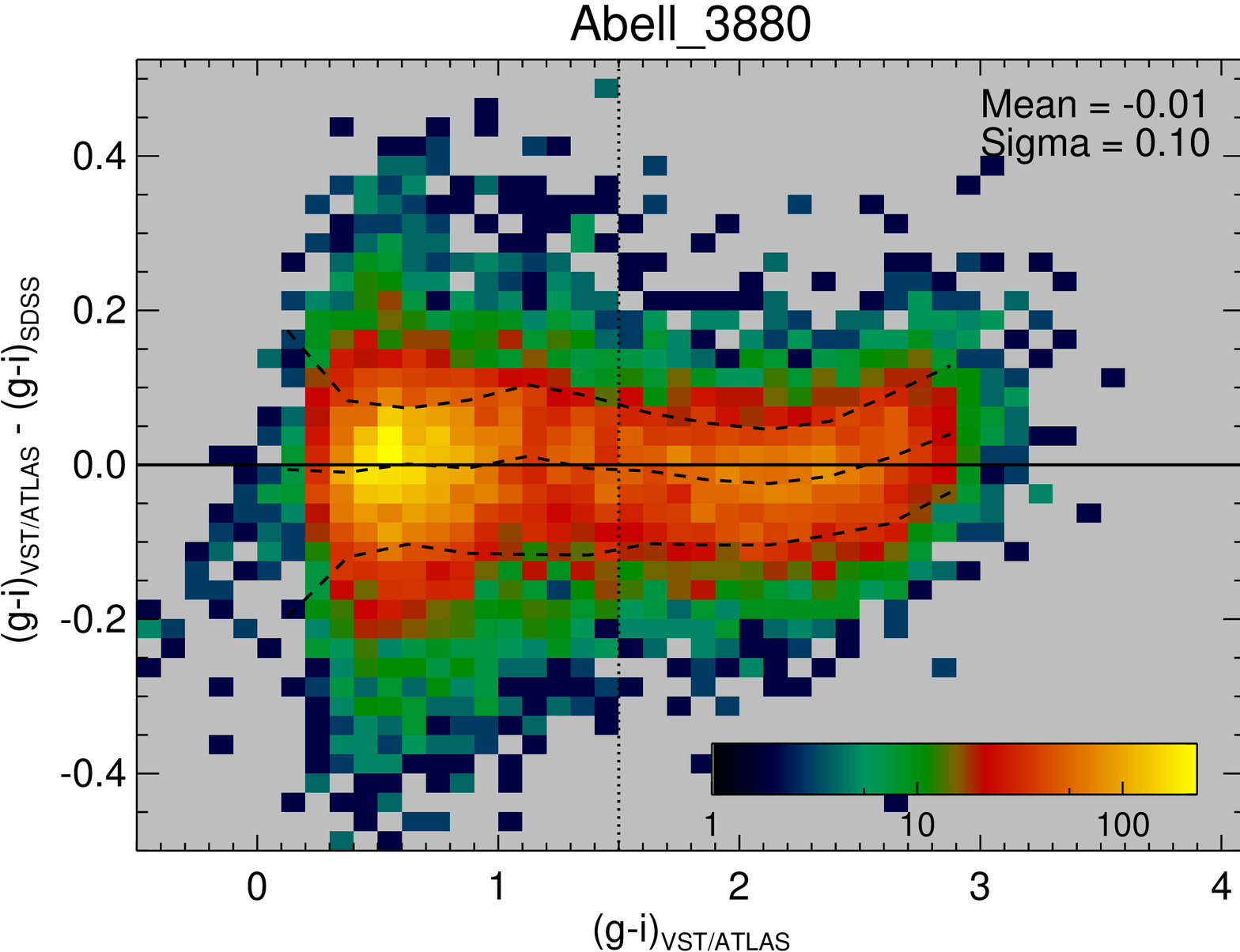}\\
\includegraphics[angle=0,width=.45\textwidth]{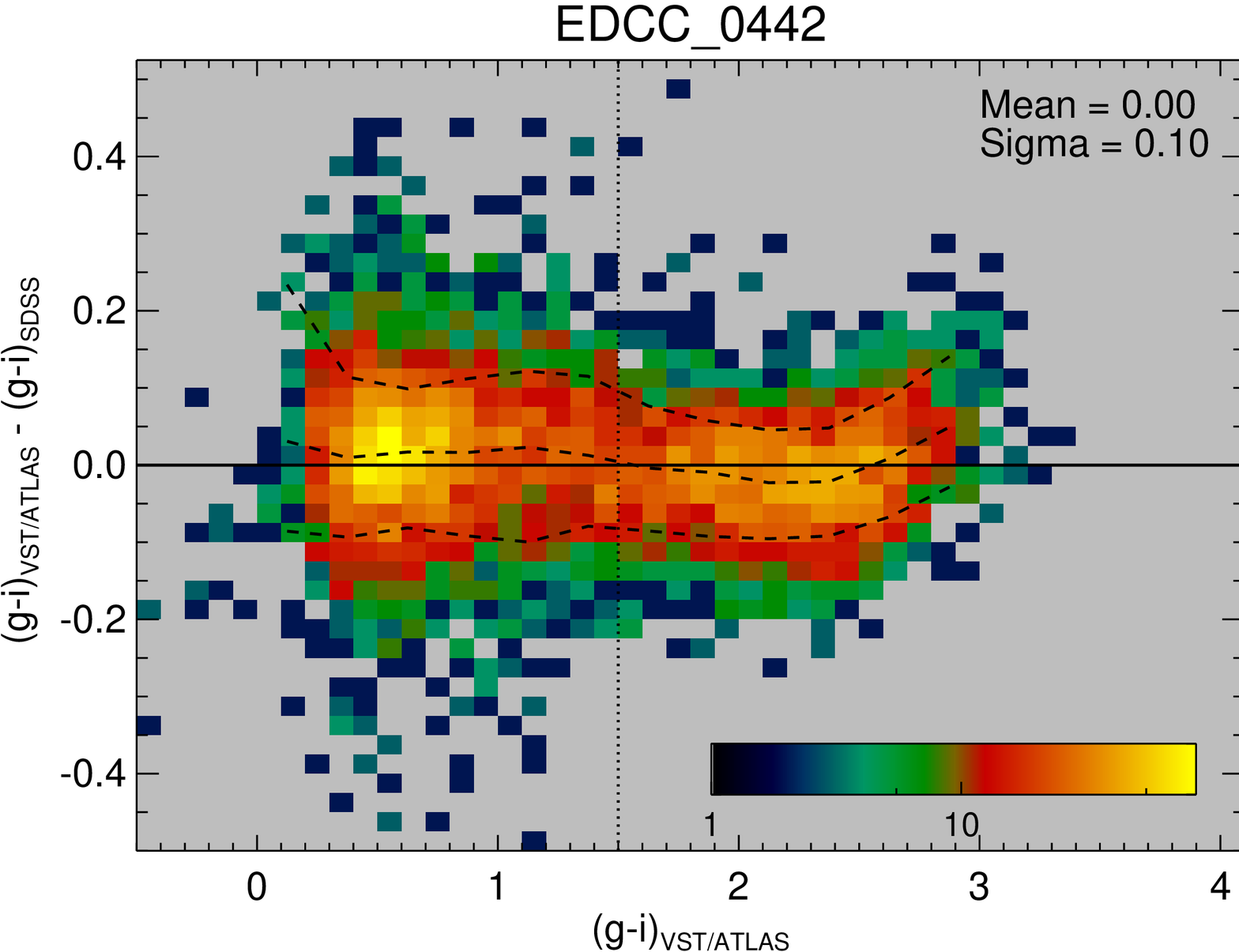}
\includegraphics[angle=0,width=.45\textwidth]{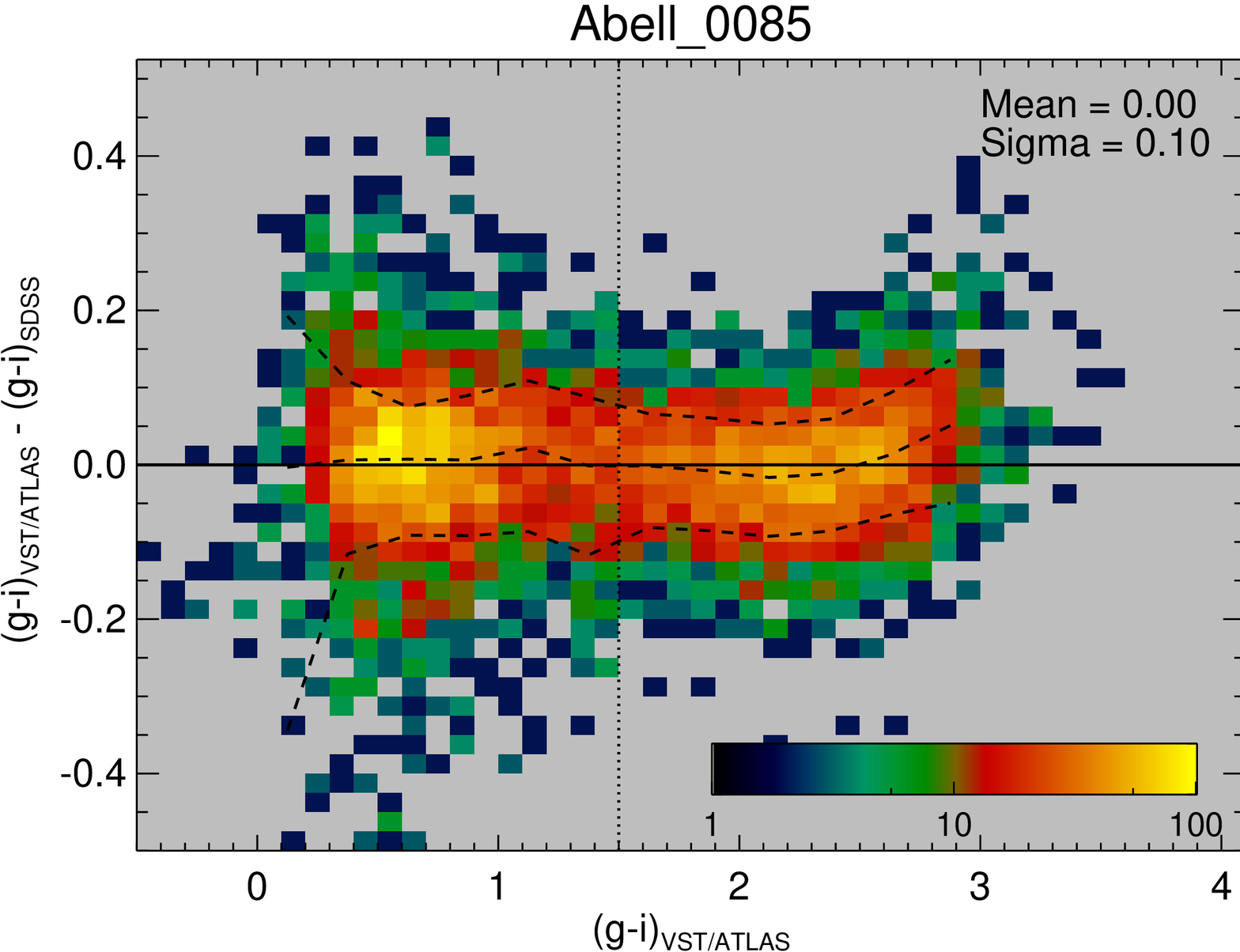}
\caption{The residual offset in $(g-i)_{\rm VST/ATLAS}$ colour after subtracting the \citet{covey2007}-defined stellar locus shown in Figure~\ref{gmr_rmi_cols} as a function of colour. The colour scale gives the number of stars in each bin, as indicated by the colour bar shown in the upper left corner of each panel. The black dashed lines show the median and $68$th percentiles as a function of $(g-i)_{\rm VST/ATLAS}$ colour. There are only very small offsets, consistent with zero, as a function of colour and the dispersion is generally less than 0.1 indicating that the VST/ATLAS-derived colours are not systematically different from those derived from SDSS photometry. The vertical dot-dashed line shows the upper limit for the $(g-i)_{\rm VST/ATLAS}$ colour of the cluster red-sequence.
\label{gmi_rmi_resids}}
\end{figure*}

We conducted a number of tests to ensure there are no systematic offsets between the SDSS and VST/ATLAS photometry which may bias the target selection. First, we make use of the well-known stellar locus produced by main-sequence stars in colour-colour space. In Figure~\ref{gmr_rmi_cols}, the four panels show the galactic-extinction-corrected $(g-r)$ versus $(r-i)$ diagrams for stars with COG-corrected aperture magnitudes $15 < r < 20$. Overplotted as thick black lines are the median stellar locus trends determined from high-quality SDSS stars presented by \citet{covey2007}. 
For all panels, the median SDSS stellar locus tracks very closely with the peak in the number density distribution. This is revealed in more detail in Figure~\ref{gmi_rmi_resids} where the residuals in $(g-i)_{\rm VST/ATLAS}$ colours after subtracting the \citet{covey2007} stellar loci for the SDSS stars are plotted as a function of VST/ATLAS colour. The difference is consistent with zero, with scatter generally less than 0.1mag. This indicates that there are no systematic offset in the colours measured for the VST/ATLAS stars when compared with the SDSS measurements.

\begin{figure*}
\includegraphics[angle=0,width=.45\textwidth]{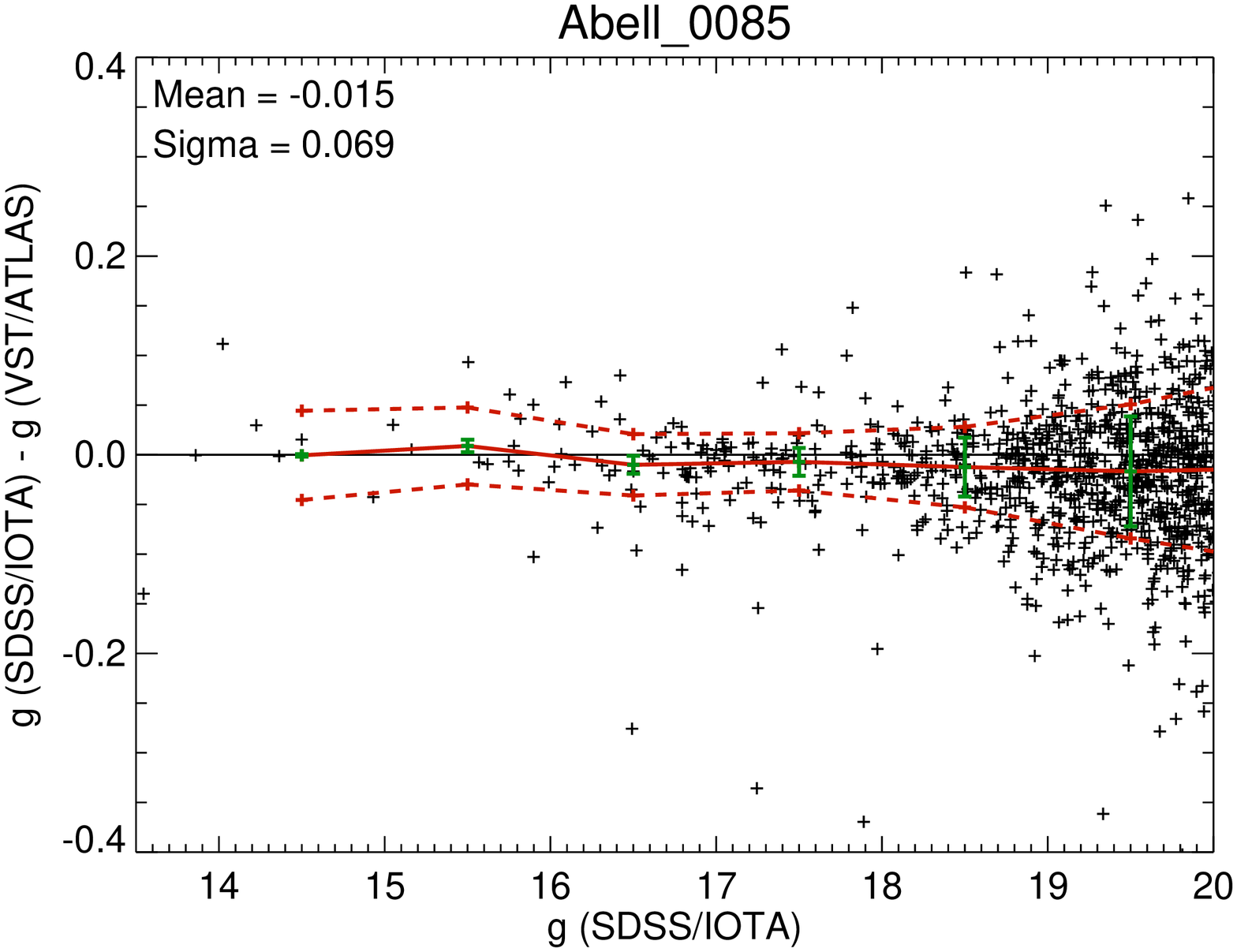}
\includegraphics[angle=0,width=.45\textwidth]{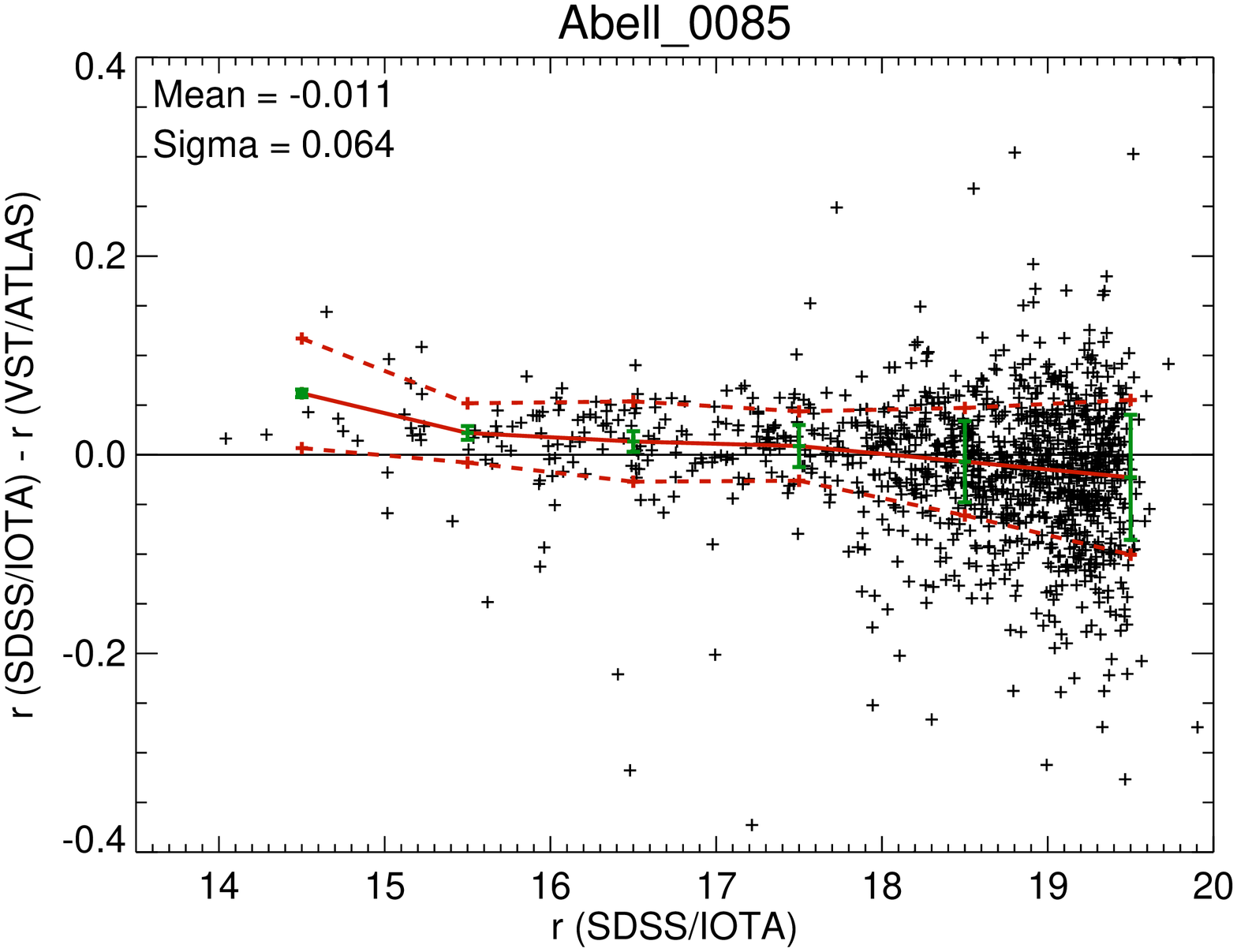}\\
\includegraphics[angle=0,width=.45\textwidth]{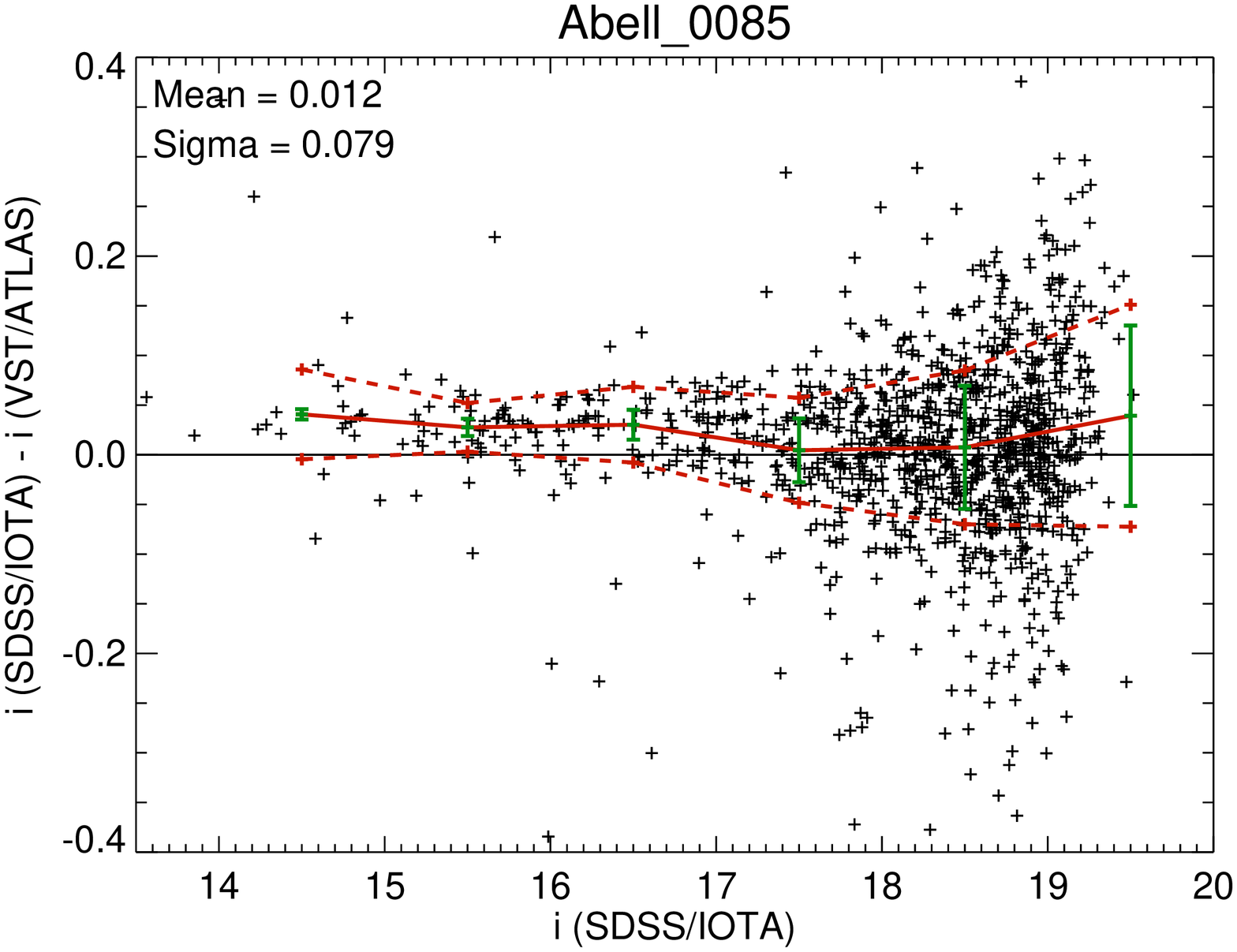}
\includegraphics[angle=0,width=.45\textwidth]{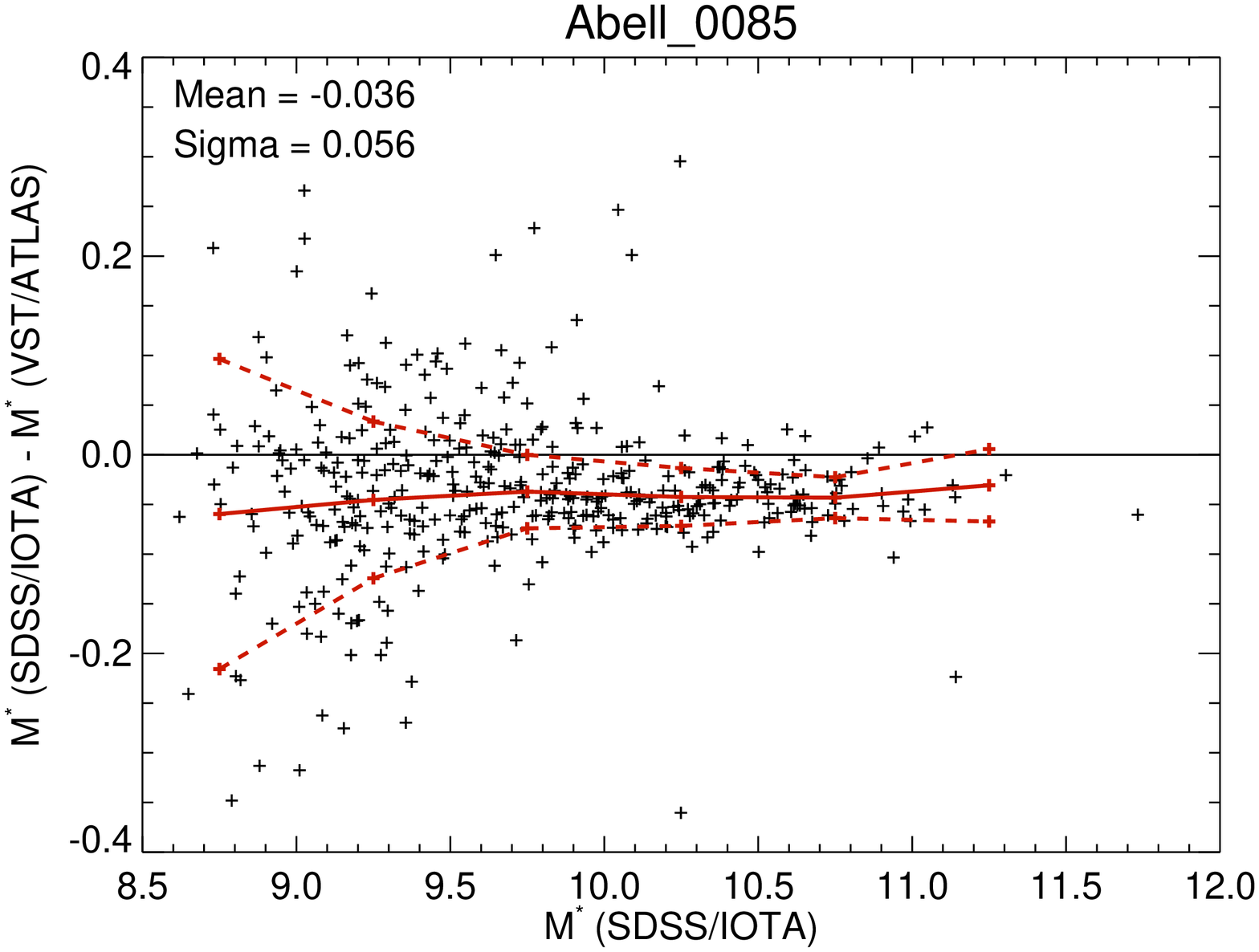}
\caption{The comparison between $gri$ magnitudes and stellar mass estimates for galaxies that had both SDSS/{\sf IOTA} and VST/ATLAS photometry. The top left, top right and bottom left plots show the distribution of SDSS/{\sf IOTA}-VST/ATLAS differences for the $g$, $r$ and $i$-bands, respectively, plotted against the VST/ATLAS measurement. The bottom right panel shows the difference in the $M^*_{\rm approx}$ estimates determined using Equation~\ref{mass_proxy}. The red solid and dashed lines in each plot show the median and scatter (as measured by the median absolute deviation) as a function of $g$, $r$ and $i$-band magnitude, as well as $M^*_{\rm approx}$. Only small differences exist between SDSS/{\sf IOTA} and VST/ATLAS-derived quantities. The green bars in the top left, top right and bottom left panels show the quadrature sum of the median uncertainty on the VST/ATLAS and SDSS measurements in each bin (centred on the median difference in that bin). For bright magnitudes, the quadrature sum of the uncertainties does not account for the scatter in the differences, indicating that systematics dominate there.
\label{IOTA_comp}}
\end{figure*}

The second test utilises the fact that A85 cluster is covered by the SDSS photometry and also partially covered by the VST/ATLAS survey. We use this overlap to directly assess the total magnitude estimates for the galaxy photometry from VST/ATLAS by comparing with those determined by the {\sf IOTA} software in Section~\ref{new_sdss_phot}, as well as the stellar mass proxies determined in Section~\ref{mstar}. These comparisons are shown in Figure~\ref{IOTA_comp} where it can be seen that the differences in $gri$-band Kron-like measurements from the aperture and PSF-matched catalogues differ by only small amounts, typically by less than 0.015mag. The $1\sigma$ scatter around the offsets is small ($\sim 0.06-0.08$\,mag), and increases at fainter magnitudes as shown by the dashed red lines in Figure~\ref{IOTA_comp}. The random uncertainties in the $gri$-bands for the VST/ATLAS (SDSS) measurements are $\Delta (g, r, i)= (0.003-0.04, 0.002-0.03, 0.003-0.03)$\,mag ($0.006-0.05, 0.003-0.05, 0.005-0.08$\,mag) over the magnitude ranges shown in Figure~\ref{IOTA_comp}. For magnitudes fainter than $\sim 17.5$, the quadrature sum of the VST/ATLAS and SDSS uncertainties, shown as green bars in Figure~\ref{IOTA_comp}, account for the majority of the scatter in the magnitude differences. For brighter magnitudes, the random uncertainties are around a factor of 5-10 too small to account for the scatter in the magnitude differences. At these brighter magnitudes,  other sources of uncertainty such as differences in aperture definition, background subtraction and zeropoint offsets dominate over the random uncertainties. Taking the quadrature difference between the $1\sigma$ scatter around the offsets and the random uncertainties indicates that these other sources of uncertainties amount to $0.03-0.04$\,mag. \citet{shanks2015} find that the uncertainty on the zeropoints for the VST/ATLAS survey are of order $0.01-0.02$\,mag when tied to APASS photometry. This indicates that aperture definition, background subtraction and other systematic uncertainties contribute of the order $0.02-0.04$\,mag to the overall uncertainty budget. There is a marginal trend in the $i$-band difference in the sense that the difference in magnitude increases to $0.03-0.04$mag for $i<18$ with the VST/ATLAS measurements being brighter. The stellar mass determinations differ by only 0.04\,dex, in the sense that the VST/ATLAS measurements are higher than the SDSS/IOTA measurements, and have a small scatter ($\sim 0.06\,$dex). These comparisons indicate that the systematic differences between the SDSS/IOTA and the VST/ATLAS photometry are small and will not significantly bias the selection of SAMI targets. 

\subsection{Stellar mass estimates}\label{mstar}

The spectroscopically confirmed SAMI targets from Section~\ref{redshift_cat} are matched to the aperture-matched photometric catalogues using a $3''$ matching radius. Stellar masses are then estimated using the empirical proxy described by \citet{taylor2011} which uses the fact that the $(g-i)$ colour in the galaxy frame of reference strongly correlates with the mass-to-$i$-band-light ratio, $M^*/L_i$. The stellar mass proxy, $M^*_{\rm approx}$, is the same as that used in \citet{bryant2015} for the GAMA portion of the SAMI-GS selection. The relationship with observed $g$- and $i$-band magnitudes is
\begin{equation}\label{mass_proxy}
\begin{split}
M^*_{approx} = -0.4i+2{\rm log}_{10}(D_L)-{\rm log}_{10}(1+z)+\\ 
(1.2177-0.5893z)+(0.7106-0.1467z)(g-i)
\end{split}
\end{equation} 
where the first three terms effectively transform the observed, extinction-corrected $i$-band magnitude into an absolute magnitude, while the final two terms account for the redshift evolution of the correlation due to the effects of the k-correction on the observed colour. Here, $(D_L)$ is the luminosity distance (for our assumed cosmology) and it is determined using the cluster redshift, which minimises the impact of the peculiar velocity due to the cluster velocity dispersion on the $M^*_{\rm approx}$ estimate. The  three terms that involve $z$ all correct for the impact of redshift on the observed magnitude/colour with respect to the intrinsic value and, therefore, use the galaxy redshift. The $g$- and $i$-band are the aperture-matched magnitudes determined in Section~\ref{new_photo} and are corrected for Galactic extinction using the \citet{schlegel1998} dust maps. This empirical approximation produces very precise stellar mass estimates when compared with full spectral energy distribution fitting, with a $1\sigma$ error of only 0.1\,dex \citep[][]{taylor2011}.

\subsection{Cluster target selection for the SAMI-GS}\label{SAMI_TS}
\begin{figure}
\includegraphics[angle=0,width=.475\textwidth]{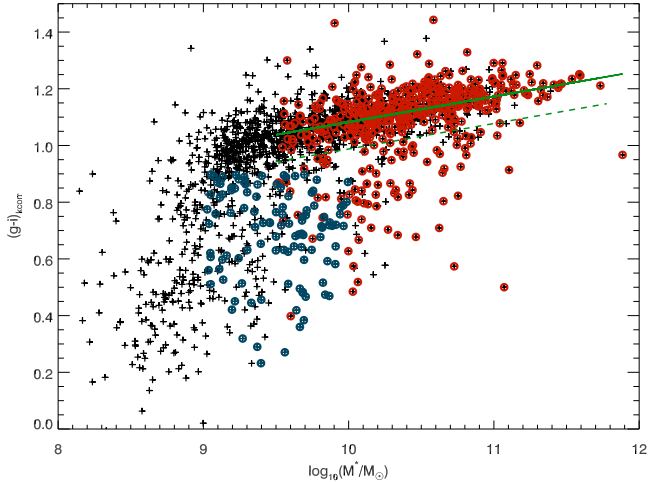}
\caption{The k-corrected $(g-i)$ colours versus log$_{10} ({\rm M}^*_{\rm approx}/$\msolar$)$ for the different SAMI target priorities. The primary targets are shown as red circles and blue fillers as blue circles. Large-radius-fillers are omitted for clarity. The black crosses show the distribution of all spectroscopically confirmed members. The green line shows the best fitting linear relation to the red-sequence for the primary SAMI-GS cluster targets. The dashed green line shows the dividing line used to separate red-sequence and blue-cloud galaxies. It is defined as being $1.6\sigma_{RS}$ below the best fitting line, where $\sigma_{RS}$ is the scatter in the red-sequence around the best fit.
\label{gmi_sel}}
\end{figure}

The primary SAMI-GS targets in the GAMA regions are selected from a series of redshift bins with an increasing stellar mass limit in higher redshift bins \citep{bryant2015}. This strategy was developed to cover a relatively broad stellar mass range as required for the science drivers for the survey, to match the target density to the SAMI instrument's multiplexing capabilities ($\sim 15$ targets per square degree), and to reach a completeness level of $\sim 90\%$ within the time allocated to the survey. A similar strategy is employed for the cluster galaxies, however, there are several additional constraints that must be considered. First, at the cluster redshifts, the limiting stellar mass used in the GAMA SAMI-GS regions produce a target density that is much higher than the $\sim 15$ targets per square degree and will require many revisits to achieve a $90\%$ completeness limit for the eight clusters. Second, while the low mass $M^* < 10^{9.5}$\,\msolar\, galaxy population in the lower-density GAMA regions is primarily composed of blue, star-forming galaxies with emission lines \citep{taylor2015}, the low-redshift cluster population at $M^* < 10^{9.5}$\,\msolar\, is dominated by red-sequence galaxies (see top left panel in Figure~\ref{gmi_sel}). These lower mass, red-sequence galaxies have small projected sizes and are faint; they have $r_{\rm petro} > 19.0$ for the clusters with $z_{clus} >0.04$. These galaxies are unlikely to produce high S/N SAMI data with a useful fraction of independently resolved spatial elements. Furthermore, as outlined in Section~\ref{spec_comp}, the SAMI-CRS begins to suffer from spectroscopic incompleteness in several clusters for these low masses (see Figure~\ref{spectro_completeness_gmi}). Third, the projected sizes of the clusters are large on the sky; the \rtwo\, values are comparable to the SAMI FOV (compare the vertical dashed and red lines in Figure~\ref{mem_allocation}). Therefore, for a galaxy to be considered as a primary target candidate, it must satisfy the following criteria based on clustercentric distance, peculiar velocity and stellar mass:
\begin{itemize}
\item $|v_{\rm pec}| < 3.5\, \sigma_{200}$\,\kms\, \& $R<\,$\rtwo with $v_{\rm pec}$ measured with respect to $z_{\rm clus}$,
\item for $z_{\rm clus} < 0.045$: $M^*_{\rm approx} > 10^{9.5}$\,\msolar,
\item for $0.045 < z_{\rm clus} < 0.06$: $M^*_{\rm approx} > 10^{10}$\,\msolar.
\end{itemize}

Table~\ref{sami_target_numbers} lists the number of primary targets based on these criteria for each cluster. The cut in $v_{\rm pec}$ is shown as a blue dashed line in Figure~\ref{mem_allocation} and is less conservative than the velocity cut used to allocate cluster membership as described in Section~\ref{subsect:memsel}. This less-conservative cut allows for the uncertainty in allocating cluster membership that, despite our best efforts, may be biased against extreme cases such as high-velocity infalling galaxies. In addition to the primry targets, we include a number of filler targets for cases where twelve primary targets cannot be allocated on a plate due to, e.g., targets being closer than the collision radius of the SAMI hexabundles (228\arcsec). We define two types of filler targets; blue filler and large-radius filler targets. The fillers must meet the following criteria:
\begin{itemize}
\item Blue filler: $M^*_{\rm lim}-0.5  < M^*_{\rm approx} < M^*_{\rm lim}$ \& $(g-i)_{\rm kcorr} < 0.9$ \& $R <\, $\rtwo.
\item Large radius filler: $M^*_{\rm approx} > M^*_{\rm lim}$ \& $R > \,$\rtwo.
\end{itemize}
The first selection criterion allows for blue galaxies which are 0.5\,dex less massive than our primary target selection limits to be included, thereby increasing the target density. Here, $(g-i)_{\rm kcorr}$ is the k-corrected colour where the k-corrections are determined using the {\sf calc\_kcor} code\footnote{http://kcor.sai.msu.ru/getthecode/} from \citet{chilingarian2010}. The second criterion allows for objects at slightly larger clustercentric distance which become necessary when the tiling software described in \citet{bryant2015} selects a plate centre that is offset from the cluster centre by large enough that part of the plate area lies beyond \rtwo. The targets are visually inspected and classified based on the criteria outlined in Table~5 of \citet{bryant2015}. The number of primary, blue- and large-radius-filler targets per cluster are outlined in Table~\ref{sami_target_numbers}.  The top left panel of Figure~\ref{gmi_sel} shows the colour-mass diagram where primary targets are shown as red circles, blue fillers as blue circles, and non-SAMI-targeted members as black crosses. 

\begin{table*}
 \centering
  \caption{Primary targets selected for the cluster portion of the SAMI-GS. Also listed are blue filler and large radius filler targets, along with the number of targets observed to date (as of 2015B). The final column lists the completeness of the primary targets observed as of semester 2015B. \label{sami_target_numbers}}
  \begin{tabular}{@{}lcccccc@{}}
  \hline
   Cluster     & $z_{\rm clus}$ & Primary  & Blue  & $R>$\rtwo\,  & Observed  & Completeness\\
   	           &                &          & Filler&Filler        & to date   & percent \\
   \hline
APMCC 917  & 0.0509 &  29 & 9 & 15 & 22/2/3 & 76\\
Abell 168  & 0.0449 &  113 & 17 & 98 & 52/0/8 & 46\\
Abell 4038  & 0.0293 &  111 & 10 & 143 & 87/1/8 & 78\\
EDCC 442 & 0.0498 &  50 & 10 & 40 & 41/3/3 & 82\\
Abell 3880 & 0.0578 &  56 & 11 & 93 &31/0/4 & 55\\
Abell 2399 & 0.0580 & 94 & 10 & 78 & 70/1/9 & 75\\
Abell 119 & 0.0442 &  259 & 23 & 220 & 107/0/0 & 41\\
Abell  85 & 0.0549 & 171 & 20 & 59 & 82/0/0 & 48\\
\hline
Total & -- & 883 & 110 & 746 & 492/7/35 & 56\\
\hline
\end{tabular}
\end{table*}

\section{Survey progress and completeness}\label{survey_progress}

As for the GAMA portion of the SAMI-GS, the aim for the cluster regions is to reach $90\%$ completeness for the primary targets. However, the input catalogues for the SAMI-GS cluster targets were not finalised in the early stages of the survey for two reasons. First, the SAMI-CRS observations were performed after the first cluster SAMI-GS run in September 2013. Second, improvements continued to be made to the VST/ATLAS photometry during the first two years of the survey. The impact of this was twofold. Initially, redshifts and therefore confirmed cluster membership was only available for bright, high stellar mass galaxies. This meant that only those galaxies could be targeted for the SAMI-GS in the first round of observations. Second, the uncertain photometry meant that a hard cut at the lower limits in stellar mass outlined in Section~\ref{SAMI_TS} was not possible. 
To mitigate the impact of this, the priorities for targets in the first two years of the survey were adjusted so that higher stellar mass objects were preferentially selected for observing. The combination of the early incomplete redshift information and later the uncertain photometry meant that selection of targets for SAMI-GS observation were biased to higher mass objects in the early phases of the survey. This may introduce significant biases into the final catalogue if care is not taken to track the completeness as a function of various parameters.  

\begin{figure}
\includegraphics[angle=0,width=.475\textwidth]{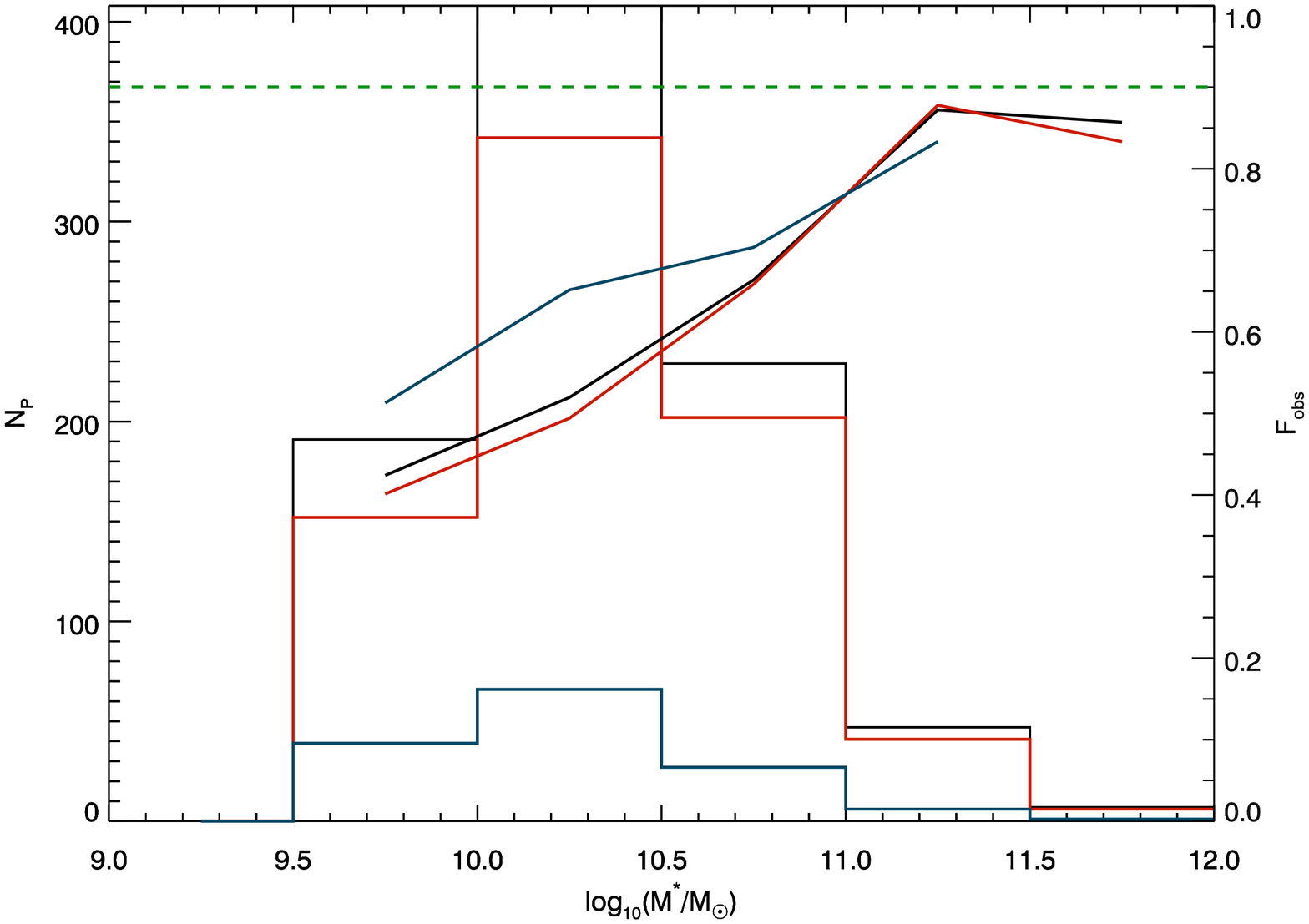}\\
\includegraphics[angle=0,width=.475\textwidth]{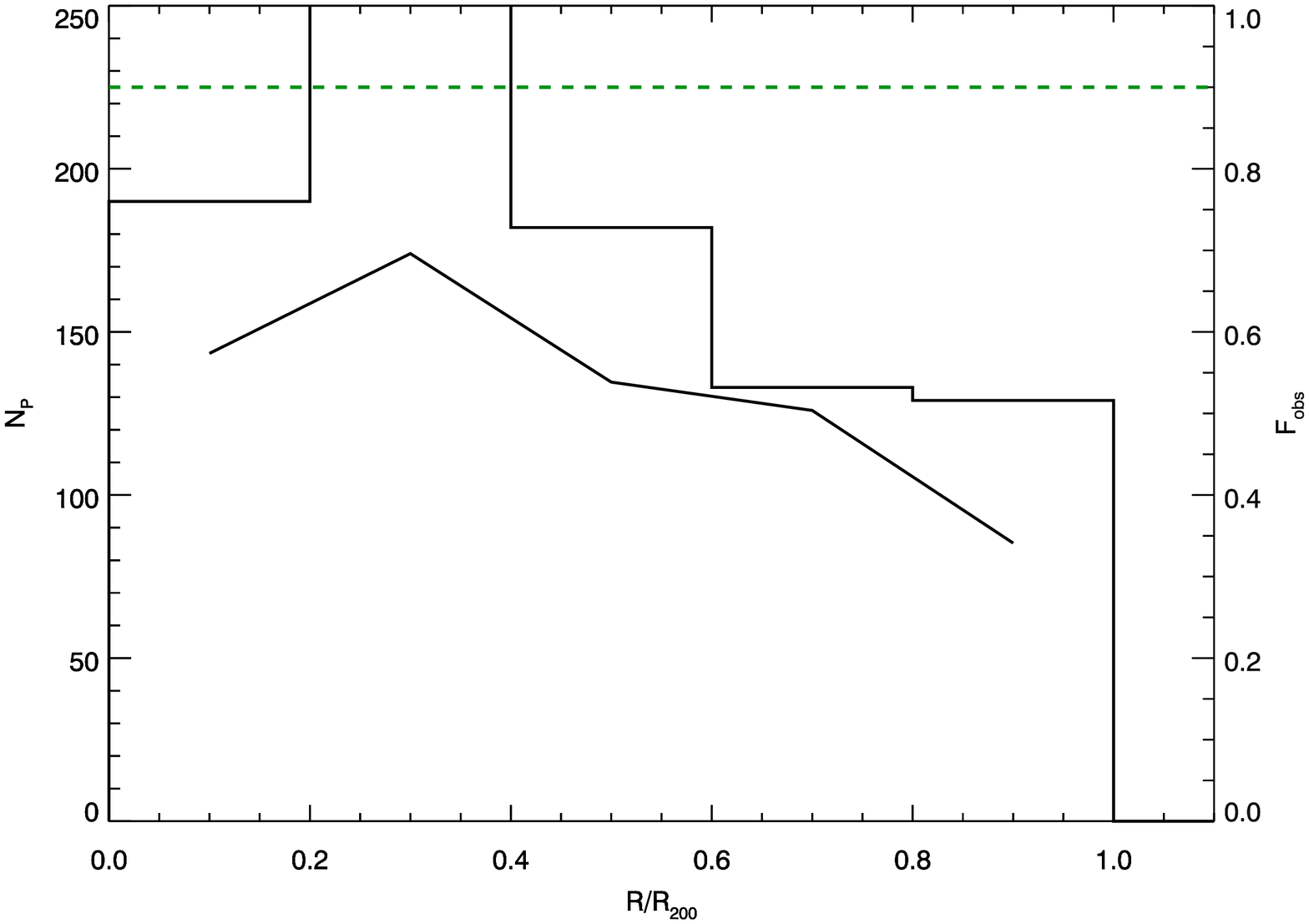}\\
\includegraphics[angle=0,width=.475\textwidth]{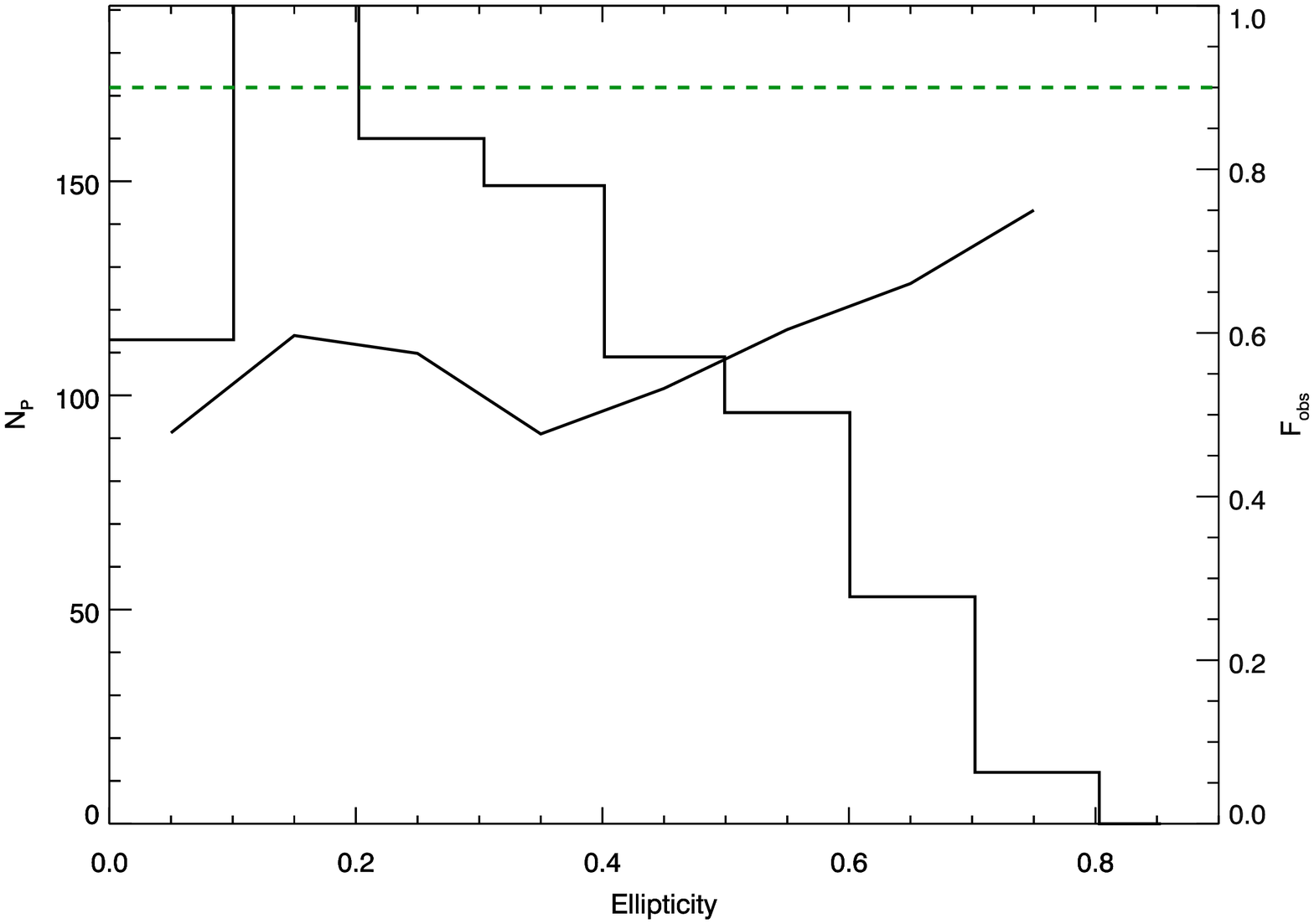}
\caption{
Completeness of current SAMI-GS observations. The top panel shows the distribution of log$_{10} ({\rm M}^*_{\rm approx}/$\msolar$)$ for primary target galaxies for the cluster portion of the SAMI-GS. The black line shows the distribution for all galaxies, the red and blue lines show the distribution for galaxies divided into red sequence and blue cloud as defined by their $(g-i)$ colour (Figure~\ref{gmi_sel}). The middle and bottom panels show the distribution of primary targets in clustercentric distance (${\rm R}/{\rm R}_{200}$) and galaxy ellipticity (coming from the SExtractor shape parameters determined during the photometric measurements in Section~\ref{new_photo}). In each panel the completeness, ${\rm F}_{\rm obs}$, determined as the ratio of observed-to-total primary targets, is plotted as a solid line where the right axis shows the relevant scale. The green dashed line shows the goal completeness for the primary targets (${\rm F}_{\rm obs} = 90\%$).
\label{SAMI_compl}}
\end{figure}

We check for potential biases in Figure~\ref{SAMI_compl} where we explore the distribution, ${\rm N}_{\rm P}$, and survey completeness, ${\rm F}_{\rm obs}$, for primary targets as a function of log$_{10} ({\rm M}^*_{\rm approx}/$\msolar$)$ (top panel), clustercentric distance (middle panel) and galaxy ellipticity (bottom panel; as measured by SExtractor during the photometric measurements outline in Section~\ref{new_photo}). The completeness is defined as the ratio of observed-to-total primary target galaxies at the end of semester 2015B. The completeness of primary targets for each cluster is listed in Table~\ref{sami_target_numbers}. We also explore the completeness as a function of log$_{10} ({\rm M}^*_{\rm approx}/$\msolar$)$ for primary SAMI-GS targets split in to red sequence and blue cloud galaxies based on their $(g-i)$ colour. The boundary for separating the red sequence and blue cloud galaxies is shown as a dashed green line in Figure~\ref{gmi_sel}, and is defined as being 1.6$\sigma_{\rm RS}$ below the best-fitting line to the red sequence galaxies (using the method outlined in Section~\ref{aaomega_obs}). Figure~\ref{SAMI_compl} shows that the observations to date are indeed biased towards galaxies with higher log$_{10} ({\rm M}^*_{\rm approx}/$\msolar$)$ with ${\rm F}_{\rm obs}$ rising from $43\%$ at log$_{10} ({\rm M}^*_{\rm approx}/$\msolar$) = 9.5$ to $85\%$ at log$_{10} ({\rm M}^*_{\rm approx}/$\msolar$) =11.5$. A similar rising trend is seen for both the red sequence and blue cloud galaxies. The ${\rm F}_{\rm obs}$ trends with ellipticity is relatively flatter, while the trend with ${\rm R}/{\rm R}_{200}$ indicates higher completeness levels closer to the cluster centres. The survey will continue until at least 2018 and by completion we aim to reach the ${\rm F}_{\rm obs}=90\%$ in each log$_{10} ({\rm M}^*_{\rm approx}/$\msolar$)$ and ${\rm R}/{\rm R}_{200}$ bin.

\section{Summary}

We have presented the selection and characterisation of eight clusters that are included in the SAMI-GS in addition to the GAMA regions. These additional regions allow the SAMI-GS to probe the full range of galaxy environments, from the low density field and groups to the high density cluster cores. In Section~\ref{sami_crs}, we provide details of a redshift survey performed for each of the eight clusters, the SAMI-CRS. In  Section~\ref{cluster_properties}, we use the data from the SAMI-CRS to characterise a number of cluster properties. The main results for Sections~\ref{sami_crs} and \ref{cluster_properties} are:

\begin{itemize}
\item We have obtained very high spectroscopic completeness (always $\ge 85\%$, $94\%$ on average) out to large clustercentric distances ($R = 2$\rtwo) and to a limiting $r$-band magnitude $r=19.4$. This allows for a large number of spectroscopically confirmed cluster members, ranging from $119-772$ members within 2\rtwo. 
\item The large number of cluster members allow very good measures of velocity dispersion, which lie in the range $492 - 1002$\kms, and virial mass measurements, which are in the range $14.24 \leq {\rm log(M_{200}/}$\msolar)$\leq 15.19$. 
\item We use the positions and velocities of the confirmed cluster members to investigate the structure of the clusters. The clusters APMCC0917, EDCC0442, A3880, and A4038 do not show strong evidence for the existence of substructure. The clusters A168 and A119 show marginal evidence for substructure while A85 and A2399 both show significant evidence for the existence of multiple substructures.
\end{itemize} 

In Section~\ref{sami_targets}, we describe the target selection for the cluster portion of the SAMI-GS. We detail our procedure for measuring aperture- and PSF-matched photometry on both the SDSS and the VST/ATLAS images. The results of Section~\ref{sami_targets} can be summarised as follows:
\begin{itemize}
\item We confirm the veracity of our VST/ATLAS photometry by both comparing the duplicated measurements of stars in the overlap regions of the different images. We find that the median difference in the duplicated $gri$ measurements is always less than 0.016 mag, with dispersion $\sim 0.05$. 
\item We compare the stellar locus of our VST/ATLAS photometry to the median SDSS stellar locus of \citet{covey2007}. We find differences consistent with zero and with 0.1dex dispersion, indicating no systematic offset in the $(g-i)$ and $(r-i)$ colours of stars in our VST/ATLAS photometry when compared with the SDSS.
\item We use the duplicate measurements of galaxies in A85, which has full SDSS and partial VST/ATLAS coverage, to show that any systematic differences between the photometric and stellar mass measurements between the two surveys are likely to be less than 0.05dex. 
\item We use our updated photometry, along with the results of the SAMI-CRS to select targets for the SAMI-GS. Our sample consists of 883 primary targets.
\end{itemize}

In Section~\ref{survey_progress}, we report on the survey progress. As of the end of 2015, 492 primary targets have been observed, and the completeness is 56\%. We present the SAMI-GS completeness for the cluster regions as a function of log$_{10} ({\rm M}^*_{\rm approx}/$\msolar$)$, clustercentric distance, and galaxy ellipticity. The analysis indicates that we have higher completeness for more massive galaxies closer to the cluster cores. Future observations will aim to reach a 90\% completeness level that is homogeneous across these three distributions.

 The SAMI-CRS provides a rich dataset for characterising the environment of the SAMI-GS cluster galaxies. Future papers will exploit this dataset in combination with the resolved spectroscopy provided by the SAMI-GS to investigate the
kinematics-morphology-density relation (Brough {\it et al. in prep}), the impact of environment on gas kinematics (Bryant {\it et al. in prep.}), the impact of merging substructure on galaxy star-formation (Owers {\it et al. in prep}), the environments of galaxies with evidence for recent truncation of star-formation (Owers {\it et al. in prep.}), the stellar ages and metallicities as a function of environment (Scott {\it et al. in prep}). The redshifts and SAMI-GS cluster input catalogues will be made public in a forthcoming data release, which is scheduled to occur in mid-2018.

\section*{Acknowledgments}

We thank Chris Lidman for useful discussions on reducing the AAOmega data. 
MSO, SB and SMC acknowledge the funding support from the Australian Research Council through a Future Fellowship (FT140100255, FT140101166, FT100100457, respectively). JTA acknowledges the award of a SIEF John Stocker Fellowship. GVK, JPM, JdJ, GS, KK and EMH acknowledge financial support from the Netherlands Research School for Astronomy (NOVA) and Target. Target is supported by Samenwerkingsverband Noord Nederland, European fund for regional development, Dutch Ministry of economic affairs, Pieken in de Delta, Provinces of Groningen and Drenthe. JdJ and EH are supported by NWO grant 614.061.610. This work was supported by the UK Science and Technology Facilities Council through the `Astrophysics at Oxford' grant ST/K00106X/1. RLD acknowledges travel and computer grants from Christ Church, Oxford and support from the Oxford Centre for Astrophysical Surveys which is funded by the Hintze Family Charitable Foundation. Support for AMM is provided by NASA through Hubble Fellowship grant \#HST-HF2-51377 awarded by the Space Telescope Science Institute, which is operated by the Association of Universities for Research in Astronomy, Inc., for NASA, under contract NAS5-26555. NS acknowledges support of a University of Sydney Postdoctoral Research Fellowship

The SAMI Galaxy Survey is based on observations made at the Anglo-Australian Telescope. The Sydney-AAO Multi-object Integral field spectrograph (SAMI) was developed jointly by the University of Sydney and the Australian Astronomical Observatory. The SAMI input catalogue is based on data taken from the Sloan Digital Sky Survey, the GAMA Survey and the VST ATLAS Survey. The SAMI Galaxy Survey is funded by the Australian Research Council Centre of Excellence for All-sky Astrophysics (CAASTRO), through project number CE110001020, and other participating institutions. The SAMI Galaxy Survey website is http://sami-survey.org/.

GAMA is a joint European-Australasian project based around a spectroscopic campaign using the Anglo-Australian Telescope.
The GAMA input catalogue is based on data taken from the Sloan Digital Sky Survey and the UKIRT Infrared Deep Sky Survey. Complementary
imaging of the GAMA regions is being obtained by a number of independent survey programmes including GALEX MIS,
VST KiDS, VISTA VIKING, WISE, Herschel-ATLAS, GMRT and ASKAP providing UV to radio coverage. GAMA is funded by the
STFC (UK), the ARC (Australia), the AAO, and the participating institutions. The GAMA website is: http://www.gama-survey.org/.

Funding for SDSS-III has been provided by the Alfred P. Sloan Foundation, the Participating Institutions, the National Science Foundation, and the U.S. Department of Energy Office of Science. The SDSS-III web site is http://www.sdss3.org/. SDSS-III is managed by the Astrophysical Research Consortium for the Participating Institutions of the SDSS-III Collaboration including the University of Arizona, the Brazilian Participation Group, Brookhaven National Laboratory, Carnegie Mellon University, University of Florida, the French Participation Group, the German Participation Group, Harvard University, the Instituto de Astrofisica de Canarias, the Michigan State/Notre Dame/JINA Participation Group, Johns Hopkins University, Lawrence Berkeley National Laboratory, Max Planck Institute for Astrophysics, Max Planck Institute for Extraterrestrial Physics, New Mexico State University, New York University, Ohio State University, Pennsylvania State University, University of Portsmouth, Princeton University, the Spanish Participation Group, University of Tokyo, University of Utah, Vanderbilt University, University of Virginia, University of Washington, and Yale University. Based on data products (VST/ATLAS) from observations made with ESO Telescopes at the La Silla Paranal Observatory under program ID 177.A-3011(A-J). This research has made use of the APASS database, located at the AAVSO web site. Funding for APASS has been provided by the Robert Martin Ayers Sciences Fund.

\noindent $^{1}$Department of Physics and Astronomy, Macquarie University, NSW, 2109, Australia\\
$^{2}$The Australian Astronomical Observatory, PO Box 915, North Ryde, NSW, 1670, Australia\\
$^{3}$Sydney Institute for Astronomy, School of Physics, University of Sydney, NSW 2006, Australia\\
$^{4}$ARC Centre of Excellence for All-sky Astrophysics (CAASTRO)\\
$^{5}$Astrophysics Research Institute, Liverpool John Moores University UK\\
$^{6}$Deptartment of Physics and Astronomy, University of North Carolina, Chapel Hill, NC 27599 USA\\
$^{7}$International Centre for Radio Astronomy Research, University of Western Australia Stirling Highway, Crawley, WA, 6009, Australia\\
$^{8}$Leiden Observatory, Leiden University, Niels Bohrweg 2, NL-2333 CA Leiden, the Netherlands\\
$^{9}$Indian Institute of Science Education and Research Mohali-IISERM, Knowledge City, Sector 81, Manauli PO 140306, India\\
$^{10}$Kapteyn Astronomical Institute, University of Groningen, PO Box 800, NL-9700 AV Groningen, the Netherlands\\
$^{11}$Research School for Astronomy \& Astrophysics, Australian National University Canberra, ACT 2611, Australia\\ 
$^{12}$Centre for Astrophysics and Supercomputing, Swinburne University of Technology, Hawthorn 3122, Australia\\
$^{13}$Astrophysics, Department of Physics, University of Oxford, Denys Wilkinson Building, Keble Rd., Oxford, OX1 3RH, UK.\\
$^{14}$School of Mathematics and Physics, University of Queensland, QLD 4072, Australia
$^{15}$Envizi Suite 213, National Innovation Centre, Australian Technology Park, 4 Cornwallis Street, Eveleigh, NSW, 2015, Australia\\
$^{16}$Cahill Center for Astronomy and Astrophysics California Institute of Technology, MS 249-17 Pasadena, CA 91125, USA\\
$^{17}$Hubble Fellow\\
$^{18}$Physics Department, University of Durham, South Road, Durham DH1 3LE, UK\\
$^{19}$School of Physics, University of Melbourne, Parkville, 3010 Australia













\bsp	
\label{lastpage}
\end{document}